\documentclass[]{article}

\usepackage[]{authblk}

\usepackage{cite}
\usepackage[margin=1.0in]{geometry}
\usepackage{amsmath}
\usepackage{amssymb}
\usepackage{braket}
\usepackage{tikz}
\usepackage{subcaption}

\usepackage{setspace}
\usepackage{abstract}

\begin{document}

\title{Richardson-Gaudin States}
\author{Paul Andrew Johnson\thanks{paul.johnson@chm.ulaval.ca}}
\affil{D\'{e}partement de chimie, Universit\'{e} Laval, Qu\'{e}bec, Qu\'{e}bec, G1V 0A6, Canada}

\date{}
\maketitle

\begin{abstract}
This chapter gives an overview of Richardson-Gaudin states which represent weakly correlated pairs of electrons. They are parametrized by sets of numbers obtained from non-linear equations. The best method to solve these equations is presented in a straightforward manner with enough detail to implement the method computationally. Optimal expressions for the density matrix elements are discussed. A simple description of 1-dimensional hydrogen chains is presented in terms of Richardson-Gaudin states. Richardson-Gaudin states are placed in the larger context of integrable models and geminal wavefunctions of quantum chemistry.
\end{abstract}

\section{Introduction}
Weakly correlated systems in quantum chemistry are well understood: the one-electron effects dominate the two-electron effects. The qualitative behavior is summarized in a mean-field model which provides a complete set of states to work with. Usually the mean-field is Hartree-Fock (HF) and the states are Slater determinants. Provided that HF is a reasonable starting point, the exact wavefunction will be monopolized by one Slater determinant with small contributions from weakly excited Slater determinants. The power of having a mean-field model is that further approximations can be made in a hierarchical manner giving model chemistries.

In strongly correlated systems the two-electron effects dominate, or are at least compete with, the one-electron effects. Hartree-Fock as a mean-field is not a qualitative description of the system and Slater determinants are a \emph{poor} basis for the exact wavefunction. \emph{Many} Slater determinants are required and it is often difficult to choose one as a starting point. Traditional methods for strong correlation, such as the complete active space self-consistent field (CASSCF), are powerful but can only handle small systems. Larger systems may be treated with state of the algorithms such as Slater determinant Monte Carlo (SDMC),\cite{booth:2009} configuration interaction using a perturbative selection made iteratively (CIPSI),\cite{huron:1973} semistochastic heat-bath configuration interaction (SHCI)\cite{sharma:2017} and the density matrix renormalization group (DMRG).\cite{chan:2002} These methods target the wavefunction directly without reference to a mean-field model.

Slater determinants can be classified by their number of unpaired electrons, their \emph{seniority}. Seniority is not preserved by orbital rotations, nor is it a symmetry of the molecular Hamiltonian. However, there is always a set of orbitals that minimize the energy of the configuration interaction (CI) of seniority-zero Slater determinants, the so-called doubly occupied configuration interaction (DOCI). In ref.\cite{bytautas:2011} it was shown that for the dissociation of N$_2$, orbital-optimized (OO-)DOCI is almost perfectly parallel to the full CI (FCI) dissociation curve. Adding Slater determinants with seniority two gave a substantial benefit, and including seniorities zero, two, and four gave quantitative agreement with full CI. As Scuseria says, \emph{strong correlation manifests in the seniority-zero channel}: OO-DOCI captures the qualitative behavior.\cite{bytautas:2015} Since DOCI costs the square root of full CI, we should aim for feasible approximations to DOCI before adding the higher seniorities.

DOCI is well approximated by wavefunctions built from pairs, \emph{geminals}.\cite{limacher:2013,stein:2014} This is not a new idea,\cite{fock:1950,silver:1969,silver:1970a,silver:1970b,silver:1970c,malrieu:2007} they are particular examples of McWeeny's group functions.\cite{mcweeny:1959,mcweeny:1960,mcweeny:1961,mcweeny:1963} These treatments again target the wavefunction without a mean-field model. The reduced Bardeen-Cooper-Schrieffer (BCS)\cite{bardeen:1957a,bardeen:1957b} Hamiltonian describes a mean-field of pairs of electrons. We have employed its eigenvectors, the Richardson\cite{richardson:1963,richardson:1964,richardson:1965}-Gaudin\cite{gaudin:1976} (RG) states, for strongly correlated problems in quantum chemistry. A single RG state agrees with DOCI for mean-field cost,\cite{fecteau:2022} and systematic improvement is possible.\cite{johnson:2023} RG states are the subject of this chapter.

In section \ref{sec:bcs} we discuss the reduced BCS Hamiltonian and its eigenvectors the RG states. It is shown that RG states are products of pairs of electrons parametrized by a set of numbers satisfying non-linear equations. In section \ref{sec:richardson} we describe step by step a fast and robust procedure to solve the required non-linear equations. Section \ref{sec:rdm} presents the density matrix elements of RG states which are required to evaluate the energy of the molecular Hamiltonian. In section \ref{sec:1dh} we present the symmetric dissociation of 1-dimensional chains of hydrogen atoms, a prototypical example of strong correlation, in terms of RG states. The description is clean and straightforward. Finally, section \ref{sec:context} places RG states in the larger context of integrable models and the traditional geminal wavefunctions of quantum chemistry.

\section{Reduced BCS Hamiltonian and RG States} \label{sec:bcs}
\subsection{Conventions}
Without exception, there are $N$ spatial orbitals that are labelled with indices $i,j,k,l$, and $M$ pairs labelled with indices $a,b$. Second-quantized operators are labelled with spatial and spin indices, e.g. $a^{\dagger}_{i\uparrow}$ creates an up-spin electron in spatial orbital $i$. These objects have the structure
\begin{align}
[a^{\dagger}_{i\sigma},a_{j,\tau}]_+ = \delta_{ij} \delta_{\sigma\tau}.
\end{align}
We use the number operator
\begin{align}
\hat{n}_i = a^{\dagger}_{i\uparrow} a_{i\uparrow} + a^{\dagger}_{i\downarrow} a_{i\downarrow}
\end{align}
which counts the number of electrons in a given spatial orbital. 

Pairs of electrons have the structure of su(2), and their representations are equivalent to the spin-$\frac{1}{2}$ representation. In a particular spatial orbital $i$, there are three objects
\begin{align} \label{eq:pair_su2}
S^+_i = a^{\dagger}_{i\uparrow}a^{\dagger}_{i\downarrow}, 
\quad S^-_i = a_{i\downarrow}a_{i\uparrow},
\quad S^z_i = \frac{1}{2}\left( a^{\dagger}_{i\uparrow}a_{i\uparrow} + a^{\dagger}_{i\downarrow}a_{i\downarrow} -1 \right).
\end{align}
A pair is created by $S^+_i$ and removed by $S^-_i$. $S^z_i$ effectively counts the number of pairs in a spatial orbital: its action on an empty orbital is $-\frac{1}{2}$ while on a doubly-occupied orbital it gives $\frac{1}{2}$. It is not difficult to verify that $\hat{n}_i = 2 S^z_i + 1$ and
\begin{align} \label{eq:su2_structure}
[S^+_i, S^-_j] &= 2\delta_{ij} S^z_i \\
[S^z_i, S^{\pm}_j] &= \pm \delta_{ij} S^{\pm}_i.
\end{align}
We denote the vacuum as $\ket{\theta}$. Usually it is the empty state, but could contain unpaired electrons, or more general states that do not participate in the pairing. The strict requirement is that for each $i$ 
\begin{align}
S^-_i \ket{\theta} &= 0.
\end{align}
Restricted HF (RHF) Slater determinants are defined by which orbitals $\{i\}$ are doubly-occupied
\begin{align}
\ket{\{i\}} = S^+_{i_1} S^+_{i_2} \dots S^+_{i_M} \ket{\theta}.
\end{align}
Unrestricted HF (UHF) Slater determinants are similarly written, with the unpaired electrons absorbed into the vacuum, e.g.
\begin{align}
\ket{\{i\},j\sigma k\tau l\mu} = S^+_{i_1} S^+_{i_2} \dots S^+_{i_M} a^{\dagger}_{j\sigma}a^{\dagger}_{k\tau}a^{\dagger}_{l\mu} \ket{\theta}
\end{align}
can be thought of an RHF Slater determinant on top of the vacuum
\begin{align}
\ket{\tilde{\theta}} = a^{\dagger}_{j\sigma}a^{\dagger}_{k\tau}a^{\dagger}_{l\mu} \ket{\theta},
\end{align}
since $\ket{\tilde{\theta}}$ is destroyed by all $S^-_i$. Notice also that 
\begin{align}
S^+_j \ket{\tilde{\theta}} &= S^-_j \ket{\tilde{\theta}} = S^z_j \ket{\tilde{\theta}} = 0 \\
S^+_k \ket{\tilde{\theta}} &= S^-_k \ket{\tilde{\theta}} = S^z_k \ket{\tilde{\theta}} = 0 \\
S^+_l \ket{\tilde{\theta}} &= S^-_l \ket{\tilde{\theta}} = S^z_l \ket{\tilde{\theta}} = 0
\end{align} 
so that the orbitals $j,k,l$ are \emph{blocked} from the pairing effects. Including unpaired electrons is therefore possible, though for the remainder of the chapter we'll assume there are no unpaired electrons to simplify the discussion.

The usual spin representation of su(2) has the three operators
\begin{align} \label{eq:spin_su2}
\tilde{S}^+_i = a^{\dagger}_{i\uparrow}a_{i\downarrow}, 
\quad \tilde{S}^-_i = a^{\dagger}_{i\downarrow}a_{i\uparrow},
\quad \tilde{S}^z_i = \frac{1}{2}\left( a^{\dagger}_{i\uparrow}a_{i\uparrow} - a^{\dagger}_{i\downarrow}a_{i\downarrow} \right)
\end{align}
which have the same structure \eqref{eq:su2_structure}, and the corresponding vacuum is ferromagnetic with all spins aligned down. RG states are understood in the pair representation \eqref{eq:pair_su2} though the entire construction can be repeated in the spin representation \eqref{eq:spin_su2} with only minor modifications. 

\subsection{Limiting cases}
RG states are the eigenvectors of the reduced BCS Hamiltonian
\begin{align} \label{eq:H_bcs}
\hat{H}_{BCS} = \frac{1}{2} \sum^N_{k=1} \varepsilon_k \hat{n}_k -\frac{g}{2} \sum^N_{k,l=1} S^+_k S^-_l
\end{align}
in which there is competition between filling the lowest single-particle levels $\{\varepsilon\}$ and a constant-strength pairing interaction $g$. Before defining RG states, it's instructive to look at particular limits of the Hamiltonian \eqref{eq:H_bcs}. First, when $g=0$, there is no interaction and the Hamiltonian is diagonal in a basis of Slater determinants labelled by which orbitals are doubly-occupied. 

As the interaction $g$ becomes large, the single-particle effects are negligible
\begin{align} \label{eq:large_g_BCS}
\lim_{g \rightarrow \infty} \hat{H}_{BCS} = -\frac{g}{2}\sum^N_{k,l=1} S^+_k S^-_l
\end{align}
and there are distinct behaviors depending on whether the interaction is attractive or repulsive. In the attractive limit ($g>0$), the ground state of \eqref{eq:large_g_BCS} is non-degenerate and it is not difficult to show that it is the repeated action of \emph{collective} pairs
\begin{align} \label{eq:collective_pair}
c^+ = \sum^N_{i=1} S^+_i,
\end{align}
which is known as extreme antisymmetrized geminal power (AGP)
\begin{align} \label{eq:eAGP}
\ket{\text{eAGP}} = (c^+)^M \ket{\theta}.
\end{align}
This state is extreme in the sense that all the weights in the sum \eqref{eq:collective_pair} are the same. It is impossible to label this state based on its occupations as each spatial orbital is equally occupied. The ground state eigenvalue scales with $g$. Variationally approximating the energy of \eqref{eq:H_bcs} with RHF leads to number-symmetry breaking (BCS instability) and projecting the resulting BCS wavefunction back onto good particle number yields eAGP. 

In the repulsive limit, the ground state of \eqref{eq:large_g_BCS} is \emph{heavily} degenerate, though again it is not difficult to show that it is a product of \emph{bonding} pairs
\begin{align} \label{eq:bonding_pairs}
b^+_{ij} = S^+_i - S^+_j
\end{align}
which is the generalized valence-bond / perfect pairing (GVB-PP) wavefunction in the dissociated limit. Trying to label the state based on occupations is again not a great idea. This state is heavily degenerate as there are many ways of forming $M$ bonding pairs \eqref{eq:bonding_pairs} among $N$ spatial orbitals. Worse, a naive listing would be linearly dependent. These problems have general solutions but this is not the present purpose. The ground state eigenvalue is independent of $g$. Variationally approximating the energy of \eqref{eq:H_bcs} with RHF leads to spin-symmetry breaking (UHF instability) and projecting the UHF wavefunction back onto good spin projection yields a product of bonding pairs.

\begin{figure} [ht!]
	\begin{subfigure}{\textwidth}
		\centering
		\includegraphics[width=0.49\textwidth]{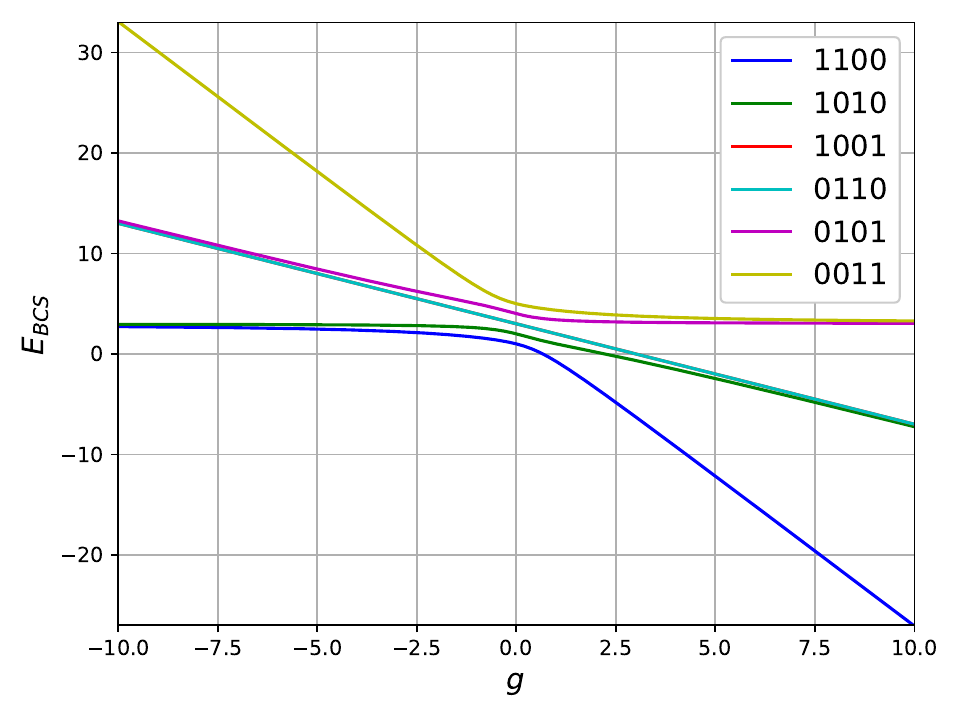} \hfill
	\end{subfigure}
	\caption{Energies of 4-site 2-pair picket-fence model.}
	\label{fig:42_bcs}
\end{figure}

\begin{figure} [ht!]
	\begin{subfigure}{\textwidth}
		\centering
		\includegraphics[width=0.32\textwidth]{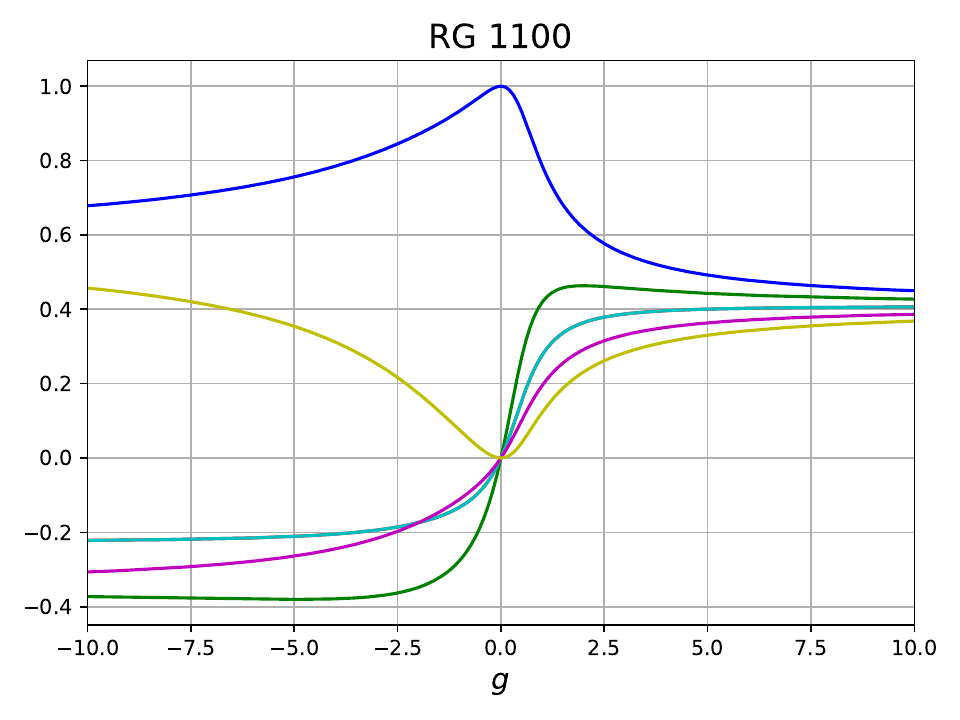} \hfill
		\includegraphics[width=0.32\textwidth]{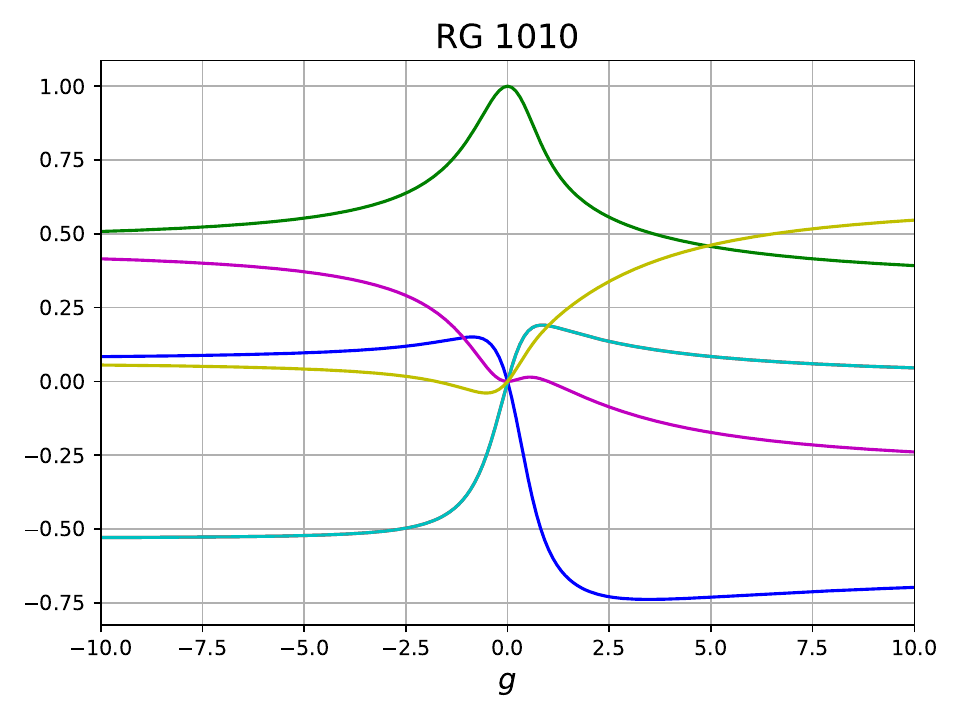} \hfill
		\includegraphics[width=0.32\textwidth]{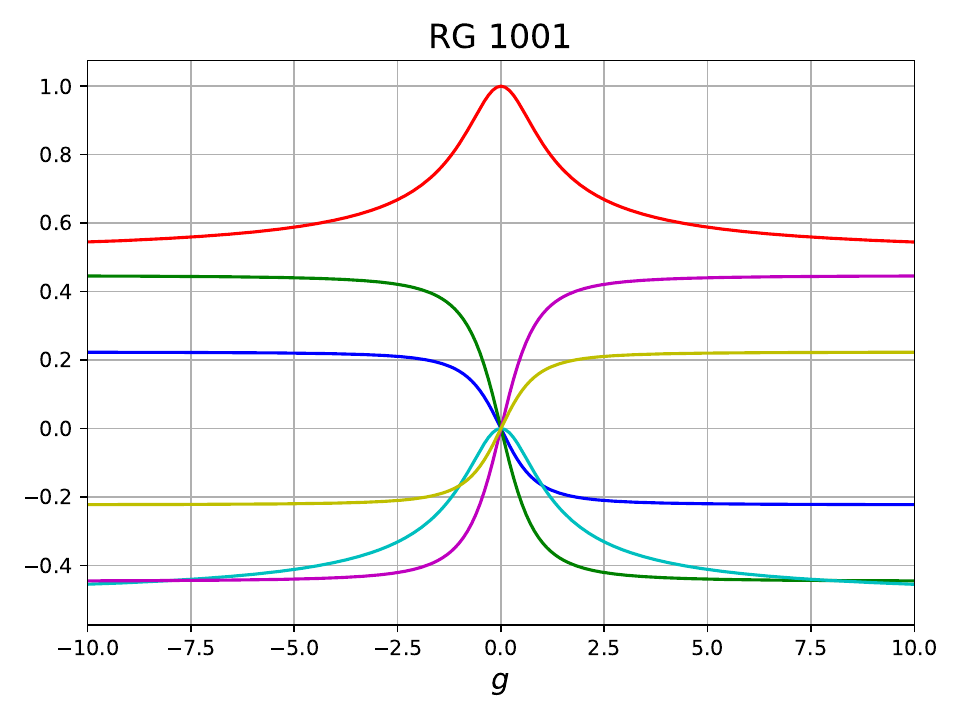} \\
		\includegraphics[width=0.32\textwidth]{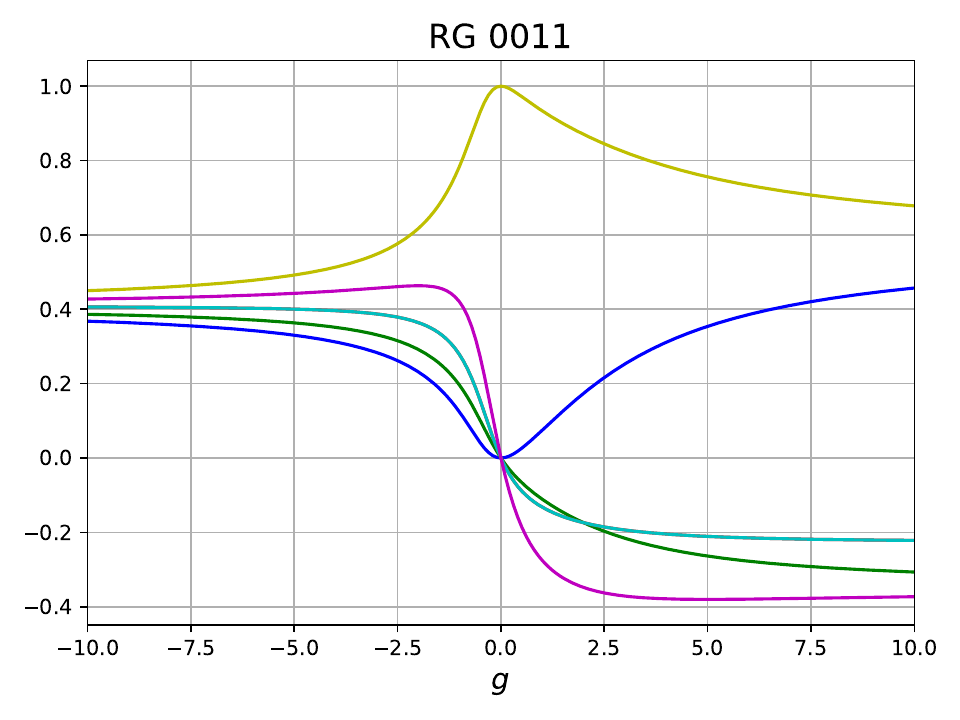} \hfill
		\includegraphics[width=0.32\textwidth]{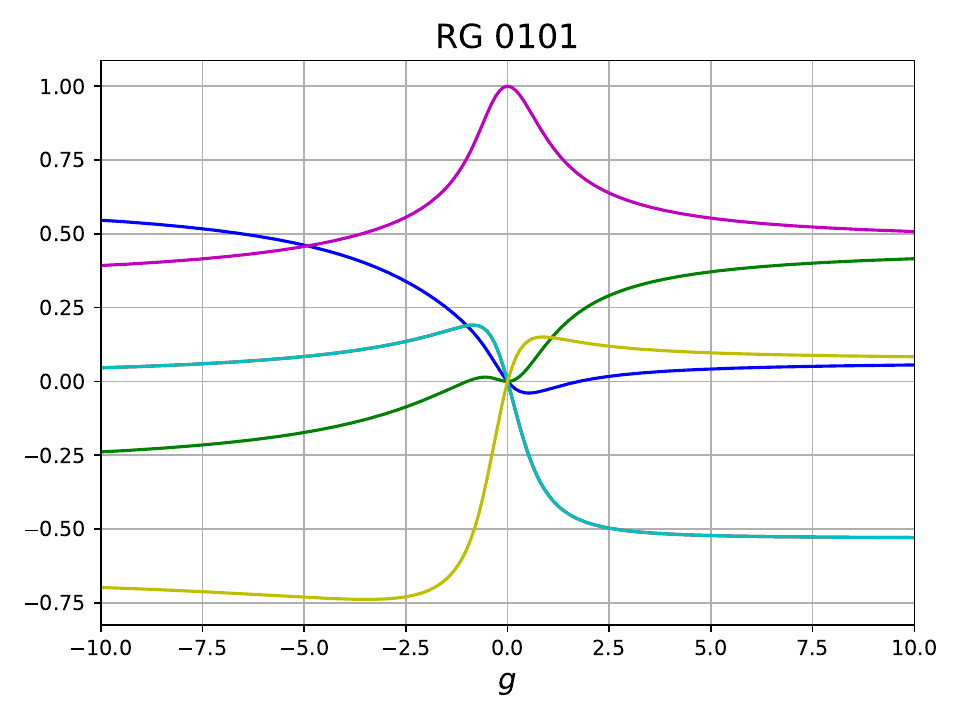} \hfill
		\includegraphics[width=0.32\textwidth]{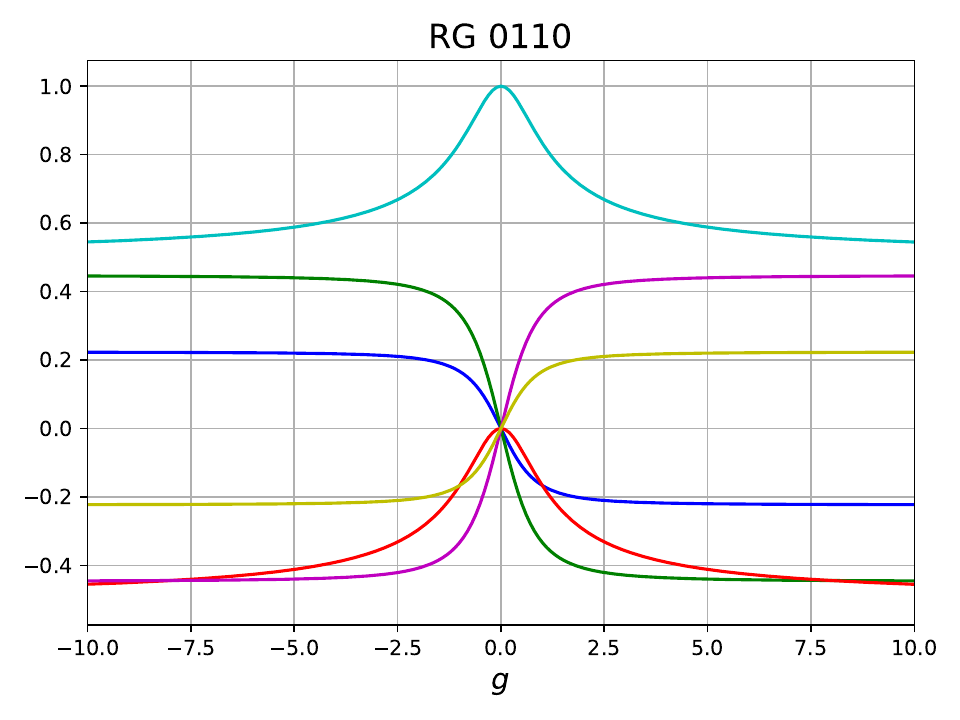} \\
		\includegraphics[scale=0.5]{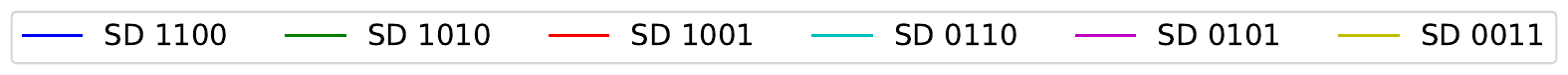} \hfill
	\end{subfigure}
	\caption{Coefficients of Slater determinants contributing to each RG state for a 4-site 2-pair picket fence model.}
	\label{fig:42_sd_bcs}
\end{figure}

Now, let's look at a small system and plot the eigenvalues as a function of $g$. In particular, we consider the \emph{picket-fence} (PF) model
\begin{align} \label{eq:h_pf}
\hat{H}_{PF} = \frac{1}{2} \sum^N_{k=1} (k-1) \Delta \varepsilon \hat{n}_k -\frac{g}{2} \sum^N_{k,l=1} S^+_k S^-_l
\end{align}
which is a reduced BCS Hamiltonian with evenly-spaced $\{\varepsilon\}$. While PF models are simplified reduced BCS Hamiltonians, the behavior of the eigenvectors and eigenvalues is indicative.  Figure \ref{fig:42_bcs} shows the eigenvalues of a 4-site, 2-pair PF model with $\Delta \varepsilon=1$ as a function of the interaction $g$.
There are two important points to highlight. First, the eigenvalues evolve uniquely from $g=0$. We can thus identify each of the RG states based on the RHF Slater determinant they connect to at $g=0$. These Slater determinants have no unpaired electrons and thus can be labelled as bitstrings with 1 (0) labelling a doubly occupied (empty) spatial orbital. RG states are labelled in the same way based on their $g=0$ Slater determinant, but as soon as $g\neq 0$, RG states are \emph{not} Slater determinants. To emphasize this point, figure \ref{fig:42_sd_bcs} shows the RG states for the 4-site 2-pair PF model expanded in Slater determinants.
The Slater determinant coefficients are continuous in each case: they change rapidly near $g=0$, then quickly stabilise to asymptotic values. Extending the domain of figure \ref{fig:42_sd_bcs} to $[-100,100]$ reveals no new information. The second important point from figure \ref{fig:42_bcs} is that the graph has a rotational symmetry about the average eigenvalue at $g=0$. There is a direct correspondence between states in attractive ($g>0$) and repulsive ($g<0$) regimes: an RG state with bitstring $\beta$ at $g$ is equivalent to the RG state at $-g$ with $\beta$ \emph{read backwards}. For example, 1100 at $g$ is equivalent to 0011 at $-g$. Notice that is not a particle/hole type duality as the RG state 1001 at $g$ is the same as 1001 at $-g$. The same is true for the RG state 0110. While these two RG states are everywhere degenerate, we will see later that they are distinct.

In general it is possible to label each RG state based on its corresponding RHF Slater determinant at $g=0$. The ground state of a reduced BCS Hamiltonian is \emph{always} the state of $M$ 1s followed by $N-M$ 0s, while the highest excited state is \emph{always} the state of $N-M$ 0s followed by $M$ 1s. Other states can and do cross, but it remains possible to label them uniquely. 
\begin{figure} [ht!]
	\begin{subfigure}{\textwidth}
		\centering
		\includegraphics[width=0.49\textwidth]{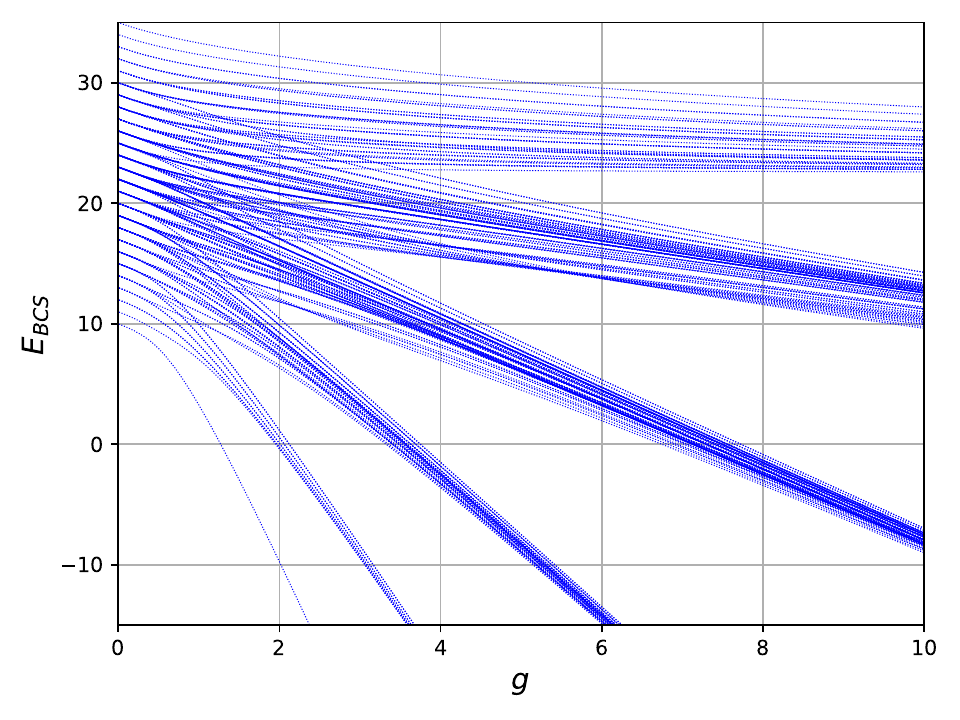} \hfill
		\includegraphics[width=0.49\textwidth]{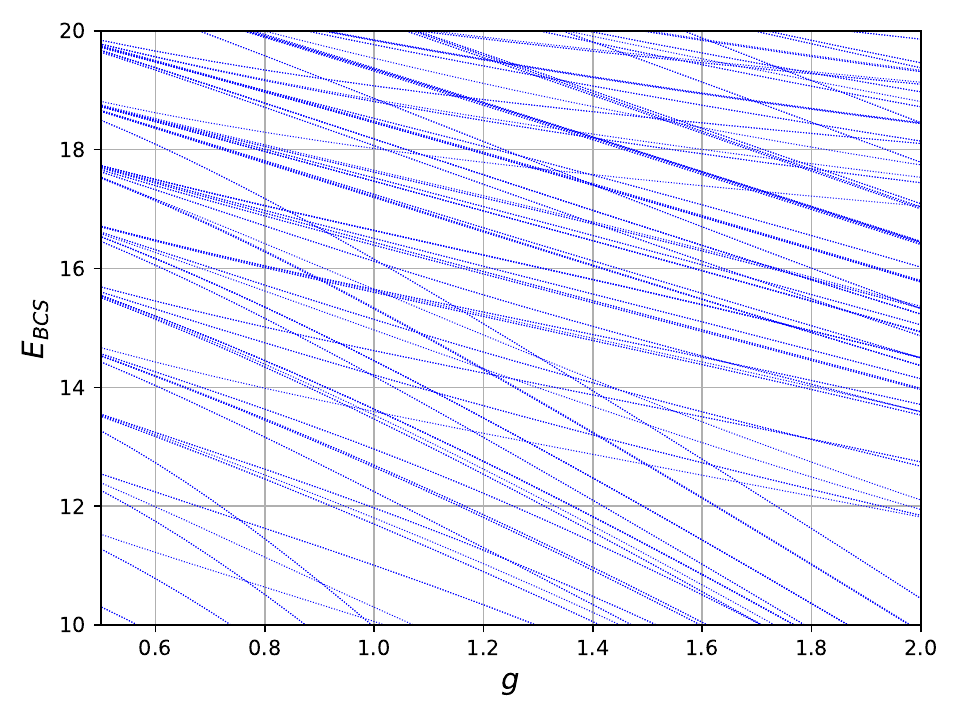}
	\end{subfigure}
	\caption{Energies of 10-site 5-pair picket-fence model.}
	\label{fig:105_bcs}
\end{figure}
A larger example of 10 sites with 5 pairs is shown in figure \ref{fig:105_bcs}. Only the attractive region is shown, as the repulsive region is obtained by rotating the graph. It is clear that at $g=0$ the energies are discrete. Since the $\{\varepsilon\}$ are evenly-spaced, there are many degeneracies at $g=0$. The large $g$ limit is reached quickly, with clear collections of degenerate states. The ground state and the highest excited state are isolated from the other states, though the highest excited state eventually becomes part of a degenerate group. It is also clear that there are \emph{many} state crossings. However, in each case it is possible to follow the states unambiguously. Ordinarily, the states would repel each other energetically due to the Wigner-von Neumann non-crossing rule and the spectrum would be a complicated mess of avoided crossings. As we will see in section \ref{sec:conserved_charges}, there is a complete set of commuting operators from which we build the reduced BCS Hamiltonian. As a result, RG states all possess different symmetries with respect to these commuting quantities and there no avoided crossings. Such behavior is a fingerprint of an integrable system.\cite{yuzbashyan:2002,yuzbashyan:2003,yuzbashyan:2005}

\subsection{Algebraic Bethe Ansatz}
Now that we have a basic idea of RG states, let's define and compute them. The construction is a specific limit of the algebraic Bethe Ansatz (ABA) for the inhomogeneous 1D Heisenberg model.\cite{bethe:1931,takhtadzhan:1979,faddeev:1980,faddeev:1980b,links:2003} RG states are products of pairs
\begin{align} \label{eq:rg_pairs}
S^+ (u) = \sum^N_{i=1} \frac{S^+_i}{u-\varepsilon_i}
\end{align}
that depend on complex numbers $\{u\}$ called \emph{rapidities}. The object \eqref{eq:rg_pairs} places a pair in each spatial orbital with a coefficient weighted by the difference of the rapidity and the corresponding single particle energy. Rapidities are not free variables: in order to be eigenvectors of \eqref{eq:H_bcs}, they satisfy non-linear equations to be solved numerically. To demonstrate this, consider the product of pairs
\begin{align} \label{eq:rg_state}
\ket{\{u\}} = S^+(u_1) S^+(u_2) \dots S^+(u_M) \ket{\theta}.
\end{align}
With the first and second commutators
\begin{align}
[\hat{H}_{BCS}, S^+(u_1)] &= u_1 S^+(u_1) - \sum^N_{i=1} S^+_i + g \sum^N_{i,j=1} \frac{S^+_i S^z_j}{u_1 - \varepsilon_j} \\
[[\hat{H}_{BCS},S^+(u_1)],S^+(u_2)] &= \frac{g}{u_2 - u_1}\sum^N_{i=1} S^+_i (S^+(u_1) - S^+(u_2)),
\end{align}
$\hat{H}_{BCS}$ is moved past each RG pair until it destroys the vacuum. The action of $\hat{H}_{BCS}$ on an RG state with arbitrary rapidities $\{u\}$ is thus
\begin{align}
\hat{H}_{BCS} \ket{\{u\}} &= \sum^M_{a=1} \prod^M_{b (\neq a)=1} S^+(u_b) [\hat{H}_{BCS},S^+(u_a)]\ket{\theta}
+ \sum^M_{a=1}\sum^M_{b=a+1} [[\hat{H}_{BCS},S^+(u_a)],S^+(u_b)] \prod^M_{c (\neq a,b)=1} S^+(u_c) \ket{\theta} \\
&= \sum^M_{a=1} u_a \ket{\{u\}} - \frac{g}{2} \sum^N_{i=1} S^+_i \sum^M_{a=1} \lambda_a (\{u\},\{\varepsilon\},g) \prod^M_{b (\neq a)=1} S^+(u_b) \ket{\theta}.
\end{align}
In the last line, there is a single term proportional to $\ket{\{u\}}$,
\begin{align}
E_{BCS} = \sum^M_{a=1} u_a
\end{align}
and a set of linearly independent undesired terms with coefficients $\lambda_a$ that are functions of the rapidities, the single particle energies and the pairing strength. Forcing these coefficients to vanish makes $\ket{\{u\}}$ an eigenvector of \eqref{eq:H_bcs} and amounts to solving the non-linear equations
\begin{align} \label{eq:rich}
\lambda_a (\{u\},\{\varepsilon\},g) = \frac{2}{g} + \sum^N_{i=1} \frac{1}{u_a - \varepsilon_i} + \sum^M_{b (\neq a)=1} \frac{2}{u_b - u_a} = 0, \qquad \forall a = 1,\dots M
\end{align}
which are called \emph{Richardson's equations}. Henceforth, the set $\{v\}$ will refer to a solution of Richardson's equations while the set $\{u\}$ remains arbitrary. An RG state $\ket{\{v\}}$ and its rapidities $\{v\}$ are referred to as \emph{on-shell} while the arbitrary RG state $\ket{\{u\}}$ and its rapidities $\{u\}$ are referred to as \emph{off-shell}. 

Richardson's equations are coupled, and thus must be solved simultaneously. A solution is a set of $\{v\}$ that satisfy the complete set of Richardson's equations. How many solutions are there? As written it's not obvious, but it is known that there are $\binom{N}{M}$ distinct solutions,\cite{richardson:1965} which coincides with the dimension of the Hilbert space of $M$ pairs among $N$ spatial orbitals. Each RG state is defined by one of the solutions of Richardson's equations. This means that the RG pairs defining one RG state are distinct from the RG pairs defining a different RG state, and thus we cannot write excitations between RG states as simple actions of second-quantized operators on a reference RG state. This is however only a notational inconvenience.

Now, do we need to solve Richardson's equations? Unfortunately we do, as the expressions for the 1- and 2-body reduced density matrices (RDM) and transition density matrices (TDM) require the RG state to be on-shell. Otherwise, computing the density matrices is completely intractable. In the next section we'll describe the best method for solving Richardson's equations. The solution is neither difficult nor expensive.

\section{Solving Richardson's Equations} \label{sec:richardson}
There are many approaches to solve Richardson's equations, but they have problems near critical points where rapidities want to coincide with single-particle energies.\cite{rombouts:2004,guan:2012,pogosov:2012,debaerdemacker:2012,claeys:2015} We won't discuss methods based on finding rapidities directly. The correct way to solve these equations is the method of Faribault and co-workers.\cite{faribault:2011,elaraby:2012}

\subsection{Eigenvalue-Based Variables}
We define the \emph{eigenvalue-based variables} (EBV) as
\begin{align} \label{eq:ebv}
U_i &= \sum^M_{a=1} \frac{g}{\varepsilon_i - u_a} \\
V_i &= \sum^M_{a=1} \frac{g}{\varepsilon_i - v_a}.
\end{align}
Richardson's equations for the rapidities are equivalent to a set of non-linear equations for the EBV
\begin{align} \label{eq:ebv_eq}
U^2_i - 2U_i -g \sum^N_{k (\neq i)=1} \frac{U_k - U_i}{\varepsilon_k - \varepsilon_i} = 0, \quad \forall i=1,\dots,N.
\end{align}
Richardson's equations are a set of $M$ coupled non-linear equations, while the EBV equations \eqref{eq:ebv_eq} are a set of $N$ coupled non-linear equations. The variables to be obtained no longer appear in denominators, and the resulting equations are much more stable numerically. Adding Richardson's equations together, one obtains that the sum of the EBV is a constant, in particular
\begin{align} \label{eq:ebv_norm}
\sum^N_{i=1} U_i = 2M
\end{align}
it is the number of electrons. It is absolutely necessary to \emph{enforce} equation \eqref{eq:ebv_norm} to keep the solution in a given sector of $M$. Given a set of $\{\varepsilon\}$ and $g$ we thus have $N+1$ equations defining the $N$ variables $\{U\}$. 

We saw in section \ref{sec:bcs} that the RG states are continuous with $g$ and are uniquely labelled based on the corresponding Slater determinant at $g=0$. The basic strategy is to solve the problem at $g=0$ then evolve the solution to any desired $g$. First, at $g=0$, the EBV equations \eqref{eq:ebv_eq} decouple,
\begin{align}
U_i (U_i -2) &= 0 \\
\sum^N_{i=1} U_i &= 2M
\end{align}
whose solutions are a set of $M$ EBV equal to 2, and the remaining $N-M$ EBV equal to zero. Again, at $g=0$, the RG states are Slater determinants and thus the non-zero EBV correspond to the spatial orbitals which are doubly-occupied. Given a bitstring, we choose which EBV to set equal to 2 at $g=0$. 

To evolve the EBV equations to the desired $g$, we proceed by taking small steps and solving the equations \eqref{eq:ebv_eq} using the EBV at the current $g$. A straightforward approach would be to use Newton-Raphson at each step, but this is incredibly wasteful: the step-size must be small enough that Newton-Raphson doesn't cause any problems. We can take much larger steps if we instead expand the EBV as a Taylor series in $g$ to obtain a guess at the new $g+\delta g$ before solving with Newton-Raphson. Derivatives of the EBV with respect to $g$ can be calculated from the EBV equations. In particular, they are the solutions to an over-determined set of linear equations
\begin{align}
A \frac{\partial \textbf{U}}{\partial g} = \textbf{r}_1
\end{align}
where $A$ is the $(N+1)\times N$ matrix
\begin{align} \label{eq:ebv_jac}
A_{ij} &= \begin{cases}
2U_i - 2 + \sum^N_{k (\neq i) =1} \frac{g}{\varepsilon_k - \varepsilon_i}, &\quad i = j \\
-\frac{g}{\varepsilon_j - \varepsilon_i}, &\quad i \neq j
\end{cases}, \quad  &1 \leq i,j \leq N, \nonumber \\
A_{N+1,j} &= 1
\end{align}
and the right-hand side (RHS) is
\begin{align}
(\textbf{r}_1)_i &= \sum^N_{k (\neq i)=1} \frac{U_k - U_i}{\varepsilon_k - \varepsilon_i}, \quad 1 \leq i \leq N \nonumber \\
(\textbf{r}_1)_{N+1} &= 0. 
\end{align}
Higher derivatives are computed in the same manner \emph{and all share the same matrix} $A$. The $p$th-order derivatives are solutions of the linear equations
\begin{align}
A \frac{\partial^p \textbf{U}}{\partial g^p} = \textbf{r}_p
\end{align}
with the $p$th-order RHS
\begin{align}
(\textbf{r}_p)_i &= p \sum^N_{k (\neq i) =1} \frac{\frac{\partial^{p-1}U_k}{\partial g^{p-1}}  - \frac{\partial^{p-1}U_i}{\partial g^{p-1}}}{\varepsilon_k - \varepsilon_i}
- \sum^{p-1}_{m=1} \binom{p}{m} \frac{\partial^m U_i }{\partial g}\frac{\partial^{p-m}U_i}{\partial g^{p-m}}, \quad 1 \leq i \leq N \nonumber \\
(\textbf{r}_p)_{N+1} &= 0. 
\end{align}
Since $A$ is rectangular, it is optimal to employ QR-factorization such that $A=QR$ and solve the linear equations
\begin{align}
R\frac{\partial^p \textbf{U}}{\partial g^p} = Q^{\dagger} \textbf{r}_p
\end{align}
by backward substitution. In principle, the QR-factorization gives the least-squares solution to a set of over-determined linear equations, but in this particular case we know that there is always a genuine solution. As $A$ is common for all the derivatives, a single QR-factorization is necessary to compute the Taylor series. Usually we compute the Taylor expansion to 4th-order though a systematic study has not been done to optimize the order.

Now, the Taylor-series gives an approximation for the EBV at $g+\delta g$ that we can use to solve with Newton-Raphson. The Jacobian of the EBV equations is again $A$, which we QR-factor $A=QR$ to solve
\begin{align} \label{eq:newton-raphson}
R  (\textbf{U}_{n+1} - \textbf{U}_n) = - Q^{\dagger} \textbf{f}(\textbf{U}_n)
\end{align}
for the update to the EBV $(\textbf{U}_{n+1} - \textbf{U}_n)$. Newton-Raphson  usually converges these equations in very few iterations. In \eqref{eq:newton-raphson}, $\textbf{f}$ is the EBV equations \eqref{eq:ebv_eq} and the sum rule \eqref{eq:ebv_norm} evaluated at the $n$th iteration
\begin{align}
f_i (\textbf{U}_n) &= 
\begin{cases}
U^2_{i,n} - 2U_{i,n} -g \sum^N_{k (\neq i) =1} \frac{U_{k,n} - U_{i,n}}{\varepsilon_k - \varepsilon_i} , &1 \leq i \leq N  \\
\sum^N_{k=1} U_{k,n} - 2M , &i = N+1.
\end{cases}
\end{align}
A step in the procedure evolving the EBV from $g=0$ to $g$ consists of a Taylor series approximation followed by a Newton-Raphson solution. 

The last problem is to decide how large a step can be taken. First, to be sure that we follow the correct solution from $g=0$, the initial step cannot be larger than the smallest difference in in single particle energies $\Delta \varepsilon$. Near-degeneracies in $\{\varepsilon\}$ lead to an increase in the number of steps required. When two $\varepsilon$ are degenerate, the EBV equations change.\cite{elaraby:2012} The situation on paper becomes more complicated, but in the end it is faster computationally. How to decide when two single-particle energies are close enough to be considered degenerate is an open problem we are considering.

To minimize the number of steps, Faribault suggested a dynamic step-size procedure: a step $\delta g$ is attempted, but rejected if either the norms of the terms in the Taylor series grow, or if the norm of the EBV changes too much in one step. A 25\% change in the norm of the EBV is considered too much presently as tighter criteria only slowed down the computation. If rejected, the step is reattempted with a step-size of $\frac{1}{2}\delta g$. We observed that with this procedure the required number of steps scales logarithmically with $g$.\cite{fecteau:2022} 

\begin{figure} [ht!]
	\begin{subfigure}{\textwidth}
		\centering
		\includegraphics[width=0.32\textwidth]{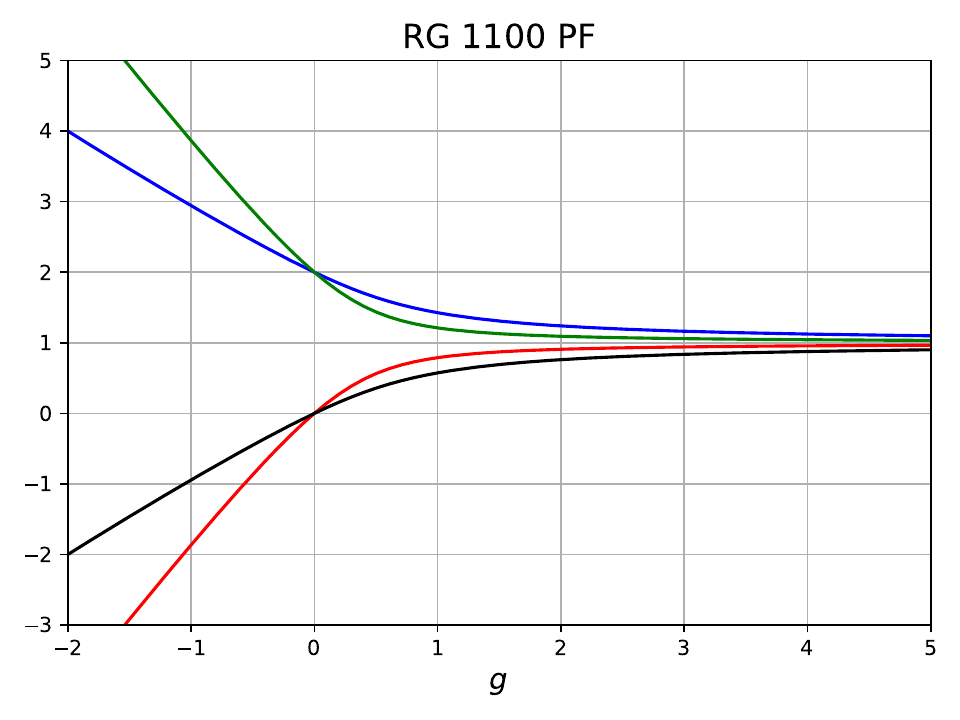} \hfill
		\includegraphics[width=0.32\textwidth]{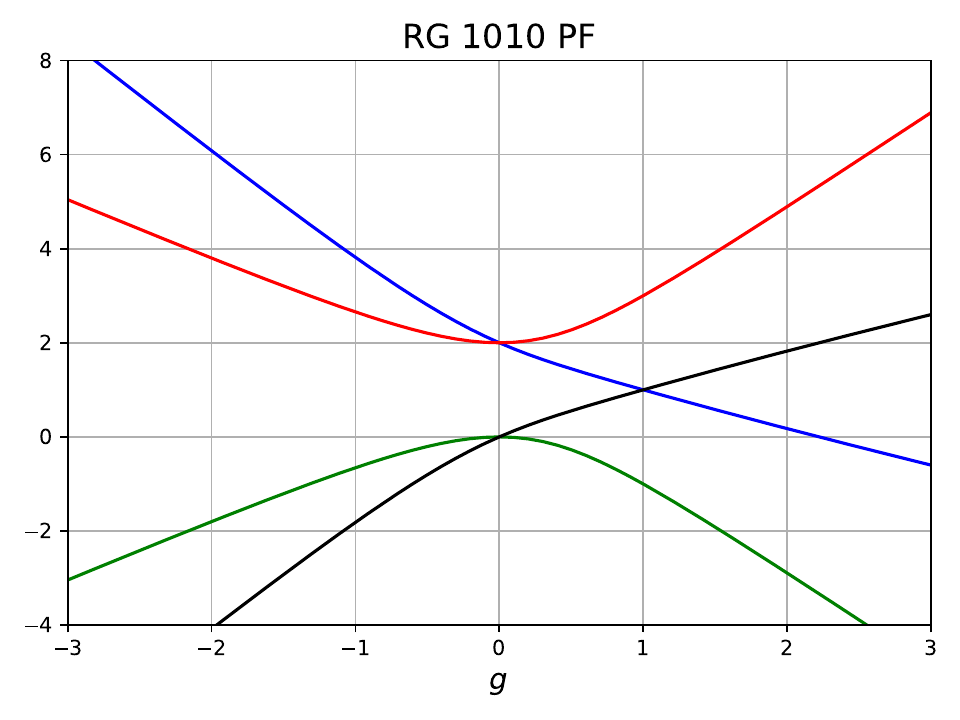} \hfill
		\includegraphics[width=0.32\textwidth]{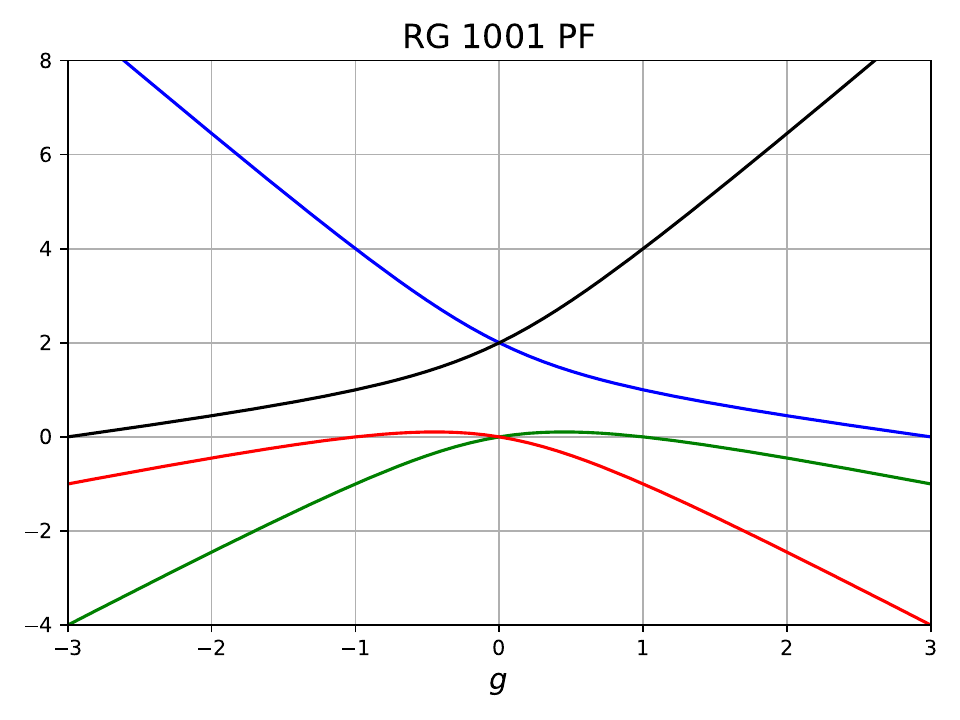} \\
		\includegraphics[width=0.32\textwidth]{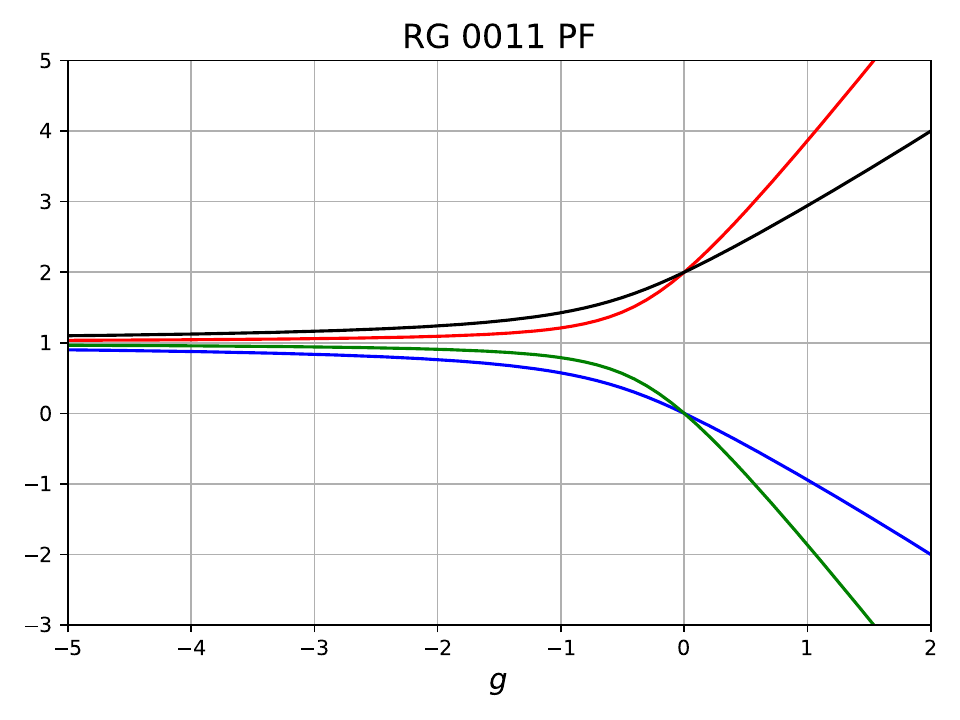} \hfill
		\includegraphics[width=0.32\textwidth]{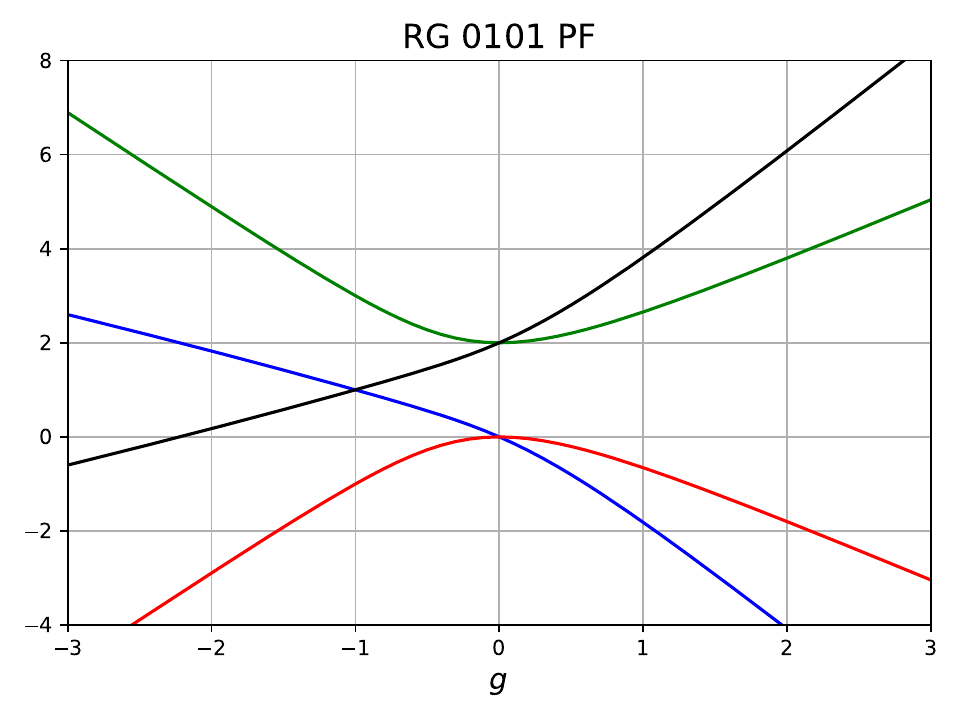} \hfill
		\includegraphics[width=0.32\textwidth]{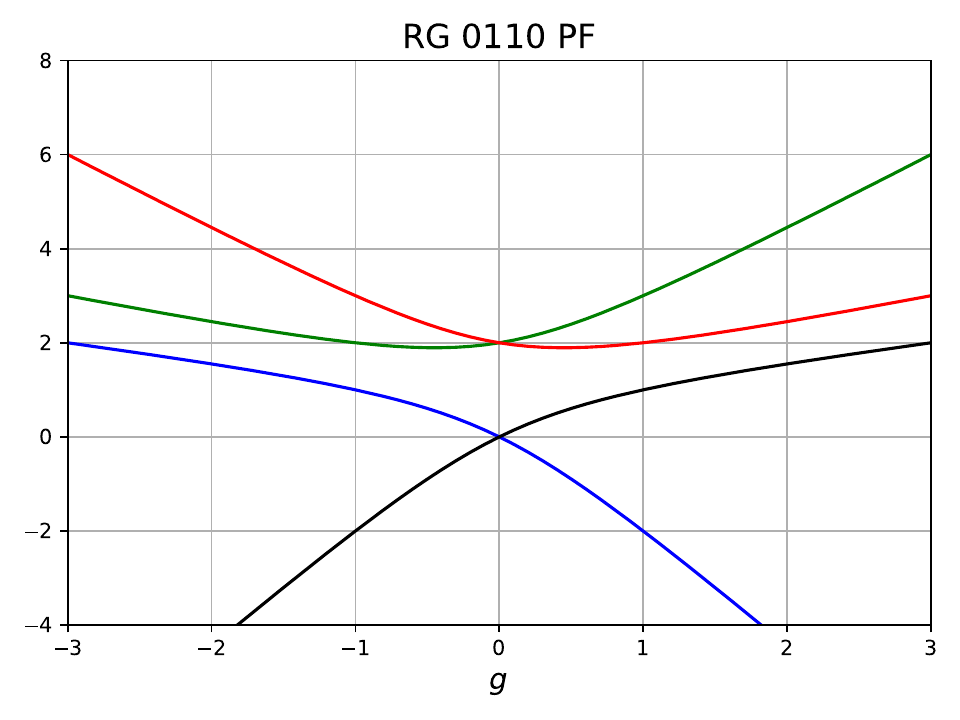} \\
		\includegraphics[scale=0.5]{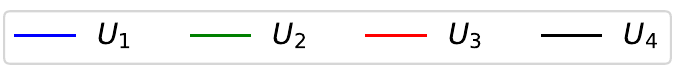} \hfill
	\end{subfigure}
	\caption{Evolution of eigenvalue-based variables for each RG state in 4-site 2-pair picket-fence model.}
	\label{fig:ebv_42_pf}
\end{figure}
The simplest concrete examples are PF models \eqref{eq:h_pf} with $\Delta \varepsilon =1$. For 4 sites and 2 pairs, the EBV for each RG state evolve as shown in figure \ref{fig:ebv_42_pf}. Again, as the EBV equations are coupled, the values of the EBV differ for each RG state. In almost all cases, the EBV behave linearly once $\vert g\vert > 1$. The exception is the ground state in the attractive regime, in which all EBV go to 1. (The same is true for the highest state in the repulsive regime). This limit is attained in very few steps from the solver.

\begin{figure} [ht!]
	\begin{subfigure}{\textwidth}
		\centering
		\includegraphics[width=0.49\textwidth]{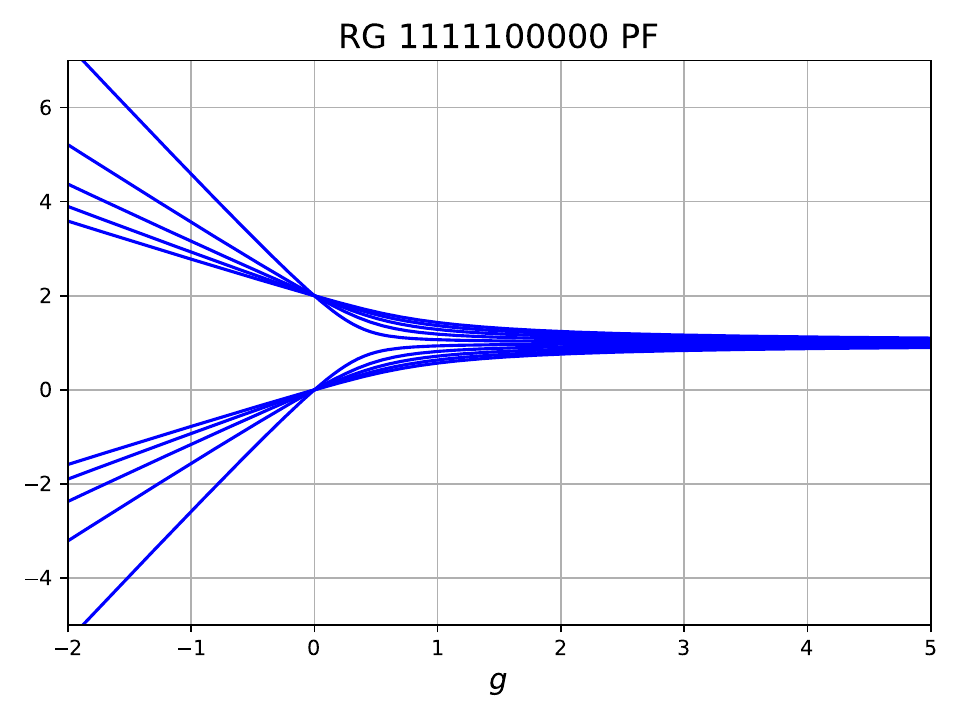} \hfill
		\includegraphics[width=0.49\textwidth]{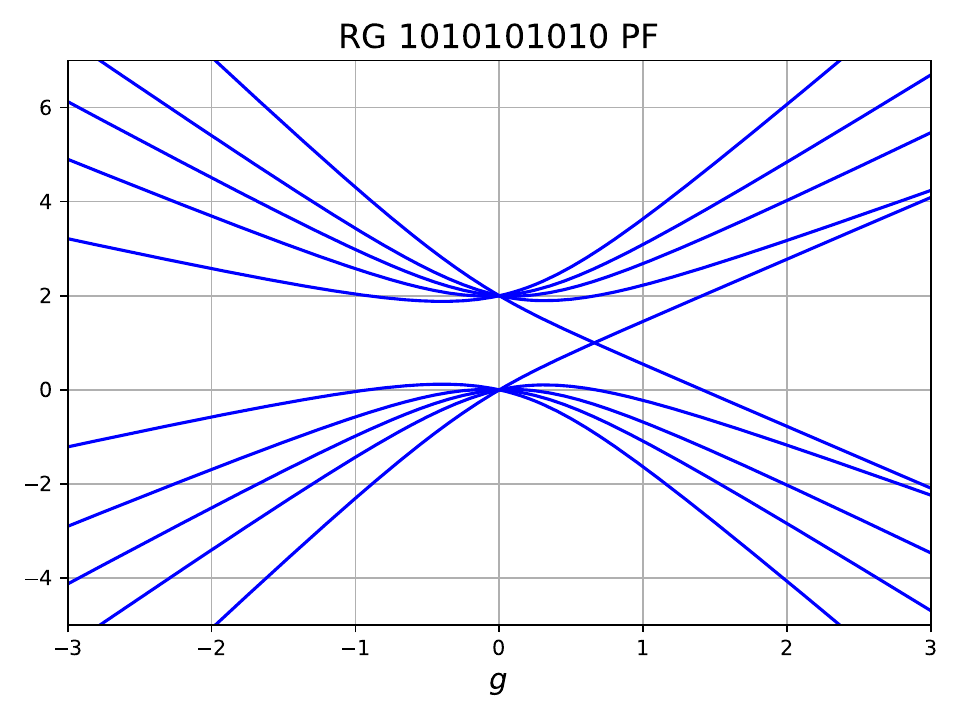}
	\end{subfigure}
	\caption{Evolution of eigenvalue-based variables of ground and N\'{e}el state in 10-site 5-pair picket-fence model.}
	\label{fig:ebv_105_pf}
\end{figure}
Larger PF models provide no complication: the EBV evolution for 5 pairs in 10 sites is presented in figure \ref{fig:ebv_105_pf}. Whereas the energies of all $\binom{10}{5}=252$ states are presented in figure \ref{fig:105_bcs}, we now concentrate on two particular states: the ground state 1111100000 and the state with bitstring 1010101010. We have referred to the latter state as the N\'{e}el RG state. At $g=0$, its Slater determinant representative has orbital occupations alternating between double and empty. Interpreted in a spin representation \eqref{eq:spin_su2}, this would be the antiferromagnetic N\'{e}el state  $\ket{\uparrow\downarrow\uparrow\downarrow\uparrow\downarrow\uparrow\downarrow\uparrow\downarrow}$, hence the name.

\begin{figure} [ht!]
	\begin{subfigure}{\textwidth}
		\centering
		\includegraphics[width=0.49\textwidth]{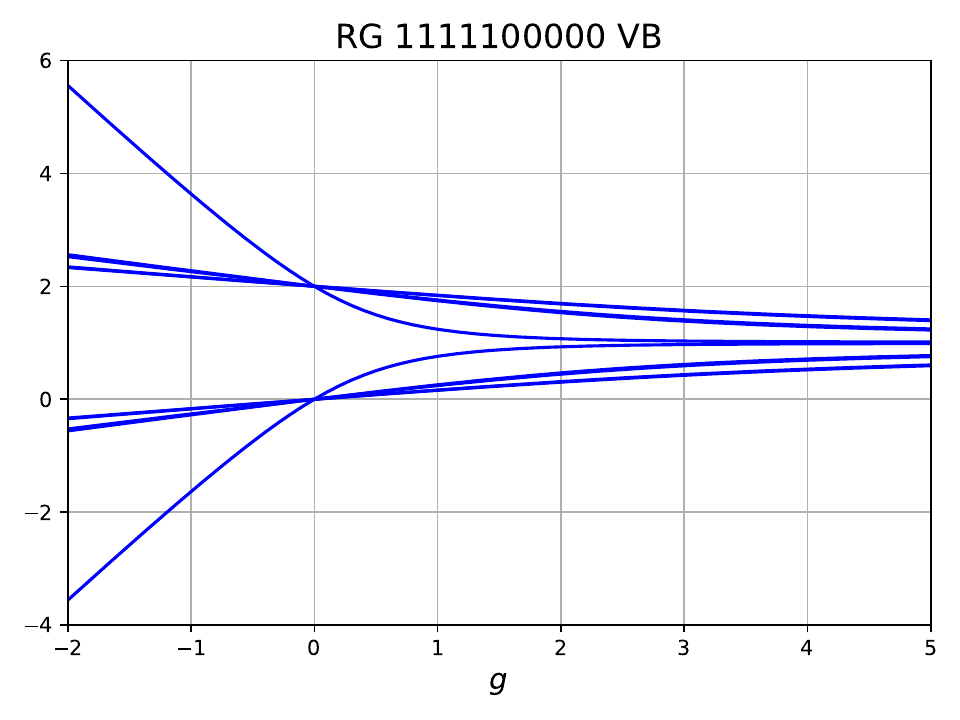} \hfill
		\includegraphics[width=0.49\textwidth]{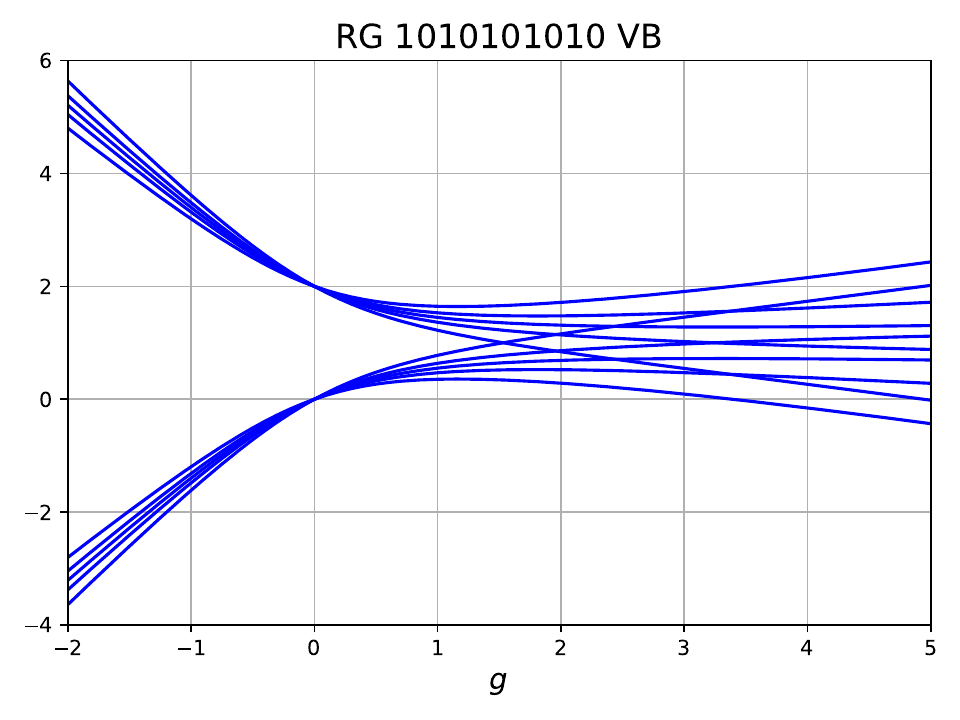}
	\end{subfigure}
	\caption{Evolution of eigenvalue-based variables of ground and N\'{e}el state in 10-site 5-pair valence-bond model.}
	\label{fig:ebv_105_vb}
\end{figure}

For the molecular systems we have considered, PF models were not so relevant. Instead, the $\{\varepsilon\}$ arranged into near-degenerate pairs that are well separated in energy from one another. We've called this a valence-bond (VB) type BCS Hamiltonian. In particular, the Hamiltonian
\begin{align} \label{eq:H_VB}
\hat{H}_{VB} = \frac{1}{2} \sum^{N/2}_{k=1} (k-1) \xi \hat{n}_{2k} + ((k-1) \xi + \Delta \varepsilon) \hat{n}_{2k+1} - \frac{g}{2}\sum^N_{k,l=1} S^+_k S^-_l
\end{align}
has $N/2$ pairs of levels, separated by the large spacing $\xi$. Within each pair, the levels are separated by the smaller spacing $\Delta \varepsilon$. 
\begin{figure} [ht!]
	\begin{subfigure}{\textwidth}
		\centering
		\includegraphics[width=0.32\textwidth]{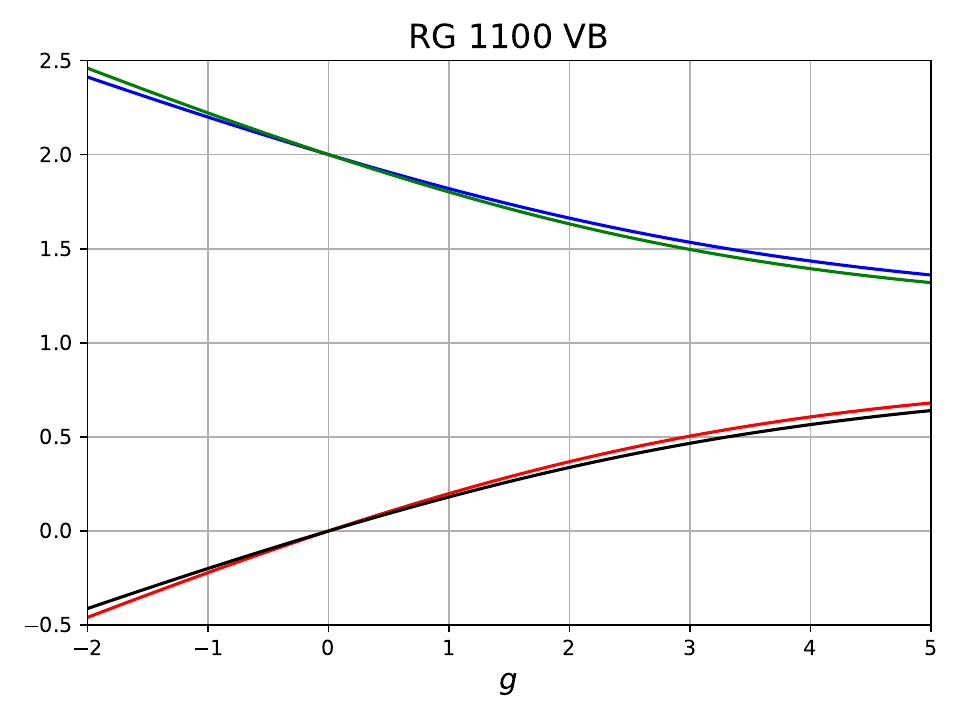} \hfill
		\includegraphics[width=0.32\textwidth]{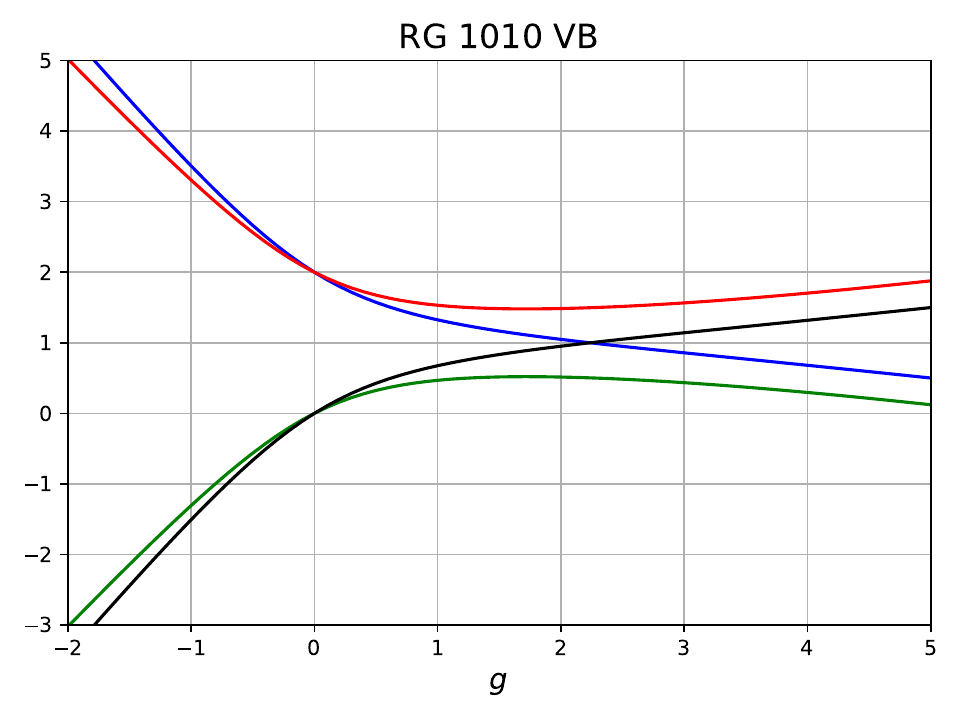} \hfill
		\includegraphics[width=0.32\textwidth]{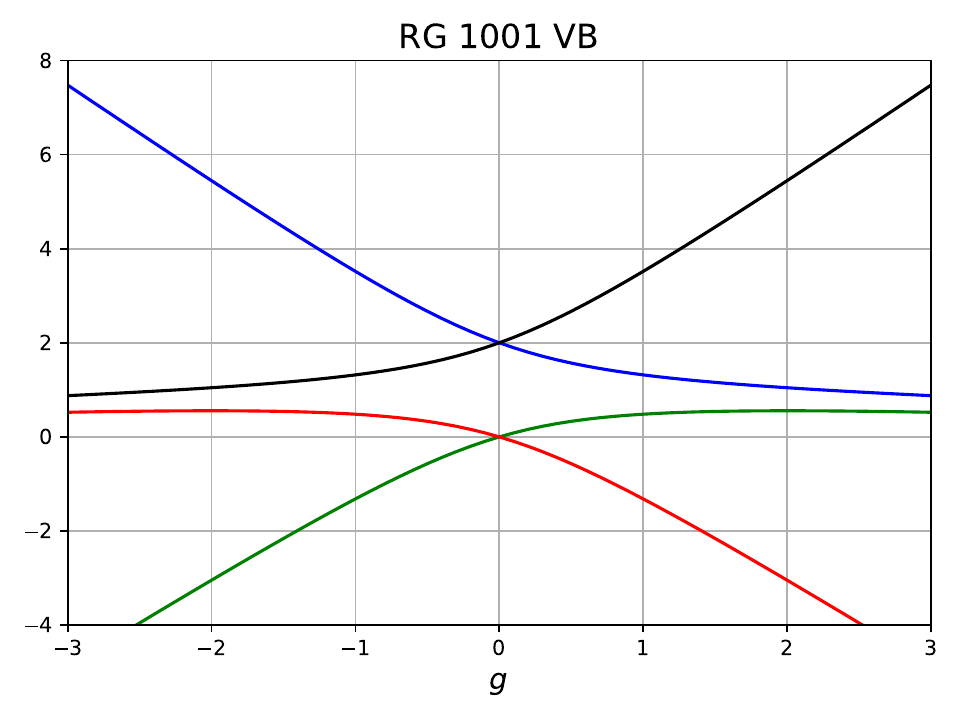} \\
		\includegraphics[width=0.32\textwidth]{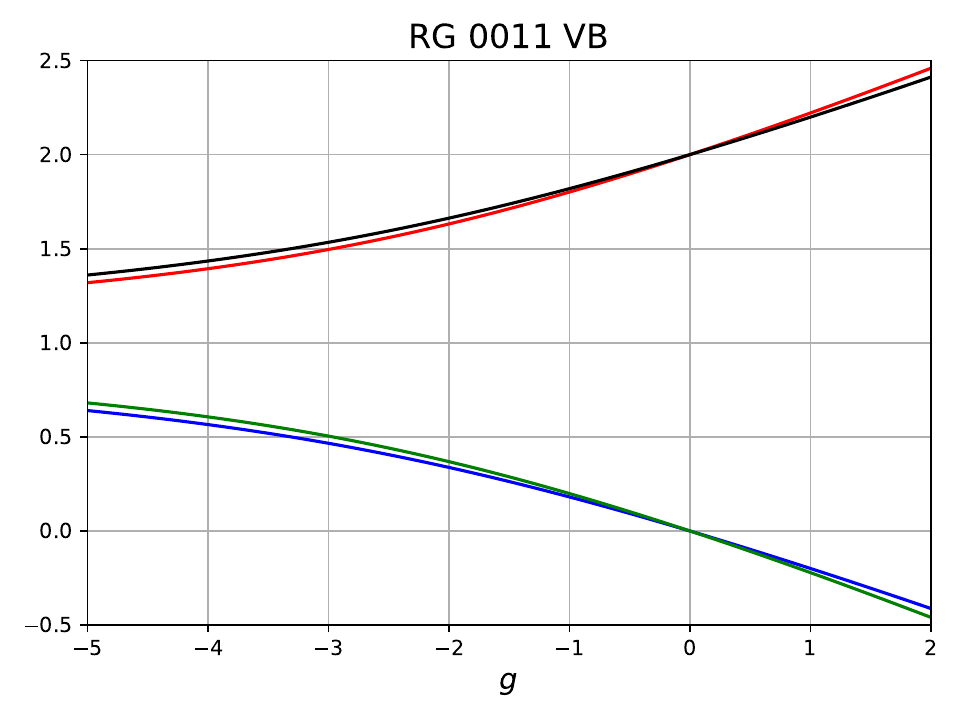} \hfill
		\includegraphics[width=0.32\textwidth]{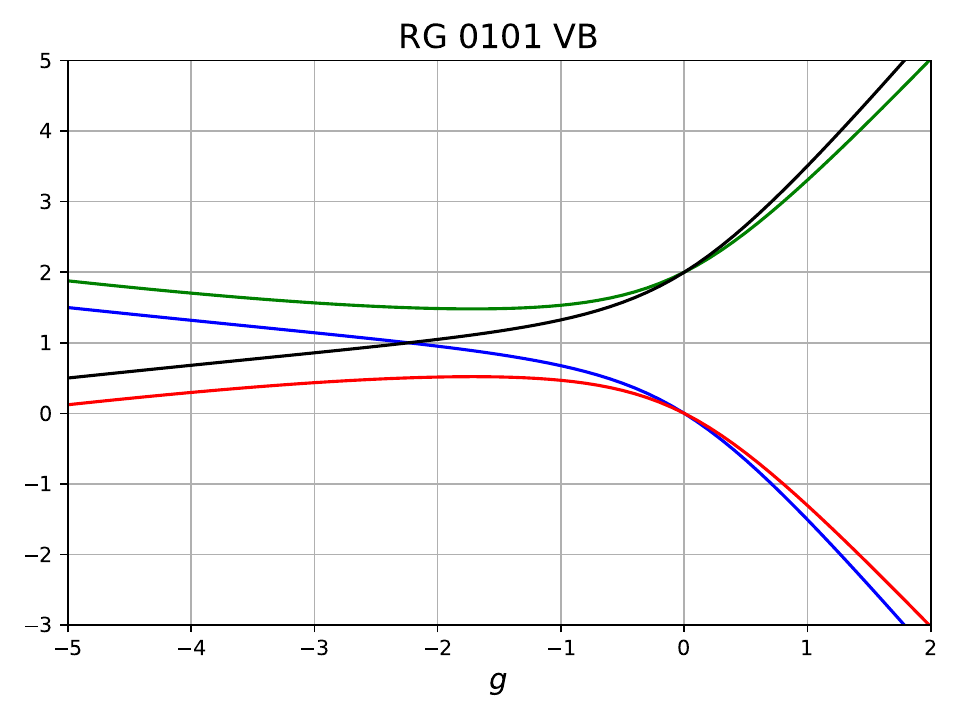} \hfill
		\includegraphics[width=0.32\textwidth]{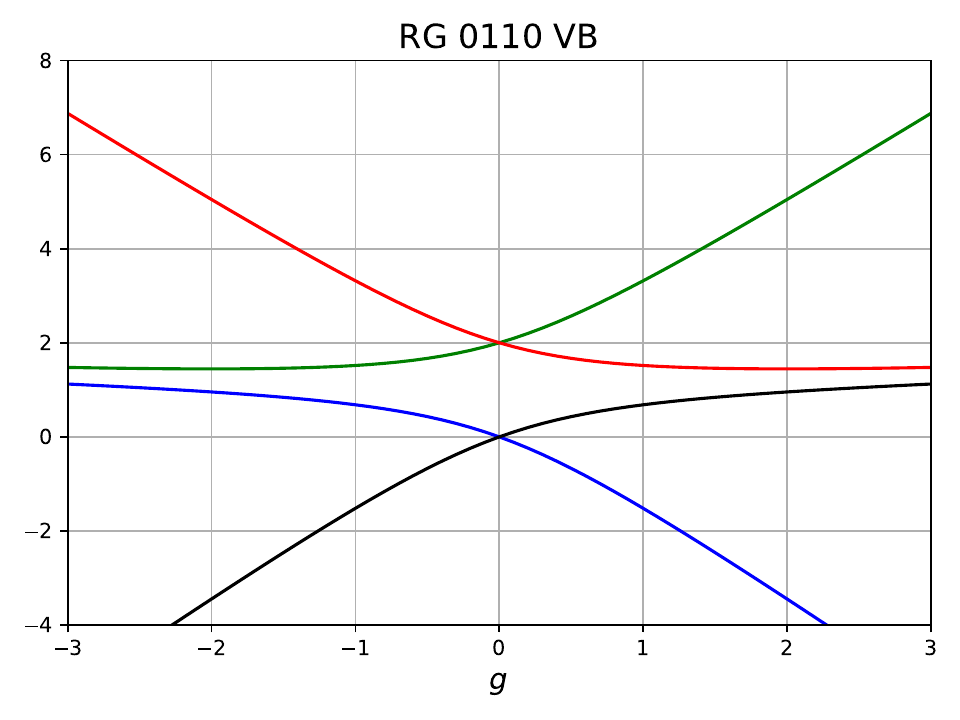} \\
		\includegraphics[scale=0.5]{img/ebv_legend.pdf} \hfill
	\end{subfigure}
	\caption{Evolution of eigenvalue-based variables for each RG state in 4-site 2-pair valence-bond model.}
	\label{fig:ebv_42_vb}
\end{figure}
The EBV for a VB Hamiltonian with $\xi =10$ and $\Delta \varepsilon = 1$ are shown in figure \ref{fig:ebv_42_vb}. As can be seen, the EBV reach their asymptotic behavior quickly, though not as quickly as in a PF model. Larger models again pose no difficulty: ground and N\'{e}el states for 5 pairs in 10 sites are shown in figure \ref{fig:ebv_105_vb}.

\subsection{Rapidity Extraction}
The RDM and TDM elements can now be computed without rapidities\cite{faribault:2022} so generally they are no longer needed. They are however useful for verifying expressions for matrix elements that are yet to be computed. 

Given a solution of the EBV equations $\{V\}$, we can compute the rapidities $\{v\}$. They are obtained as the roots of an interpolated polynomial. Define $P(z)$ whose roots are the rapidities
\begin{align}
P(z) = \prod^M_{a=1} (z-v_a).
\end{align}
The logarithmic derivative of $P(z)$ evaluated at $\varepsilon_i$ is just
\begin{align}
\frac{P'(\varepsilon_i)}{P(\varepsilon_i)} = \frac{1}{g} V_i.
\end{align}
Richardson's equations \eqref{eq:rich} are equivalent to the second-order ODE
\begin{align} \label{eq:ODE}
P''(z) - F(z) P'(z) + G(z) P(z) = 0,
\end{align}
with
\begin{align}
F(z) &= \frac{2}{g} + \sum^N_{i=1} \frac{1}{z-\varepsilon_i} \\
G(z) &= \frac{1}{g} \sum^N_{i=1} \frac{V_i}{z-\varepsilon_i}.
\end{align}
The ODE is solved by Lagrange interpolating $P(z)$ on a grid of points which reduces to linear equations for the corresponding weights. One root of this polynomial is found by Laguerre's method, the root is factored out, and the procedure is repeated until all roots are found. The grid points must be a reasonable guess for the roots or the procedure will fail. The grid is chosen at $g=0$ and evolved along with the EBV. This does not need to be done at each step. At small g, it's more important to keep more steps. Once g is large enough, the rapidities are either stable or scale linearly with $g$.

And now the details. In the Lagrange basis, the polynomial $P(z)$ is
\begin{align}
P(z) = l(z) \sum^{M+1}_{a=1} \frac{w_a}{z-z_a}
\end{align} 
with $\{w\}$ the weights and
\begin{align}
l(z) = \prod^{M+1}_{a=1} (z-z_a).
\end{align}
The derivatives of $P(z)$ scaled by $l(z)$ are simple
\begin{align}
\frac{P'(z)}{l(z)}  &= \sum^{M+1}_{a=1} \sum^{M+1}_{b (\neq a) =1} \frac{w_a}{(z-z_a)(z-z_b)} \\
\frac{P''(z)}{l(z)} &= \sum^{M+1}_{a=1} \sum^{M+1}_{b (\neq a) =1} \sum^{M+1}_{c (\neq a,b) =1} \frac{w_a}{(z-z_a)(z-z_b)(z-z_c)}.
\end{align}
Using these expressions in the ODE \eqref{eq:ODE} at the grid points defines a set of linear equations for the weights $\{w\}$
\begin{align}
\sum^{M+1}_{b (\neq a) =1} \sum^{M+1}_{c(\neq a,b) =1} \frac{w_a + 2w_b}{(z_a-z_b)(z_a-z_c)} - F(z_a) \sum^{M+1}_{b (\neq a)} \left( \frac{w_a +w_b}{z_a-z_b} \right) + G(z_a) w_a = 0, \quad 1 \leq a \leq M+1.
\end{align}
The weights must also be normalized
\begin{align}
\sum^{M+1}_{a=1} w_a = 1,
\end{align}
which is once again an over-determined set of linear equations
\begin{align}
B \textbf{w} = \textbf{b}
\end{align}
for the weights to be solved by $B=QR$ as
\begin{align} \label{eq:linear_weight}
R \textbf{w} = Q^{\dagger} \textbf{b}.
\end{align}
The first $M+1$ rows of $B$ are
\begin{align}
B_{ab} &= \begin{cases}
\sum^{M+1}_{c (\neq a) =1} \sum^{M+1}_{c \neq d (\neq a)} \frac{1}{(z_a-z_c)(z_a-z_d)} - \sum^{M+1}_{c(\neq a)=1} \frac{F(z_a)}{z_a-z_c}  +G(z_a), &\quad a = b \\
\sum^{M+1}_{c (\neq a,b)=1} \frac{2}{(z_a-z_b)(z_a-z_c)} - \frac{F(z_a)}{z_a-z_b}  , &\quad a \neq b
\end{cases},
\end{align}
and the last row is
\begin{align}
B_{M+2,b} &= 1.
\end{align}
The RHS $\textbf{b}$ is $M+1$ zeros followed by a single entry 1. The solution of the linear equations \eqref{eq:linear_weight} is the polynomial $P(z)$ that solves the ODE \eqref{eq:ODE}. A root is found by Laguerre's method, and factored out of the polynomial by removing the grid point $z_{\alpha}$ closest to the root $\alpha$. In terms of weights, this means removing $w_{\alpha}$ and modifying the other weights
\begin{align}
w_b \rightarrow w_b \frac{z_b - z_{\alpha}}{z_b - \alpha}.
\end{align}
These weights represent a polynomial of lower rank, and the approach is iterated until all the roots have been found. 

How do we choose the grid at $g=0$? There are $M+1$ points, of which $M$ are assigned as the occupied single-particle energies, based on the bitstring, plus some small random noise. The last grid point is set to ten times the largest $\varepsilon$. Each time the rapidities are computed, the first $M$ grid points are replaced with the rapidities from the previous iteration \emph{except} if the difference between a grid point and a rapidity is less than a threshold. The final grid point is updated to be ten times the largest rapidity.

Finally, Newton-Raphson applied directly to Richardson's equations can improve the precision of the rapidities at the final value of $g$ if required. 

\begin{figure} [ht!]
	\begin{subfigure}{\textwidth}
		\centering
		\includegraphics[width=0.32\textwidth]{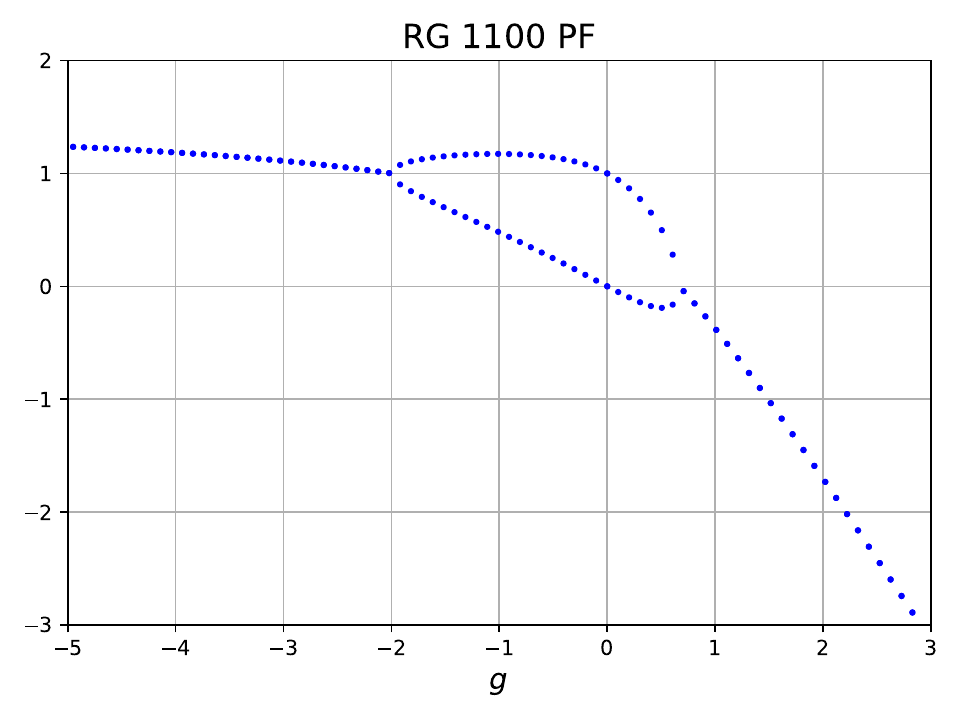} \hfill
		\includegraphics[width=0.32\textwidth]{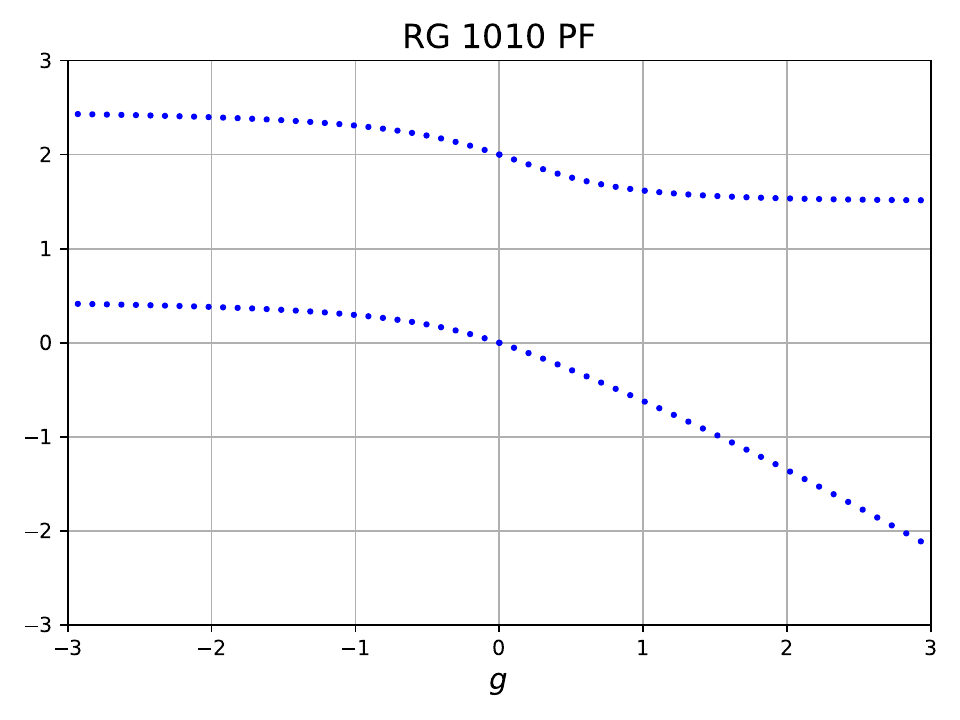} \hfill
		\includegraphics[width=0.32\textwidth]{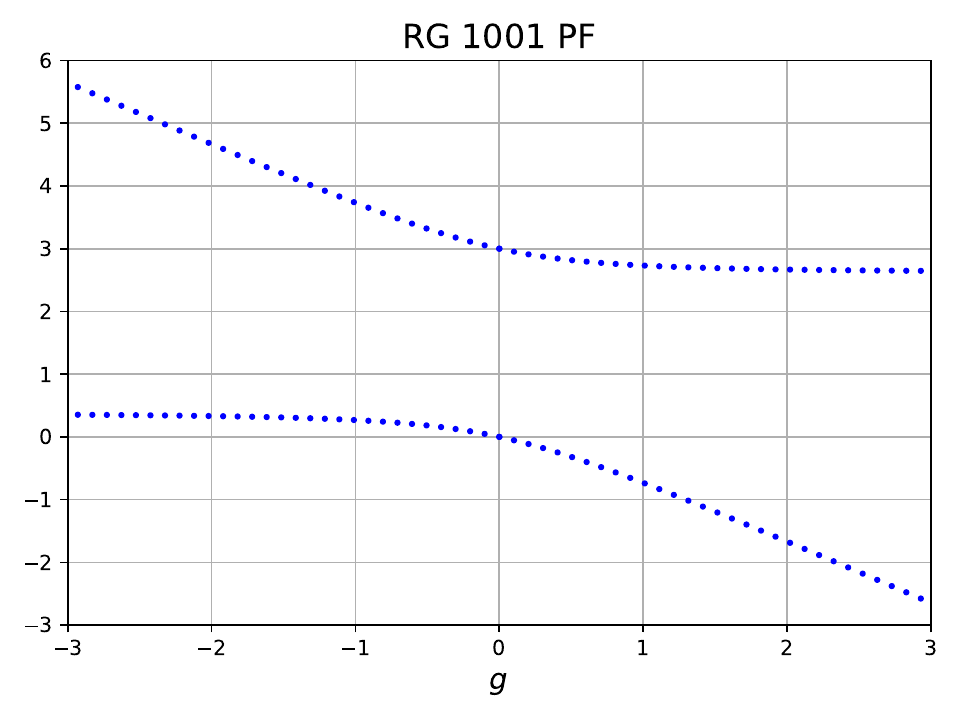} \\
		\includegraphics[width=0.32\textwidth]{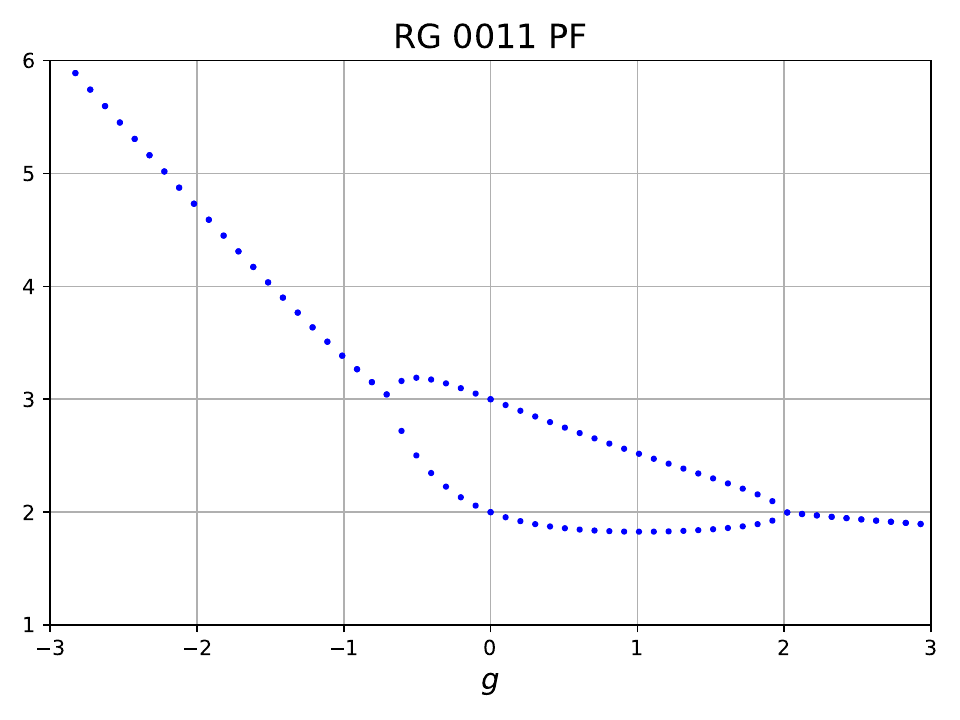} \hfill
		\includegraphics[width=0.32\textwidth]{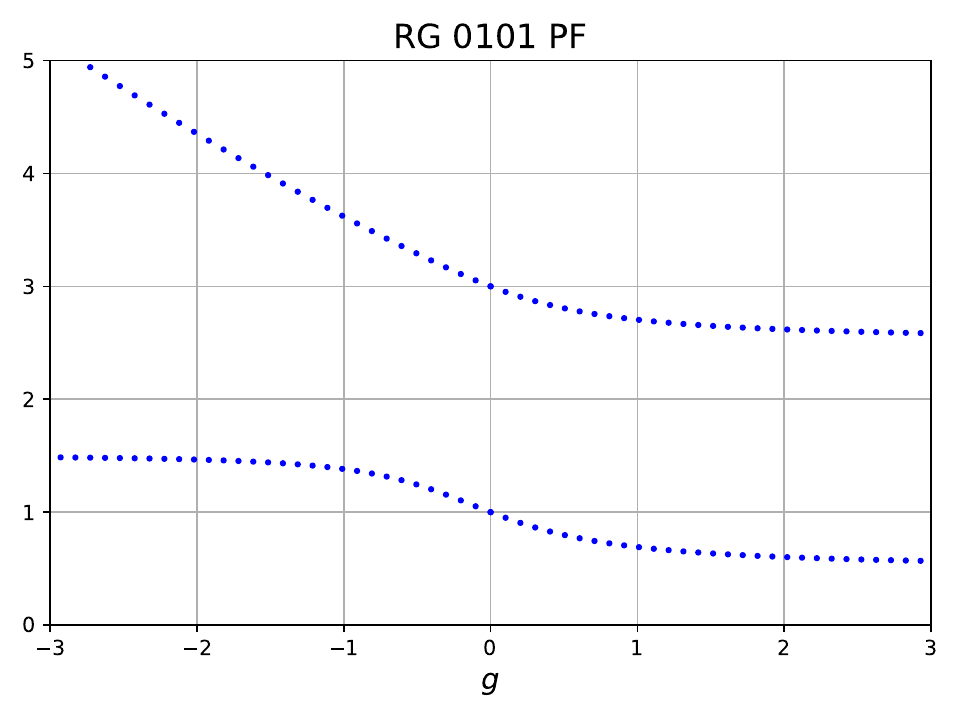} \hfill
		\includegraphics[width=0.32\textwidth]{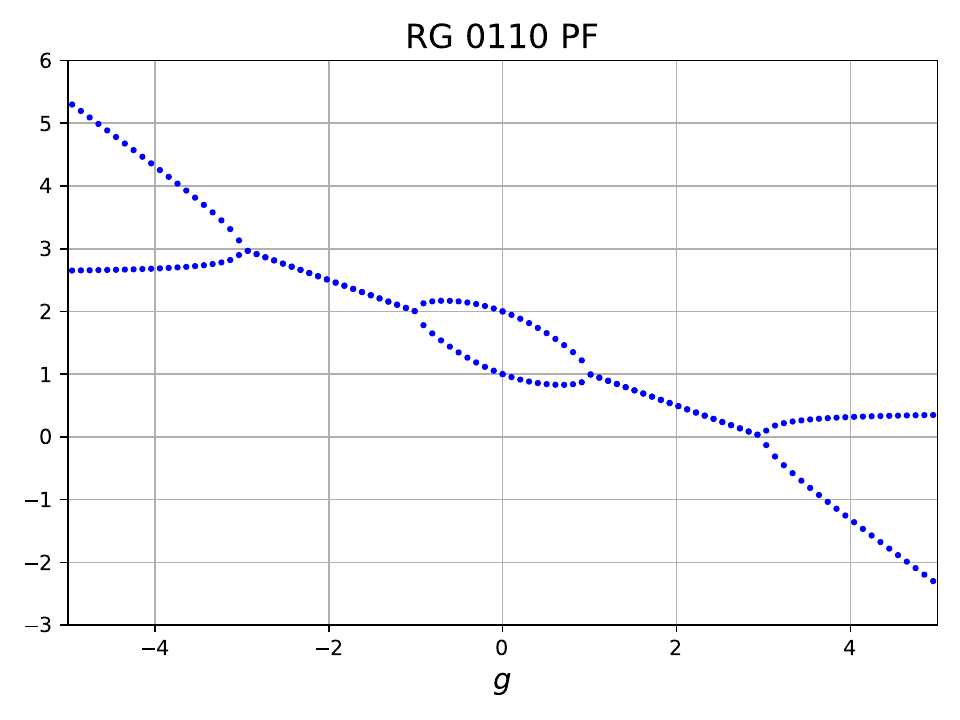} 
	\end{subfigure}
	\caption{Evolution of real parts of rapidities of all RG states in 4-site 2-pair picket-fence model.}
	\label{fig:rap_real_42_pf}
\end{figure}
\begin{figure} [ht!]
	\begin{subfigure}{\textwidth}
		\centering
		\includegraphics[width=0.32\textwidth]{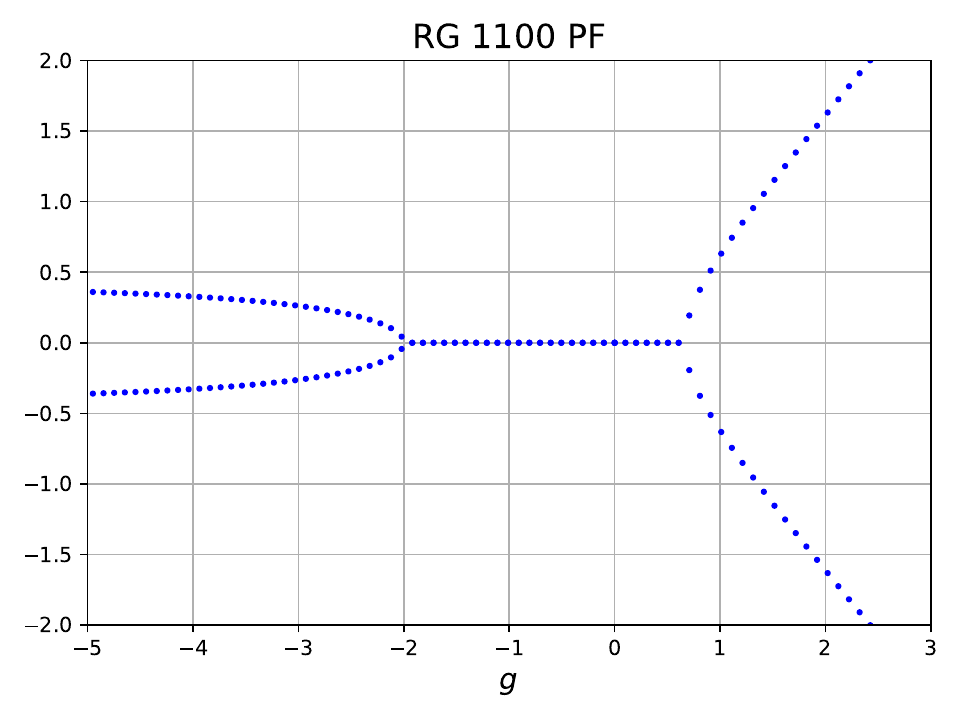} \hfill
		\includegraphics[width=0.32\textwidth]{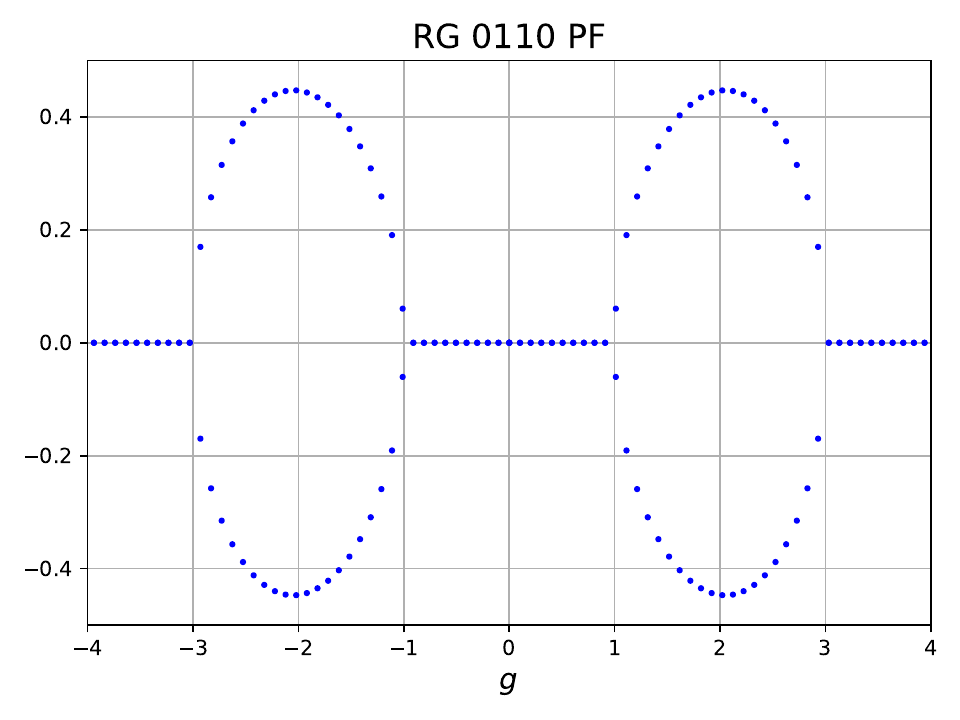} \hfill
		\includegraphics[width=0.32\textwidth]{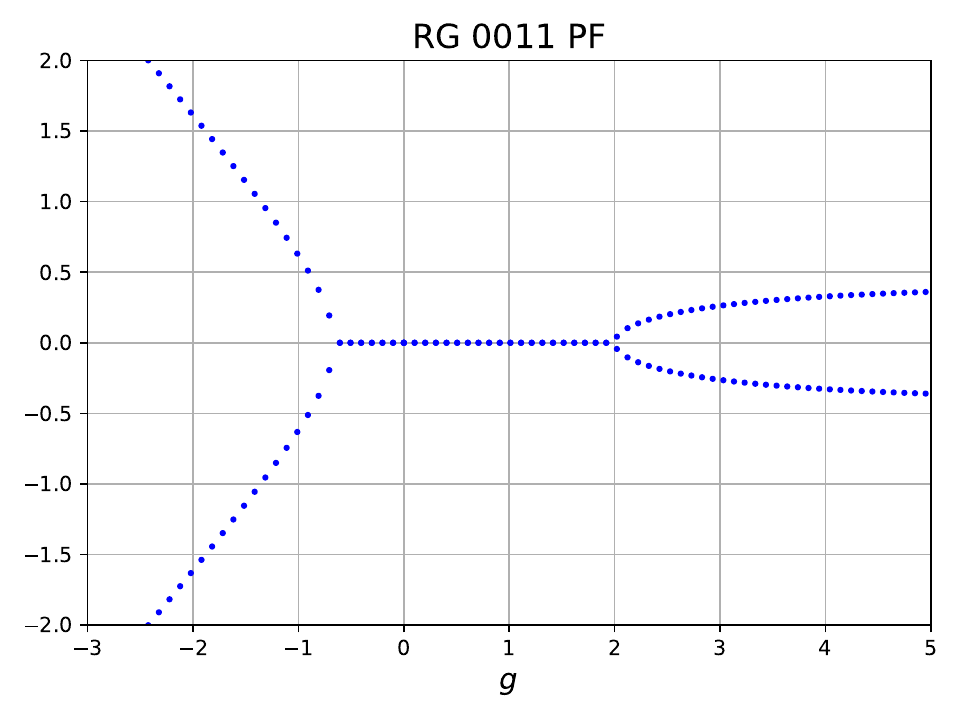}
	\end{subfigure}
	\caption{Evolution of imaginary parts of rapidities of 1100, 0100, and 0011 RG states in 4-site 2-pair picket-fence model. The rapidities of the 1010, 1001, and 0101 RG states are real everywhere.}
	\label{fig:rap_imag_42_pf}
\end{figure}
The real parts of the rapidities obtained for a PF model of 2 pairs in 4 sites are shown in figure \ref{fig:rap_real_42_pf}, while the corresponding imaginary parts are shown in figure \ref{fig:rap_imag_42_pf}. From figure \ref{fig:rap_real_42_pf}, characteristic behavior of the rapidities is clear. First, as $g$ becomes large, each rapidity either grows linearly with $g$, or remains trapped between two single particle energies. Second, critical points appear where two rapidities are exactly equal to a single particle energy. Starting from $g=0$, this process can occur both ways: two real rapidities meet and form a complex conjugate pair, or a complex conjugate pair splits into two real rapidities. In either case, the derivatives $\frac{\partial u_a}{\partial g}$ are undefined and the numerical solution becomes more tedious. The numerical solution of the EBV equations passes through these critical points without issue. The root-finding procedure to extract the rapidities from the EBV fails \emph{precisely} at the critical points, though any small perturbation in either direction of $g$ fixes the issue. \emph{At} the critical points the troublesome rapdities are equal to a particular single particle energy. Finally, the states 1001 and 0110 are everywhere degenerate, but their corresponding rapidities demonstrate that they are quite different.

\begin{figure} [ht!]
	\begin{subfigure}{\textwidth}
		\centering
		\includegraphics[width=0.32\textwidth]{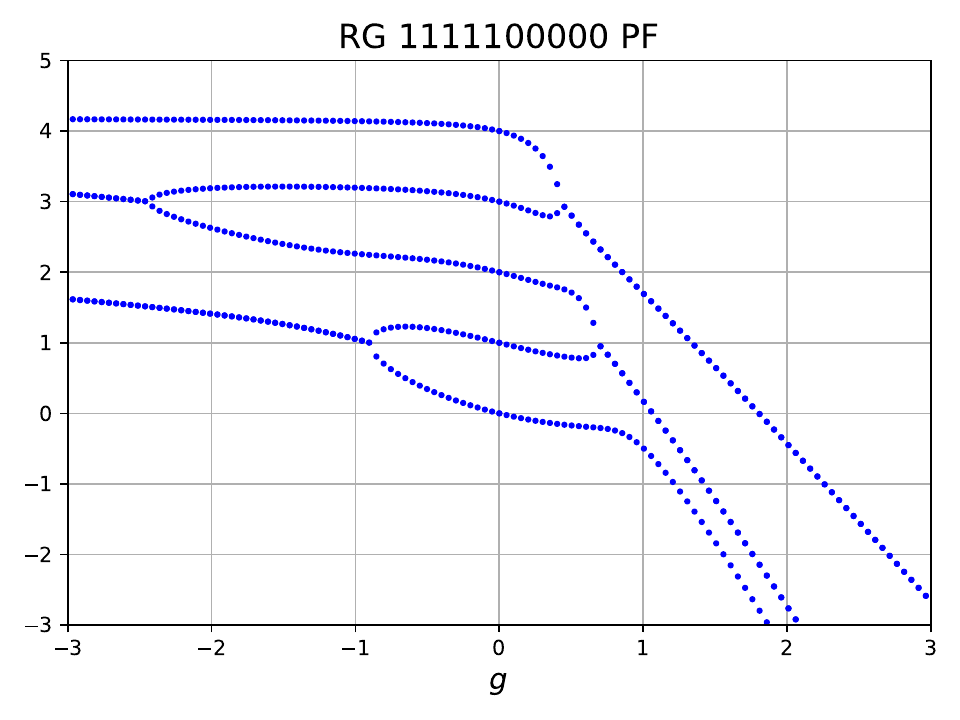} \hfill
		\includegraphics[width=0.32\textwidth]{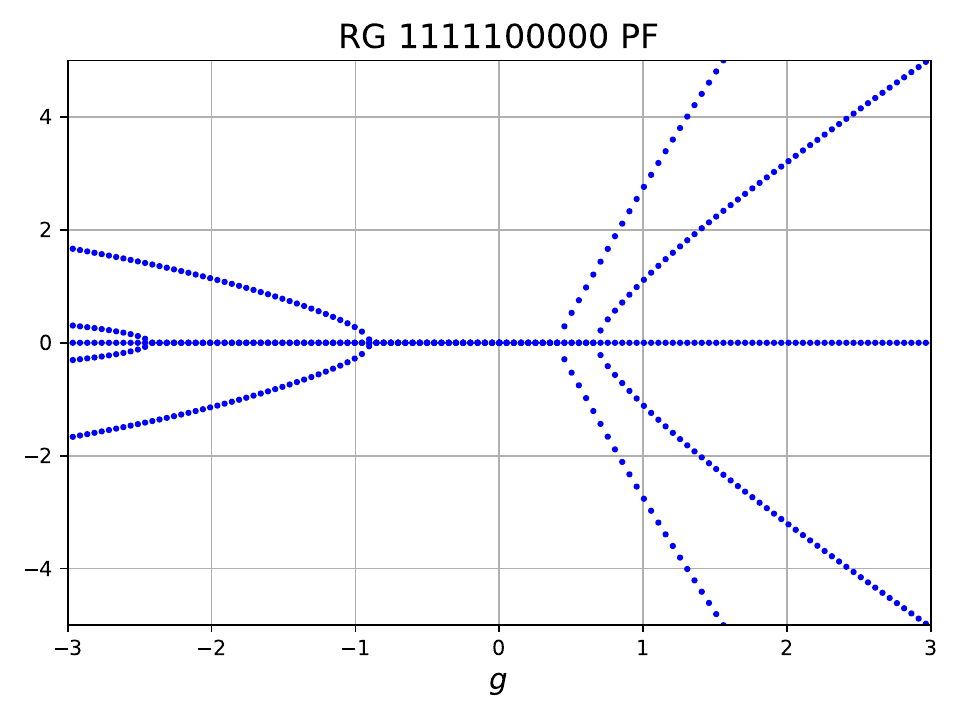} \hfill
		\includegraphics[width=0.32\textwidth]{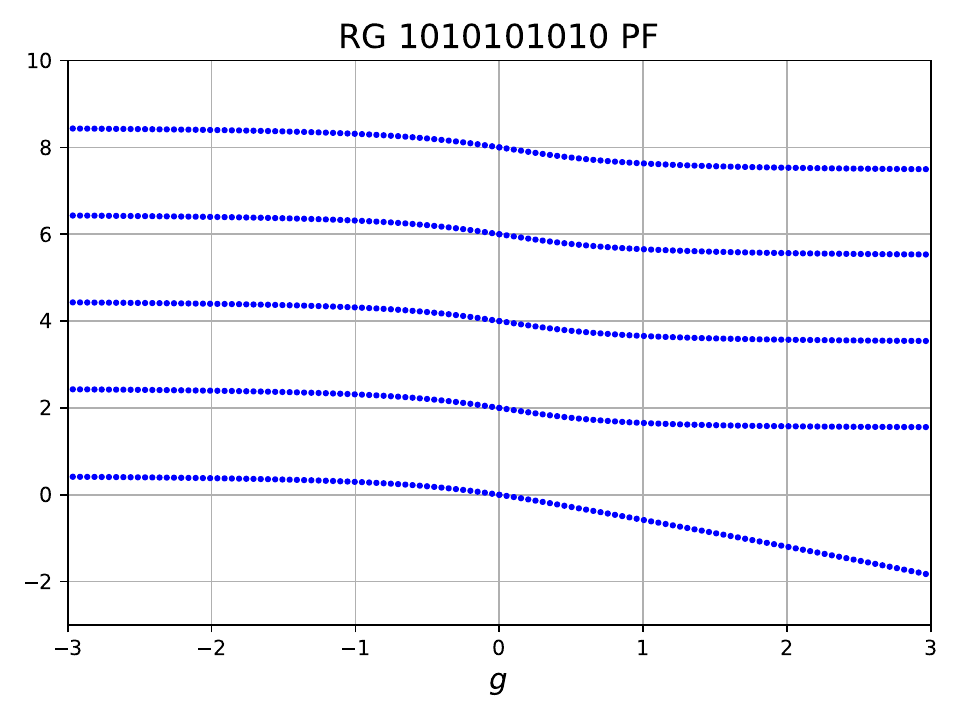} \\
		\includegraphics[width=0.32\textwidth]{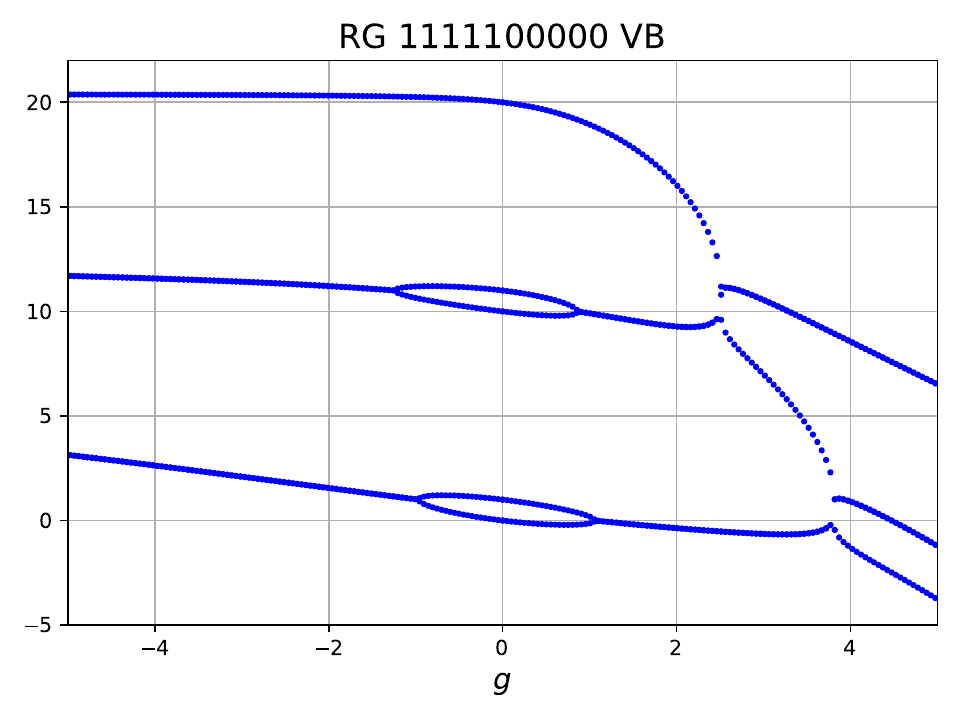} \hfill
		\includegraphics[width=0.32\textwidth]{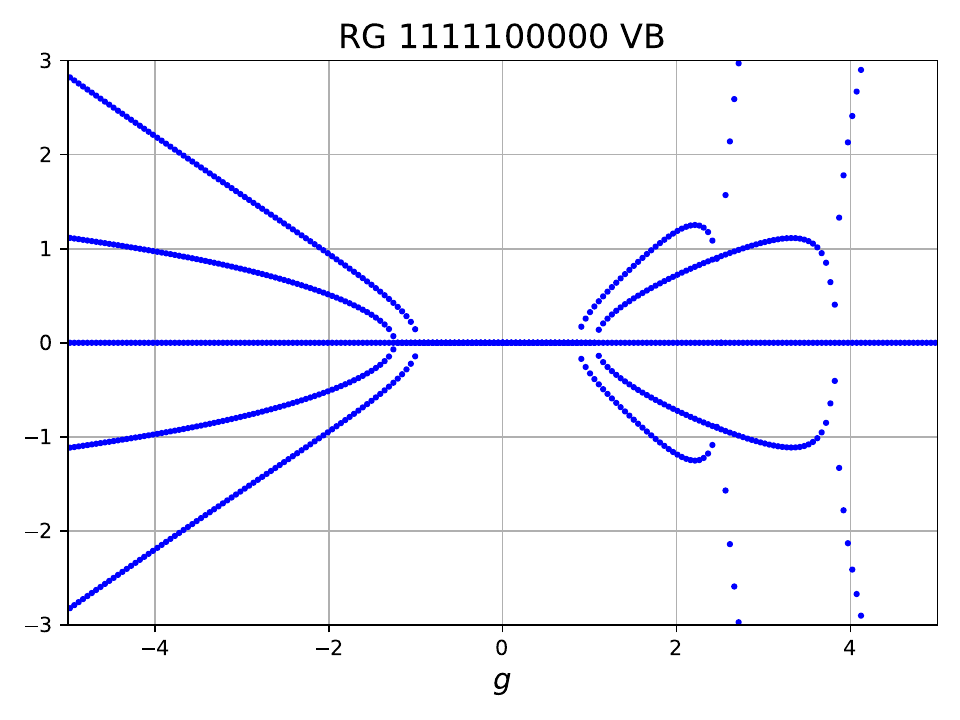} \hfill
		\includegraphics[width=0.32\textwidth]{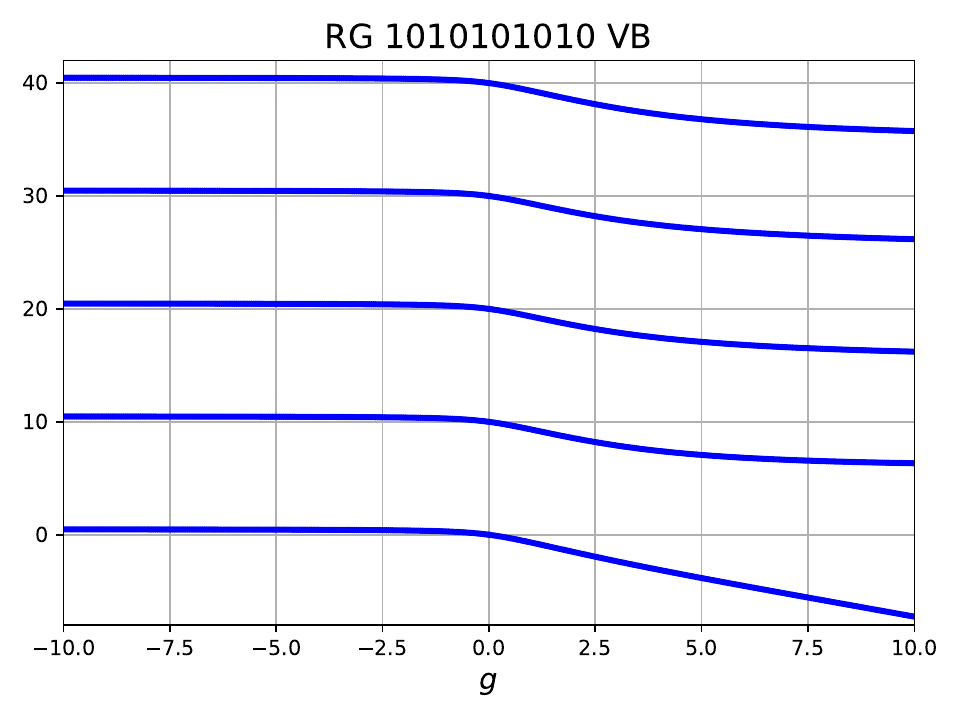}
	\end{subfigure}
	\caption{Evolution of rapidities in 10-site 5-pair picket-fence (top row) and valence-bond (bottom row) models for ground and N\'{e}el RG states: real parts of ground state rapidities (left column), imaginary parts of ground state rapidities (center column), and real parts of N\'{e}el state rapidities (right column).}
	\label{fig:rap_105}
\end{figure}
Moving to larger systems again poses no problem. Rapidities for the ground and N\'{e}el states of 10-site, 5-pair PF and VB models are shown in figure \ref{fig:rap_105}. The rapidities of the ground state have more critical points than for the 4-site 2-pair case, but the large $g$ behavior is the same: all the rapidities are linear with $g$ in the attractive limit, and are trapped between single particle energies in the repulsive limit. The rapidities of the N\'{e}el state are everywhere real. One rapidity is linear with $g$ in the attractive limit. All other rapidities stay between single particle energies. The VB models have solutions that are more stretched out. For the ground state, this leads to a more involved set of critical points in the attractive regime. For the N\'{e}el state there is no qualitative difference in the rapidity pattern. We have presented results for small systems principally to keep the figures simple. Solving the EBV equations for hundreds of levels and pairs is easily done on a laptop computer.

With solutions of Richardson's equations, we can now move on to computing the density matrix elements of RG states.

\section{Scalar Products and Correlation Functions} \label{sec:rdm}
The point is to use RG states as a basis for the physical wavefunction. We will therefore need to evaluate matrix elements with the Hamiltonian we care about, in this case a 2-body Coulomb interaction
\begin{align} \label{eq:C_ham}
\hat{H}_C = \sum^N_{i,j=1} h_{ij} \sum_{\sigma} a^{\dagger}_{i \sigma} a_{j \sigma} + \frac{1}{2} \sum^N_{i,j,k,l=1} V_{ijkl} \sum_{\sigma \tau} a^{\dagger}_{i \sigma} a^{\dagger}_{j \tau} a_{l \tau} a_{k \sigma}
\end{align}
in which $\sigma$ and $\tau$ represent spin labels, and the 1- and 2-electron integrals are computed in a basis $\{\phi\}$
\begin{align}
h_{ij} &= \int d\mathbf{r} \phi^*_i (\mathbf{r}) \left( - \frac{1}{2} \nabla^2 - \sum_I \frac{Z_I}{| \mathbf{r} - \mathbf{R}_I |} \right) \phi_j (\mathbf{r}) \\
V_{ijkl} &= \int d\mathbf{r}_1 d\mathbf{r}_2 \frac{\phi^*_i(\mathbf{r}_1)  \phi^*_j(\mathbf{r}_2)  \phi_k(\mathbf{r}_1)  \phi_l(\mathbf{r}_2)  }{| \mathbf{r}_1 - \mathbf{r}_2|}.
\end{align}
RG states have zero seniority, meaning that the energy of \eqref{eq:C_ham} computed with an RG state $\ket{\{u\}}$
\begin{align} \label{eq:sz_energy}
E [\{\varepsilon\},g] = 2 \sum^N_{k=1} h_{kk} \gamma_k + \sum^N_{k=1} \sum^N_{l (\neq k) =1} (2V_{klkl} - V_{kllk})D_{kl} + \sum^N_{k,l=1} V_{kkll} P_{kl}
\end{align}
depends only on the  1-RDM elements $\gamma_k$ along with the non-zero elements of the 2-RDM: the diagonal-correlation function $D_{kl}$ and the pair-correlation function $P_{kl}$. Explicitly, these elements are
\begin{subequations} \label{eq:sz_dm}
\begin{align}
\gamma_k &= \frac{1}{2} \frac{\braket{ \{u\} | \hat{n}_k | \{u\} }}{\braket{ \{u\} | \{u\} }} \\
D_{kl} &= \frac{1}{4}   \frac{\braket{ \{u\} | \hat{n}_k \hat{n}_l | \{u\} }} {\braket{ \{u\} | \{u\} }} \\
P_{kl} &= \frac{\braket{\{u\} | S^+_k S^-_l | \{u\}}} {\braket{\{u\} | \{u\} }}.
\end{align}
\end{subequations}
It is important to note that as written, $D_{kk}$ and $P_{kk}$ correspond to the same element of the 2-RDM. As a convention, we assign this element to $P_{kk}=\gamma_k$ while taking $D_{kk}=0$. The energy \eqref{eq:sz_energy} will be minimized with respect to the single-particle energies $\{\varepsilon\}$ and the pairing strength $g$. Again, a particular choice of $\{\varepsilon\}$ and $g$ dictates the values of the rapidities $\{u\}$ through Richardson's equations \eqref{eq:rich}.

\subsection{Slater Determinant Expansion}
It's important to understand that expansion in Slater determinants \emph{is not productive}. If we wanted to work with Slater determinants there is no reason to use RG states at all. Slater determinants are a basis for the Hilbert space, so we can project against a complete set of Slater determinants labelled with indices $\{i\}$
\begin{align}
\ket{\{u\}} =  \sum_{\{i\}} \ket{\{i\}}\bra{\{i\}} \ket{\{u\}} = \sum_{\{i\}} C^{\{i\}}_{\{u\}} \ket{\{i\}}.
\end{align}
The coefficients $C^{\{i\}}_{\{u\}} = \braket{\{i\}|\{u\}}$ are symmetric sums (permanents)
\begin{align}
C^{\{i\}}_{\{u\}} = \underset{a,b} {\text{per}} \; \left( \frac{1}{u_a - \varepsilon_{i_b}}  \right) = 
\sum_{\sigma \in \mathfrak{S}_M} \frac{1}{u_1 - \varepsilon_{i_{\sigma(1)}}} \frac{1}{u_2 - \varepsilon_{i_{\sigma(2)}}}  \dots
\frac{1}{u_M - \varepsilon_{i_{\sigma(M)}}}.
\end{align}
The summation is over the $M!$ elements of the symmetric group $\mathfrak{S}_M$. Computing a permanent is in general intractable, though in this particular case they may computed as a ratio of determinants 
\begin{align}
\underset{a,b} {\text{per}} \; \left( \frac{1}{u_a - \varepsilon_{i_b}}  \right) \underset{a,b} {\det} \; \left( \frac{1}{u_a - \varepsilon_{i_b}}  \right) =
\underset{a,b} {\det} \; \left( \frac{1}{(u_a - \varepsilon_{i_b})^2}  \right)
\end{align}
which is a result of Borchardt.\cite{borchardt:1857} It is also a simple determinant in terms of the EBV:\cite{gaudin_book,faribault:2012}
\begin{align}
C^{\{i\}}_{\{u\}} = \det Q
\end{align}
with
\begin{align}
Q_{ij} =
\begin{cases}
\frac{1}{g}U_i + \sum^N_{k (\neq i)=1} \frac{1}{\varepsilon_k - \varepsilon_i}, \quad & i = j \\
\frac{1}{\varepsilon_i - \varepsilon_j}, \quad & i \neq j.
\end{cases}
\end{align}
However, there are $\binom{N}{M}$ permanents to evaluate unless we could choose a smaller basis of Slater determinants to represent an RG state.\cite{fecteau:2021} In this case, we should give up on RG states and just work directly with Slater determinants.

The scalar product between two RG states is
\begin{align} 
\braket{ \{v\} | \{u\} } &= \sum_{\{i\}} \sum_{\{j\}} C^{\{v\}}_{\{j\}} C^{\{i\}}_{\{u\}} \braket{ \{j\} | \{i\} } \\
&= \sum_{\{i\}} C^{\{v\}}_{\{i\}} C^{\{i\}}_{\{u\}} \label{eq:apig_slavnov_sd}
\end{align}
since the Slater determinants $\ket{\{i\}}$ form an orthonormal basis. The 1-body density matrix elements are
\begin{align}
\frac{1}{2} \braket{\{v\} | \hat{n}_k | \{u\}} &= \sum_{\{i\}} C^{\{v\}}_{\{i\}} C^{\{i\}}_{\{u\}} \; \delta(k\in \{i\}) \label{eq:apig_ww_1rdm}
\end{align}
are sums over permanents including the $k$th column. The indicator function $\delta (k \in \{i\})$ ensures that index $k$ is an element of the set $\{i\}$ Similarly, 
\begin{align}
\frac{1}{4} \braket{\{v\} | \hat{n}_k \hat{n}_l | \{u\}} = \sum_{\{i\}} C^{\{v\}}_{\{i\}} C^{\{i\}}_{\{u\}} \; \delta(k,l\in \{i\}) \label{eq:apig_ww_Dij}
\end{align}
is a summation over Slater determinants involving both $k$ and $l$, and
\begin{align} \label{eq:apig_ww_Pij}
\braket{ \{v\} | S^+_k S^-_l | \{u\}} &= \sum_{\{i\}} C^{\{v\}}_{\{i\}} C^{(\{i\} \setminus k \cup l  )}_{\{u\}} \; \delta(k\in \{i\}) \delta (l \notin \{i\})
\end{align}
is a sum over Slater determinants that differ by one doubly-occupied orbital. The notation $\{i\} \setminus k \cup l$ is understood as the set $\{i\}$ in which $k$ is replaced by $l$. These expressions are valid for any $\{u\}$ and $\{v\}$ regardless of whether they are solutions of Richardson's equations or not. They are computationally intractable as they are essentially the same expressions as for DOCI.\cite{weinhold:1967a,weinhold:1967b}

\subsection{Rapidity Expressions}
As expressions for the RDM and TDM elements are now known in terms of the EBV, expressions in terms of rapidities are obsolete from a numerical viewpoint. However, they are instructive to work through, and are not at all complicated to compute. The key result is that the scalar product $\braket{\{v\}|\{u\}}$ for $\{v\}$ on-shell and arbitrary $\{u\}$ is a single determinant. Slavnov proved this result for the ABA of the 6-vertex model,\cite{korepin:1982,slavnov:1989,korepin_book,belliard:2019} and RG states are a limiting case.\cite{zhou:2002} In particular
\begin{align} \label{eq:slavnov}
\braket{\{v\}|\{u\}} = K (\{v\},\{u\} ) \det R(\{v\},\{u\})
\end{align}
with
\begin{align} \label{eq:K_fac}
K (\{v\},\{u\} ) = \frac{\prod^M_{a,b=1} (v_a-u_b)}{\prod^M_{a=1} \prod^M_{b=a+1} (u_a-u_b)(v_b-v_a)}
\end{align}
and
\begin{align} \label{eq:slv_mat}
R_{ab} = \frac{1}{(v_a - u_b)^2} \left( 2 \alpha (u) - \sum^M_{c (\neq a)=1} \frac{2}{(u_b-v_c)} \right).
\end{align}
Notice that the single-particle information does not appear explicitly. Instead, the matrix $R$ only requires the rapidities $\{u\}$ and the elements
\begin{align}
\alpha (u) = \frac{1}{g} + \frac{1}{2} \sum^N_{i=1} \frac{1}{u - \varepsilon_i}.
\end{align}
The square of the norm is obtained by taking the limit $\{u\} \rightarrow \{v\}$, or
\begin{align}
\braket{\{v\}|\{v\}} = \det G
\end{align}
where $G$ is usually called the Gaudin matrix. $G$ is the Jacobian of Richardson's equations
\begin{align} \label{eq:gmat}
G_{ab} = 
\begin{cases}
\sum^N_{i=1} \frac{1}{(v_a - \varepsilon_i)^2} -\sum^M_{c (\neq a)=1} \frac{2}{( v_a - v_c )^2} , &\quad a = b \\
\frac{2}{(v_a - v_b )^2}, &\quad a \neq b.
\end{cases}
\end{align}

The strategy to evaluate the expectation of an operator $\hat{O}$ is simple: $\hat{O}$ is moved past each $S^+(u)$ until it acts on the vacuum. This gives a sum of scalar products that may be evaluated as limits of \eqref{eq:slavnov}. In particular,
\begin{align}
[\hat{n}_k,S^+(u_a)] = \frac{2S^+_k}{u_a - \varepsilon_k}
\end{align}
gives 
\begin{align}
\frac{1}{2}\braket{\{v\} | \hat{n}_k | \{u\}} &= \sum^M_{a=1} \frac{1}{u_a - \varepsilon_k}\braket{ \{v\} | S^+_k | \{u\}_a} \\
\frac{1}{4}\braket{\{v\} | \hat{n}_k \hat{n}_l | \{u\}} &= \sum^M_{a=1}\sum^M_{b (\neq a) =1} \frac{1}{(u_a-\varepsilon_k)(u_b-\varepsilon_l)} \braket{ \{v\} | S^+_k S^+_l | \{u\}_{a,b}}
\end{align} 
where $\{u\}_a$ is the set $\{u\}$ without $u_a$ etc. Moving $S^-_l$ past $S^+(u_a)$ gives an $S^z_l$, which must also be moved to the right giving 
\begin{align}
\braket{ \{v\} | S^+_k S^-_l | \{u\}} &= \sum^M_{a=1} \frac{1}{u_a - \varepsilon_l} \braket{ \{v\} | S^+_k | \{u\}_a} 
- \sum^M_{a=1}\sum^M_{b (\neq a) =1} \frac{1}{(u_a-\varepsilon_l)(u_b-\varepsilon_l)}  \braket{ \{v\} | S^+_k S^+_l | \{u\}_{a,b}}.
\end{align}
Notice that this treatment will give an asymmetric expression for $P_{kl}$ as the state on the right is treated differently than the state on the left. The expression for $P_{kl}$ looks different than $P_{lk}$, but numerically they are identical. 

The scalar products $\braket{\{v\}|S^+_k|\{u\}_a}$ and $\braket{\{v\}|S^+_k S^+_l|\{u\}_{a,b}}$ are what we call form factors. The individual $S^+_k$ are the residues of the RG pairs when the rapidity coincides with $\varepsilon_k$
\begin{align}
S^+_k = \lim_{u\rightarrow \varepsilon_k} (u-\varepsilon_k) S^+(u).
\end{align}
The form factors can thus be evaluated as the corresponding residues of the scalar product \eqref{eq:slavnov}
\begin{align}
\braket{\{v\}|S^+_k|\{u\}_a} = \lim_{u_a \rightarrow \varepsilon_k} (u_a - \varepsilon_k) \braket{\{v\}|\{u\}}.
\end{align}
The simple pole $\frac{1}{u_a - \varepsilon_k}$ occurs only in the $a$th column. Further, the pole appears explicitly in $\alpha (u_a)$
\begin{align}
\lim_{u_a \rightarrow \varepsilon_k} (u_a-\varepsilon_k) \alpha (u_a) = \frac{1}{2} + 
 \lim_{u_a \rightarrow \varepsilon_k} (u_a-\varepsilon_k) \left(\frac{1}{g} + \frac{1}{2} \sum^N_{i (\neq k)=1} \frac{1}{u_a - \varepsilon_i} \right) = \frac{1}{2}
\end{align}
so only the term proportional to $\alpha (u_a)$ survives. In the matrix $R$, the $a$th column is replaced by the vector
\begin{align} \label{eq:b_RHS}
\textbf{b}_k = 
\begin{pmatrix}
\frac{1}{(v_1 - \varepsilon_k)^2} \\
\vdots \\
\frac{1}{(v_M - \varepsilon_k)^2}
\end{pmatrix}.
\end{align}
$K$ is modified, but when we take the limit $\{u\} \rightarrow \{v\}$ we get simply
\begin{align}
\braket{\{v\}| S^+_k | \{v\}_a} = (v_a - \varepsilon_k ) \det G^k_a
\end{align}
where $G^k_a$ is the Gaudin matrix whose $a$th column has been replaced with the vector \eqref{eq:b_RHS}. The elements of the 1-RDM are computed 
\begin{align}
\gamma_k  = \frac{1}{\det G} \sum^M_{a=1} \frac{\braket{\{v\}|S^+_k|\{v\}_a}}{v_a - \varepsilon_k} = \sum^M_{a=1} \frac{\det G^k_a}{\det G}.
\end{align}
Computing a ratio of two determinants is wasteful as Cramer's rule ensures that the elements of the sum on the RHS are the solutions of a set of linear equations with matrix $G$ and RHS $\textbf{b}_k$. It is not difficult to show that the derivatives $\frac{\partial \textbf{v}}{\partial \varepsilon_k}$ are the solutions of the same set of linear equations
\begin{align} \label{eq:G_lin_eq}
G \frac{\partial \textbf{v}}{\partial \varepsilon_k} = \textbf{b}_k
\end{align}
so the elements of the 1-RDM are
\begin{align}
\gamma_k = \sum^M_{a=1} \frac{\partial v_a}{\partial \varepsilon_k}.
\end{align}
This result is obtainable directly from the Hellmann-Feynman theorem since $E_{BCS} = \sum_a u_a$ and $\hat{n}_k = \frac{\partial \hat{H}_{BCS}}{\partial \varepsilon_k}$, but this approach does not lead to expressions for the 2-RDM elements, nor the TDM elements at all. 

Computing the two-pair form factor presents no more difficulty, except that we obtain determinants of matrices that are $G$ with more than one replaced column. In this case we have that
\begin{align} \label{eq:col_jacobi}
\frac{\det G^{kl}_{ab}}{\det G} = \frac{\det G^k_a}{\det G} \frac{\det G^l_b}{\det G} - \frac{\det G^l_a}{\det G} \frac{\det G^k_b}{\det G}
\end{align}
in which $G^{kl}_{ab}$ is the matrix $G$ with the $a$th column replaced with $\textbf{b}_k$ and the $b$th column replaced with $\textbf{b}_l$. This result extends to any order: the scaled determinant of a matrix with $p$ column replacements is a $p \times p$ determinant of its single-column replacements. Equation \eqref{eq:col_jacobi} is valid for the updates of any non-singular matrix. It is similar enough to a theorem of Jacobi for cofactors that we have referred to it as Jacobi's theorem.\cite{vein_book} 

For $k\neq l$, the non-zero elements of the 2-RDM are
\begin{align}
D_{kl} &= \sum^M_{a=1}\sum^M_{b=a+1} \frac{(v_a - \varepsilon_k)(v_b - \varepsilon_l)+(v_a - \varepsilon_l)(v_b - \varepsilon_k)}{(\varepsilon_k - \varepsilon_l)(v_b - v_a)} \left( \frac{\partial v_a}{\partial \varepsilon_k} \frac{\partial v_b}{\partial \varepsilon_l} - \frac{\partial v_a}{\partial \varepsilon_l} \frac{\partial v_b}{\partial \varepsilon_k} \right) \\
P_{kl} &= \sum^M_{a=1} \frac{(v_a-\varepsilon_k)}{(v_a - \varepsilon_l)} \frac{\partial v_a}{\partial \varepsilon_k} -2 \sum^M_{a=1}\sum^M_{b=a+1} \frac{(v_a-\varepsilon_k)(v_b-\varepsilon_k)}{(\varepsilon_k - \varepsilon_l)(v_b - v_a)} \left( \frac{\partial v_a}{\partial \varepsilon_k} \frac{\partial v_b}{\partial \varepsilon_l} - \frac{\partial v_a}{\partial \varepsilon_l} \frac{\partial v_b}{\partial \varepsilon_k} \right).
\end{align}
The primitive elements to compute for the 2-RDM are the solutions of the $N$ sets of linear equations \eqref{eq:G_lin_eq}, one for each RHS $\textbf{b}_k$. Because the same matrix $G$ appears in each set of equations, we can LU-decompose $G$ to be more efficient. The scaling bottleneck should come from evaluating the 2-RDM: there are $\mathcal{O}(N^2)$ elements, and each requires a summation over $M^2$ precomputed objects. 

Transition density matrix elements are less clean. They are obtained in the same manner, but instead of taking the limit $\{u\}\rightarrow \{v\}$ we send $\{u\} \rightarrow \{w\}$ where $\{w\}$ is a solution of Richardson's equations \emph{distinct} from $\{v\}$. We get
\begin{align}
\gamma^{vw}_k &:= \frac{1}{2} \braket{\{v\}|\hat{n}_k | \{w\}} = K \sum^M_{a=1} \mu^k_a \det R^k_a \\
D^{vw}_{kl}   &:= \frac{1}{4} \braket{\{v\}|\hat{n}_k \hat{n}_l |\{w\}} = K \sum^M_{a=1}\sum^M_{b=a+1} \mu^k_a \mu^l_b \frac{(w_a-\varepsilon_k)(w_b-\varepsilon_l)+(w_a-\varepsilon_l)(w_b-\varepsilon_k)}{(\varepsilon_k-\varepsilon_l)(w_b-w_a)} \det R^{kl}_{ab} \\
P^{vw}_{kl}   &:=  \braket{\{v\}|S^+_kS^-_l | \{w\}}= K\sum^M_{a=1} \mu^k_a \frac{w_a-\varepsilon_k}{w_a - \varepsilon_l} \det R^k_a 
- 2K \sum^M_{a=1}\sum^M_{b=a+1} \mu^k_a \mu^l_b \frac{(w_a-\varepsilon_k)(w_b-\varepsilon_k)}{(\varepsilon_k-\varepsilon_l)(w_b-w_a)} \det R^{kl}_{ab}
\end{align}
where $K$ is understood as equation \eqref{eq:K_fac}, with $K(\{v\},\{w\})$. The common factor
\begin{align}
\mu^k_a = \frac{\prod^M_{c=1} (v_c - \varepsilon_k)}{\prod^M_{c=1} (v_c - w_a)} \frac{\prod^M_{a=1}\prod^M_{b(\neq a)=1} (w_b-w_a)}{\prod^M_{b=1} (w_b - \varepsilon_k)}
\end{align}
is dangerous. In principle rapidities for distinct states are distinct, but this is not always true. We could predict and handle zero denominators based on bitstrings and $\{\varepsilon\}$ patterns, but this becomes intractable in general. $R^{kl}_{ab}$ is the matrix $R$ whose $a$th column has been replaced by $\textbf{b}_k$, and $b$th column replaced with $\textbf{b}_l$. Notice that $R$ is singular: it has rank $M-1$ since 
\begin{align}
\sum^M_{a=1} c_a R_a = 0
\end{align}
with $R_a$ the $a$th column of $R$, and
\begin{align}
c_a = \prod^M_{b=1} (w_a - v_b)^2. 
\end{align}
As a result, Cramer's rule cannot be used in the same way as for the RDM elements.  

Sklyanin was the first to compute scalar products and correlation functions for RG states by using an exponential generating function.\cite{sklyanin:1999} His expressions were intractable in general, but were employed by Amico and Osterloh,\cite{amico:2002} and us for off-shell RG states.\cite{moisset:2022a} Faribault, Calabrese and Caux wrote the first feasible expressions in terms of determinants\cite{faribault:2008,faribault:2010} that we employed in our first variational calculations.\cite{johnson:2020} These expressions scale polynomially, but are still expensive as they require computing $M$ determinants for \emph{each} element of the 2-RDM. The best expressions, that avoid determinants altogether with Cramer's rule, were first presented by Gorohovsky and Bettelheim.\cite{gorohovsky:2011} Finally, we derived the expressions for the 2-RDM in the basis of RG pairs.\cite{fecteau:2020} As the matrix elements between Slater determinants in the MO basis are simple, we had hoped that the TDM elements between RG states would be simpler numerically in the basis of RG pairs. This turned out not to be the case, and introduced numerical problems near the critical points of Richardson's equations.\cite{johnson:2021}

\subsection{EBV expressions}
Given that we compute the EBV first, it would be better if we could just compute the matrix elements directly from them. Indeed this is possible. The RG states themselves do not seem to have a clean expression in terms of the EBV, so we will continue to label them with rapidities. The scalar product $\braket{\{v\}|\{u\}}$ has a determinant expression in terms of the EBV\cite{claeys:2017b}
\begin{align} \label{eq:ebv_scalar}
\braket{\{v\}|\{u\}} = \eta \det J
\end{align}
with the factor
\begin{align}
\eta = (-1)^{N-M} \left( \frac{g}{2} \right)^{-2M}
\end{align}
and the matrix
\begin{align}
J_{kl} =
\begin{cases}
U_k + V_k - 2 + \sum^N_{i (\neq k)=1} \frac{g}{\varepsilon_i - \varepsilon_k}, \quad & k=l, \\
\frac{g}{\varepsilon_k - \varepsilon_l}, \quad & k\neq l.
\end{cases}
\end{align}
The factor $\eta$ and the matrix $J$ differ from references\cite{claeys:2017b,faribault:2022,johnson:2023} but are consistent. Equation \eqref{eq:ebv_scalar} is valid provided either $\{U\}$ or $\{V\}$ is a solution of the EBV equations. The matrix elements are computed in the same manner as for rapidities. We take a sum over form factors, which are themselves obtained as residues of the scalar product \eqref{eq:ebv_scalar}. The difference is that the intermediate manipulations are much more complicated and tedious, so we will present only the final results. The development is presented in detail in Appendix B of reference\cite{faribault:2022}. The TDM elements will be presented first as the RDM elements are a special case. Fix $\{W\}$ a set of solutions of the EBV equations. The non-normalized 1-TDM elements are
\begin{align}
\gamma^{vw}_k := \frac{1}{2} \braket{\{v\}|\hat{n}_k|\{w\}} = \sum^N_{l=1} W_l A[J]^{l,k}
\end{align}
where
\begin{align}
A[J]^{l,k} = (-1)^{k+l} \eta \det J^{l,k}
\end{align}
is the $(l,k)$-th first cofactor of $J$ and $J^{l,k}$ is $J$ without the $l$th row and the $k$th column. We will also need second cofactors, 
\begin{align}
A[J]^{ij,kl} = (-1)^{i+j+k+l+h(i-j)+h(k-l)} \eta \det J^{ij,kl}
\end{align}
where $J^{ij,kl}$ is $J$ without the $i$th and $j$th rows, and the $k$th and $l$th columns, while $h(x)$ is the Heaviside function
\begin{align}
h(x) = 
\begin{cases}
1 & x > 0 \\
0 & x \leq 0.
\end{cases}
\end{align}
Second cofactors are antisymmetric with the exchange of $i$ and $j$ or $k$ and $l$. With
\begin{align}
L_{kl} = W_k W_l + g\frac{W_k - W_l}{\varepsilon_k - \varepsilon_l}
\end{align}
the non-normalized elements of the 2-TDM are
\begin{align}
D^{vw}_{kl} :&= \frac{1}{4} \braket{\{v\}|\hat{n}_k \hat{n}_l |\{w\}} \\
 &= L_{kl} A[J]^{kl,kl} + \sum^N_{i (\neq k,l)=1} L_{il}A[J]^{il,kl} + \sum^N_{i(\neq k,l)=1} L_{ik}A[J]^{ki,kl} \nonumber \\
 &+ \sum^N_{i (\neq k,l)=1} \sum^N_{j (\neq k,l) = i +1} \frac{(\varepsilon_k-\varepsilon_i)(\varepsilon_l-\varepsilon_j) +(\varepsilon_k-\varepsilon_j)(\varepsilon_l-\varepsilon_i)}{(\varepsilon_k-\varepsilon_l)(\varepsilon_j-\varepsilon_i)} L_{ij} A[J]^{ij,kl}
\end{align}
and
\begin{align}
P^{vw}_{kl} :&= \braket{\{v\}|S^+_k S^-_l | \{w\}} \\
&= \left(W_l + \frac{(\varepsilon_k-\varepsilon_l)}{g}(W_l W_l - W_l J_{ll}) \right) A[J]^{l,k} + \sum^N_{i (\neq k,l)=1} \frac{\varepsilon_i - \varepsilon_k}{\varepsilon_i - \varepsilon_l} W_i A[J]^{i,k} \nonumber \\
&- 2 \sum^N_{i (\neq k,l)=1} \frac{\varepsilon_k - \varepsilon_i}{\varepsilon_l - \varepsilon_i} L_{il} A[J]^{il,kl} 
 - 2 \sum^N_{i (\neq k,l)=1} \sum^N_{j (\neq k,l) = i +1} \frac{(\varepsilon_k-\varepsilon_i)(\varepsilon_k-\varepsilon_j)}{(\varepsilon_k-\varepsilon_l)(\varepsilon_j-\varepsilon_i)} L_{ij} A[J]^{ij,kl}.
\end{align}
Our published results\cite{johnson:2023} for RGCI computed these elements \emph{very poorly} by computing the $N^4$ second cofactors for each pair of states. Clearly this is unacceptable, but is certainly improvable.

When the two states are the same, the square of the norm simplifies
\begin{align}
\braket{\{v\}|\{v\}} = \eta \det \bar{J}
\end{align}
where $\bar{J}$ is the matrix $J$ where both sets of EBV are identical. Notice that $\bar{J}$ is the Jacobian of the EBV equations \eqref{eq:ebv_eq}, without the $N+1$st row. As was the case for the rapidity expressions, \emph{normalized} RDM elements are particularly simple to compute as scaled second cofactors are $2 \times 2$ determinants of first cofactors
\begin{align}
\frac{A[J]^{ij,kl}}{\det \bar{J}} = \frac{A[J]^{i,k}}{\det \bar{J}}\frac{A[J]^{j,l}}{\det \bar{J}} - \frac{A[J]^{i,l}}{\det \bar{J}}\frac{A[J]^{j,k}}{\det \bar{J}}.
\end{align}
\emph{This} is a theorem of Jacobi and extends generally: a $p$th order scaled cofactor is a $p \times p$ determinant of first cofactors.\cite{vein_book} Thus, the primitive elements to compute are the scaled cofactors of the matrix $J$. Rather than compute $N^2$ determinants, notice that the adjugate formula for the inverse gives
\begin{align}
\bar{J}^{-1}_{ij} = \frac{A[J]^{j,i}}{\det \bar{J}},
\end{align}
thus simply by inverting $\bar{J}$, we obtain all the primitives we need at once. So long as $\{\varepsilon\}$ is non-degenerate, $\bar{J}$ is well-conditioned. The normalized 1-RDM elements are
\begin{align}
\gamma_k = \sum^N_{l=1} V_l \bar{J}^{-1}_{kl}
\end{align}
while the 2-RDM elements are
\begin{align}
D_{kl} &= L_{kl} (\bar{J}^{-1}_{kk}\bar{J}^{-1}_{ll} - \bar{J}^{-1}_{lk}\bar{J}^{-1}_{kl}) 
+ \sum^N_{j (\neq l)=1} L_{jk} (\bar{J}^{-1}_{kk}\bar{J}^{-1}_{lj} - \bar{J}^{-1}_{lk}\bar{J}^{-1}_{kj})
+ \sum^N_{j (\neq k)=1} L_{jl} (\bar{J}^{-1}_{kj}\bar{J}^{-1}_{ll} - \bar{J}^{-1}_{lj}\bar{J}^{-1}_{kl}) \nonumber \\
&+ \sum^N_{i (\neq k,l)=1} \sum^N_{j (\neq k,l) = i +1} \frac{(\varepsilon_k-\varepsilon_i)(\varepsilon_l-\varepsilon_j)+(\varepsilon_k-\varepsilon_j)(\varepsilon_l-\varepsilon_i)}{(\varepsilon_k - \varepsilon_l)(\varepsilon_j - \varepsilon_i)}L_{ij} (\bar{J}^{-1}_{ki}\bar{J}^{-1}_{lj} - \bar{J}^{-1}_{li}\bar{J}^{-1}_{kj})
\end{align}
and
\begin{align}
P_{kl} &= \left(2V_l + \sum^N_{i (\neq k,l)=1} \frac{\varepsilon_i - \varepsilon_k}{\varepsilon_i - \varepsilon_l} V_i - 2M \right) \bar{J}^{-1}_{kl} \nonumber \\
&+ \sum^N_{i (\neq k,l)=1} \frac{\varepsilon_i - \varepsilon_k}{\varepsilon_i - \varepsilon_l} (V_i \bar{J}^{-1}_{ki} -2L_{il} (\bar{J}^{-1}_{ki}\bar{J}^{-1}_{ll} - \bar{J}^{-1}_{li}\bar{J}^{-1}_{kl}))
\nonumber \\
&- 2\sum^N_{i (\neq k,l)=1} \sum^N_{j (\neq k,l) = i +1} \frac{(\varepsilon_k - \varepsilon_l)(\varepsilon_k-\varepsilon_j)}{(\varepsilon_k-\varepsilon_l)(\varepsilon_j-\varepsilon_i)}L_{ij}
(\bar{J}^{-1}_{ki}\bar{J}^{-1}_{lj} - \bar{J}^{-1}_{li}\bar{J}^{-1}_{kj}).
\end{align}

Computing the 2-RDM in terms of the EBV requires $N^2$ double sums of known objects. Unless there are many more spatial orbitals than pairs, there is thus no reason to compute rapidities at all.

\subsection{Consistency Checks}
The numerical solution of the EBV equations can have problems if the single particle energies get too close to one another. For pairwise degeneracy, this is not a problem as the solution is incredibly stable, though the number of required $g$ steps becomes large. When using an RG state as a variational wavefunction ansatz, we check a few consistency conditions on the 1- and 2-RDM elements. $\gamma_k$ and $D_{kl}$ have sum rules
\begin{align}
\sum^N_{k=1} \gamma_k &= M \\
\sum^N_{k,l=1} D_{kl} &= M(M-1).
\end{align}
The trace of $P_{kl}$ is the off-diagonal long range order (ODLRO)\cite{yang:1962,dunne:1994} parameter, but a sum rule can be found by evaluating the energy of the reduced BCS Hamiltonian in 2 ways. It can be evaluated with $\gamma_k$ and $P_{kl}$, but there is also a direct expression in terms of the EBV
\begin{align}
E_{BCS} &= \sum^N_{k=1} \varepsilon_k \gamma_k - \frac{g}{2} \sum^N_{k,l=1} P_{kl} \\
&= \frac{g}{2}M(M-N-1) + \frac{1}{2} \sum^N_{k=1} \varepsilon_k V_k.
\end{align}
This second expression can be obtained from the eigenvalue $E_{BCS} = \sum_a v_a$ in terms of the rapidities and using Richardson's equations. Rearranging, we get a sum rule
\begin{align}
\sum^N_{k,l=1} P_{kl} = \frac{1}{g} \sum^N_{k=1} \varepsilon_k \left( 2\gamma_k - V_k \right) + M(N-M+1).
\end{align}
If one of these conditions fails by more than 10$^{-6}$, the point is rejected. In the final results, the conditions are usually satisfied to much higher precision (less than 10$^{-12}$).

When using many RG states, the above conditions are verified for \emph{each} RG state in addition to off-diagonal conditions for each pair of states:
\begin{align}
\sum^N_{k=1} \gamma^{vw}_k &= 0 \\
\sum^N_{k,l=1} D^{vw}_{kl} &= 0 \\
\sum^N_{k,l=1} P^{vw}_{kl} &= \frac{2}{g} \sum^N_{k=1} \varepsilon_k \gamma^{vw}_k.
\end{align}
In the computations we've performed thus far,\cite{johnson:2020,moisset:2022a,fecteau:2022,johnson:2023} the largest violation we've observed is on the order of 10$^{-10}$, but is usually much smaller ($\approx$ 10$^{-13}$).

\section{1D Hydrogen Chains} \label{sec:1dh}
In this section we summarize the results we've computed over several papers\cite{johnson:2020,moisset:2022a,fecteau:2022} concerning 1-dimensional chains equally spaced hydrogen atoms. Thus far, the RG states we've employed are all seniority-zero. This means that they are strictly variational approximations to DOCI. While FCI is invariant to orbital transformations, DOCI is not. The optimal orbitals for DOCI generally resemble bonding/antibonding pairs localised on two centres.\cite{limacher:2014a,ward_thesis} However, the OO-DOCI landscape has many local minima leading to difficulties. FCI results were computed with \textsc{Psi4},\cite{sherrill:1999,parrish:2017} while the OO-DOCI results were computed using the multi-configurational self-consistent field (MCSCF) method in the doubly occupied space using GAMESS(US).\cite{barca:2020}

\begin{figure} [ht!]
	\begin{subfigure}{\textwidth}
		\centering
		\includegraphics[width=0.49\textwidth]{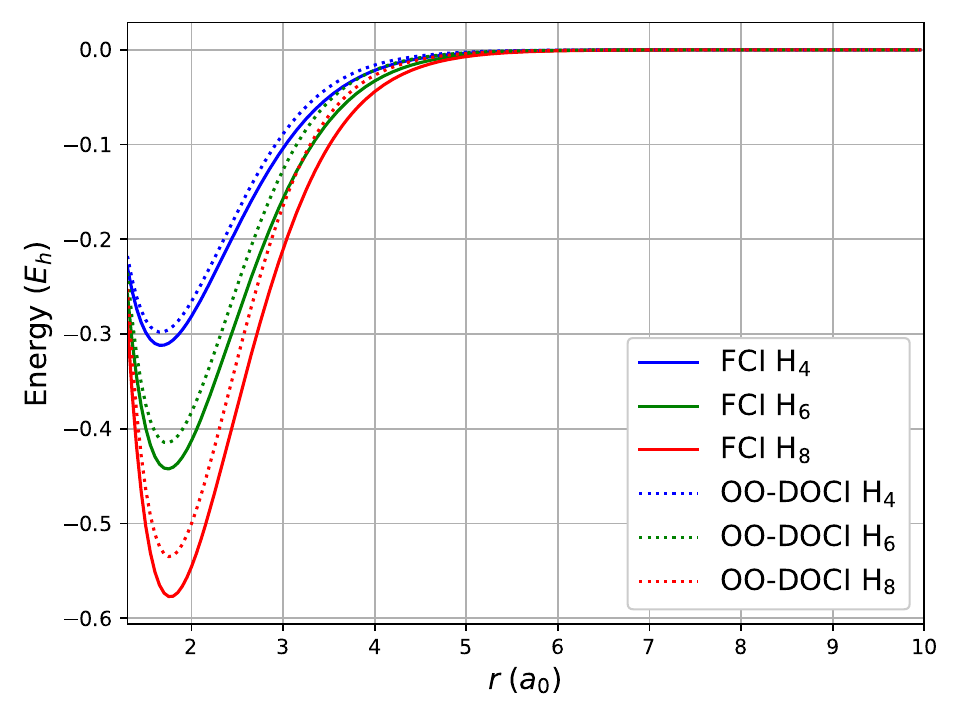} \hfill
		\includegraphics[width=0.49\textwidth]{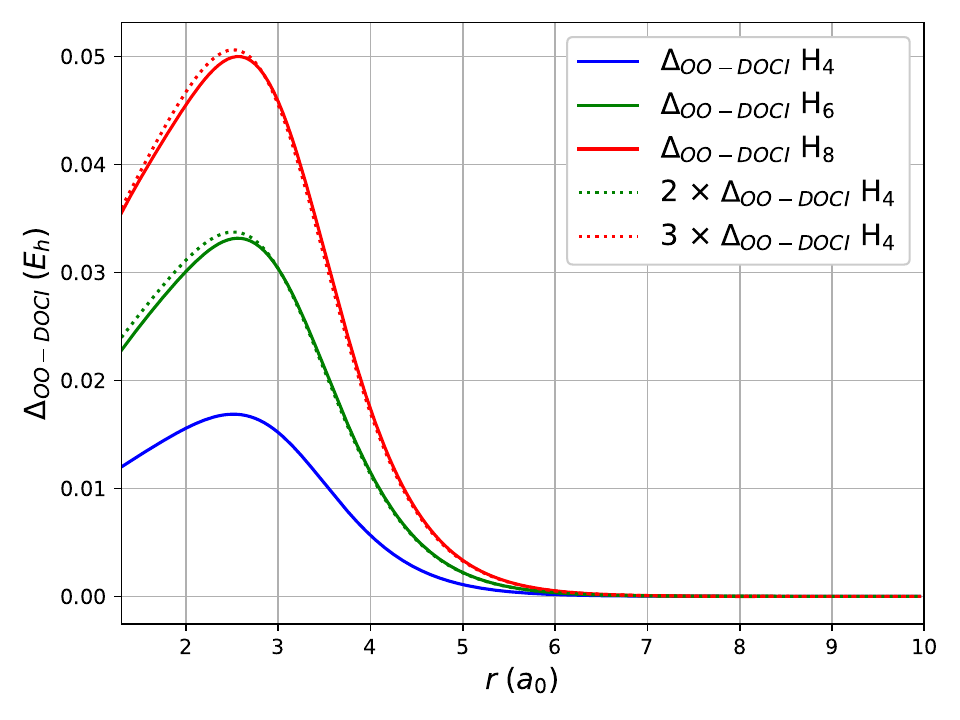}
	\end{subfigure}
	\caption{(left) FCI and OO-DOCI energies for linear H$_4$, H$_6$, and H$_8$. The FCI curves are shifted so that they all dissociate to zero. (right) Errors of OO-DOCI with respect to FCI. Computations performed in the STO-6G basis. Dotted lines represent multiples of the error for H$_4$.}
	\label{fig:doci_fci_h468}
\end{figure}
FCI and OO-DOCI energies for small 1D hydrogen chains are plotted in figure \ref{fig:doci_fci_h468}. The energies have been shifted so that the FCI curves match at dissociation. OO-DOCI is qualitatively the same as FCI everywhere, but misses weak correlation at equilibrium where weakly excited Slater determinants with seniorities two and four will contribute. As $r$ is increased, the error 
\begin{align}
\Delta_{OO-DOCI} = E_{OO-DOCI} - E_{FCI}
\end{align}
goes to zero quite rapidly. Also in figure \ref{fig:doci_fci_h468}, one can see that $\Delta_{OO-DOCI}$ scales with system size for 1D hydrogen chains: the error for H$_6$ is twice the error for H$_4$, while the error for H$_8$ is three times the error for H$_4$. It is thus reasonable to develop approximate functionals to account for the weak contributions from non-zero seniorities.

\begin{figure} [ht!]
	\begin{subfigure}{\textwidth}
		\centering
		\includegraphics[width=0.49\textwidth]{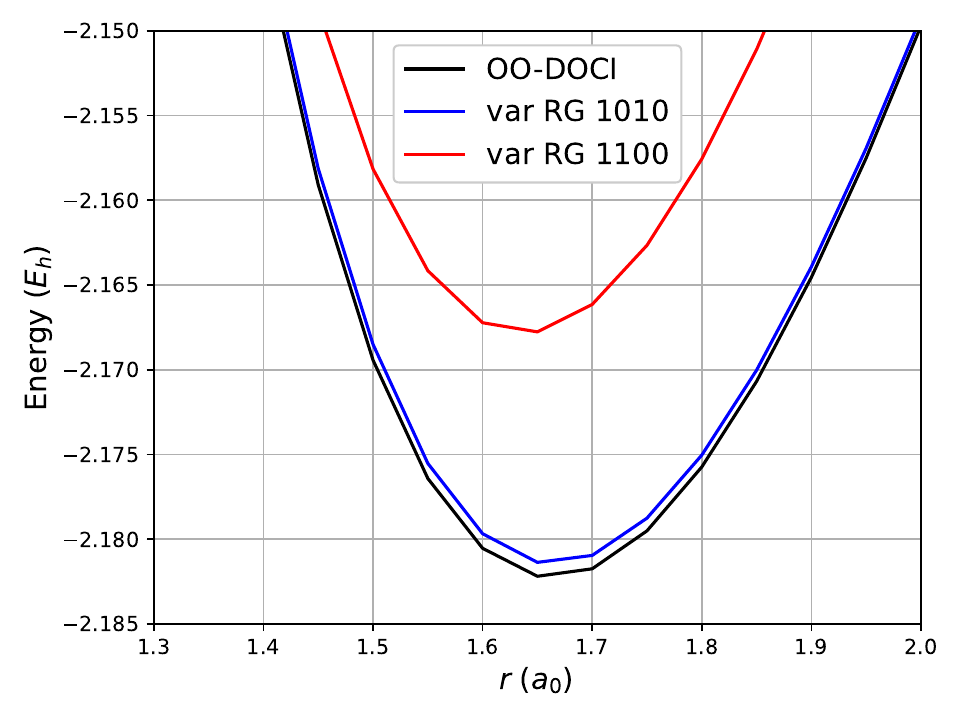} \hfill
		\includegraphics[width=0.49\textwidth]{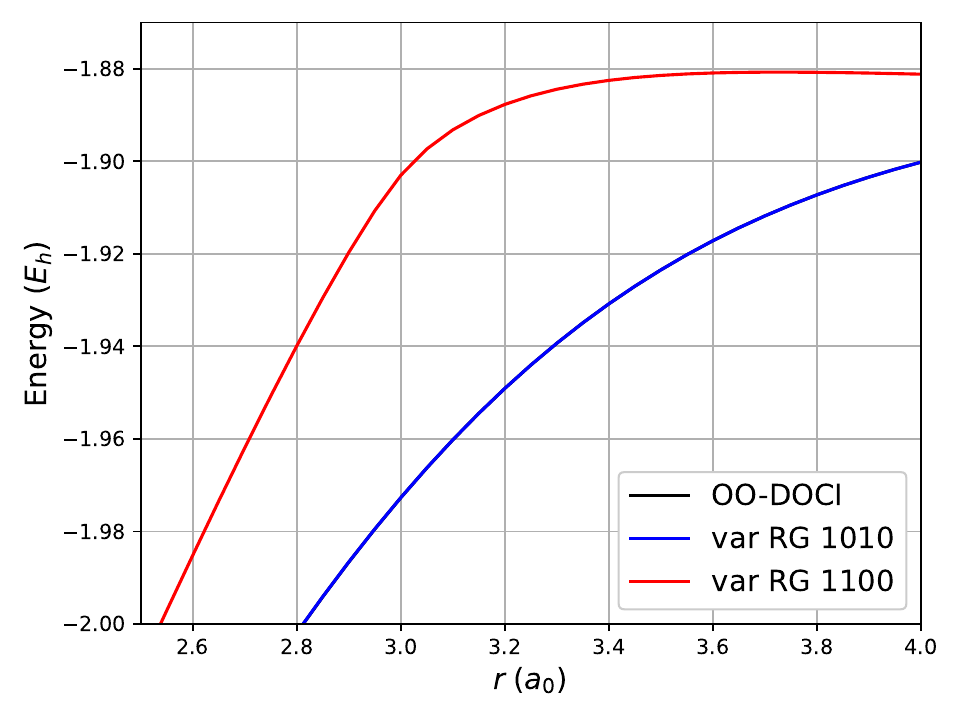}
	\end{subfigure}
	\caption{Symmetric dissociation of linear H$_4$ treated variationally with 1100 and 1010 RG states: near equilibrium (left) and re-coupling region (right). All three curves dissociate to the same limit.\cite{fecteau:2022} Results computed with the OO-DOCI orbitals in the STO-6G basis set.}
	\label{fig:h4_energies}
\end{figure}
Variational computations with RG states for 1D hydrogen chains were first performed in ref. \cite{johnson:2020} In that contribution, the ground state of a reduced BCS Hamiltonian was employed, and the parameters $\{\varepsilon\}$ and $g$ were optimized to minimize the energy of the Hamiltonian \eqref{eq:C_ham} in the basis of OO-DOCI orbitals. For H$_4$, in figure \ref{fig:h4_energies}, using the 1100 RG state gave qualitatively incorrect results: in the re-coupling region, the variational RG 1100 curve looks to have an avoided crossing. The results were \emph{worse} for H$_6$ and H$_8$. Brute-force computations with off-shell RG states\cite{moisset:2022a} agreed very well with OO-DOCI which led us to consider other RG states. We found that the 1010 RG state was near-perfect for H$_4$.\cite{fecteau:2022} 

\begin{figure} [ht!]
	\begin{subfigure}{\textwidth}
		\centering
		\includegraphics[width=0.32\textwidth]{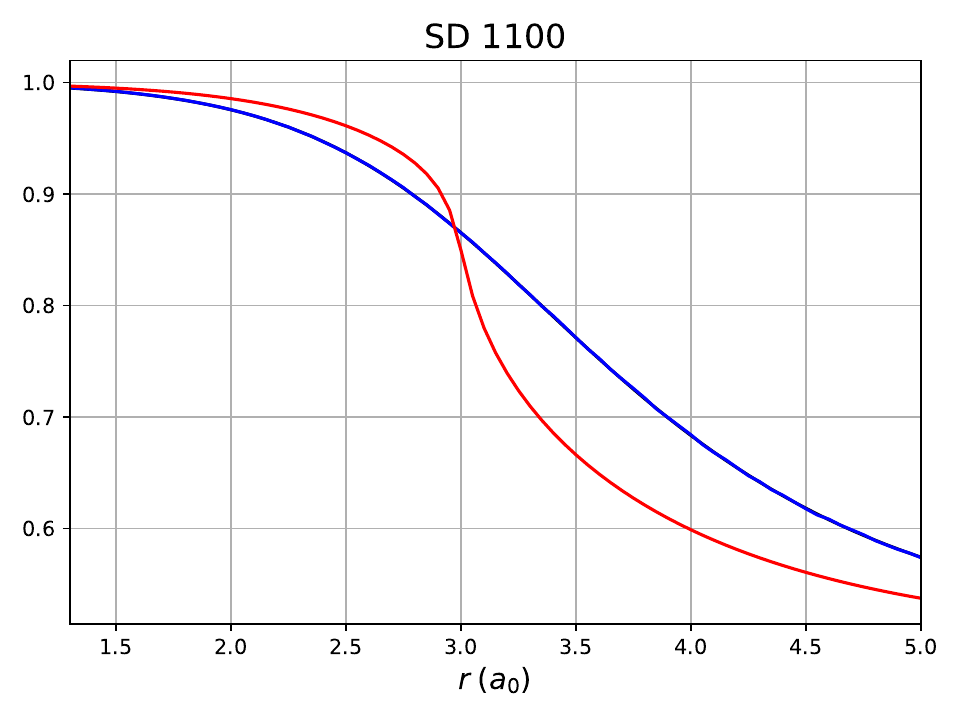} \hfill
		\includegraphics[width=0.32\textwidth]{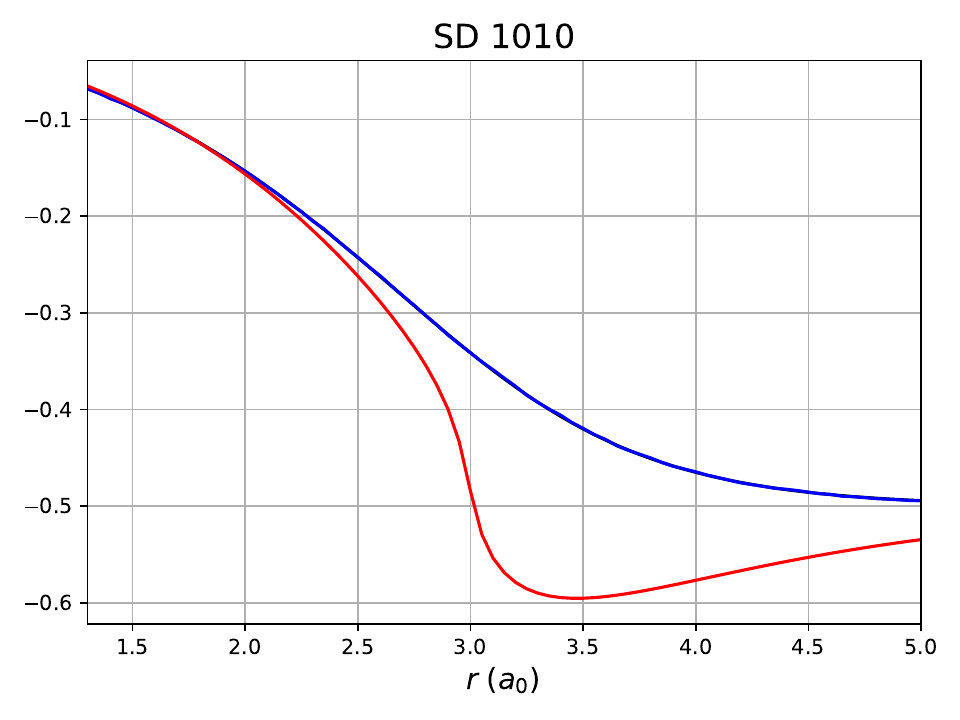} \hfill
		\includegraphics[width=0.32\textwidth]{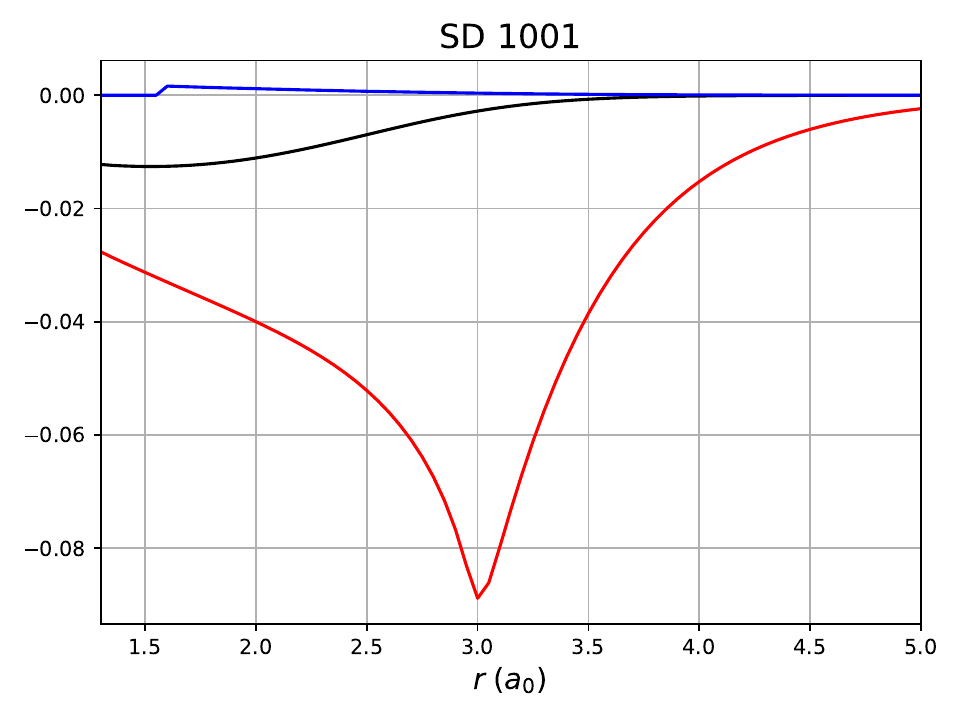} \\
		\includegraphics[width=0.32\textwidth]{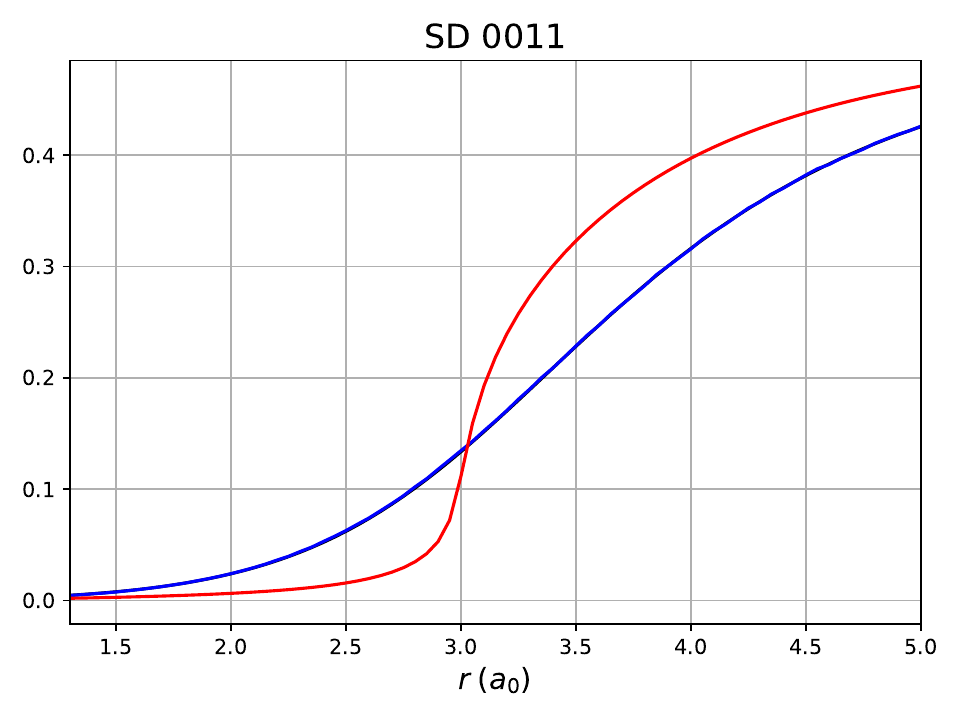} \hfill
		\includegraphics[width=0.32\textwidth]{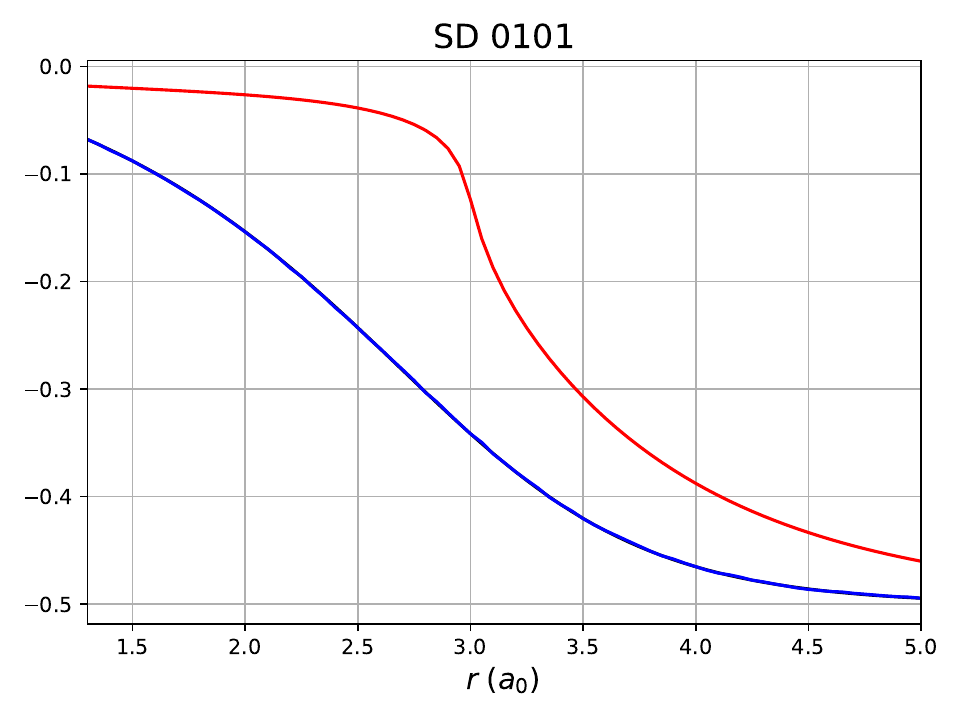} \hfill
		\includegraphics[width=0.32\textwidth]{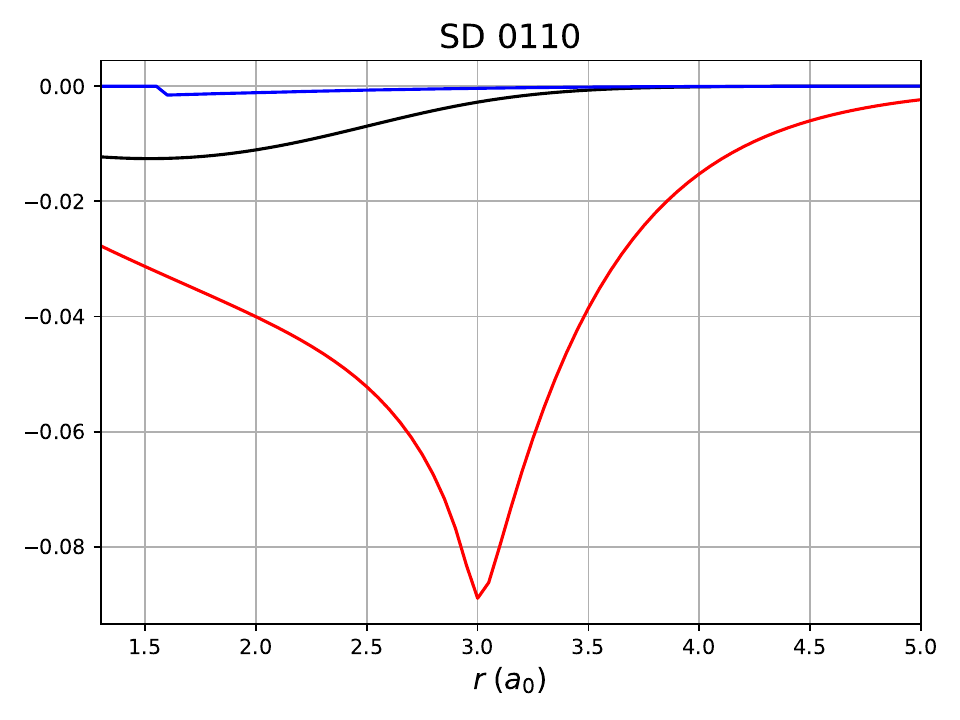} \\
		\includegraphics[scale=0.5]{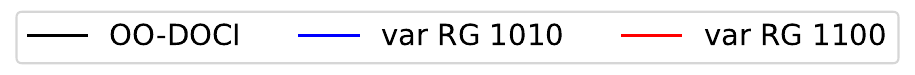} \hfill
	\end{subfigure}
	\caption{Coefficients of Slater determinants contributing to variationally optimized 1100 and 1010 RG states compared with OO-DOCI coefficients.}
	\label{fig:h4_sd}
\end{figure}
The Slater determinant expansions of OO-DOCI, as well as the variational 1100 and 1010 RG states are shown in figure \ref{fig:h4_sd}. One can see that the 1010 RG state is quite similar to OO-DOCI, missing only contributions from the 1001 and 0110 Slater determinants at equilibrium. One can also see that the variationally found 1100 RG state is quite different from OO-DOCI. 

\begin{figure} [ht!]
	\begin{subfigure}{\textwidth}
		\centering
		\includegraphics[width=0.49\textwidth]{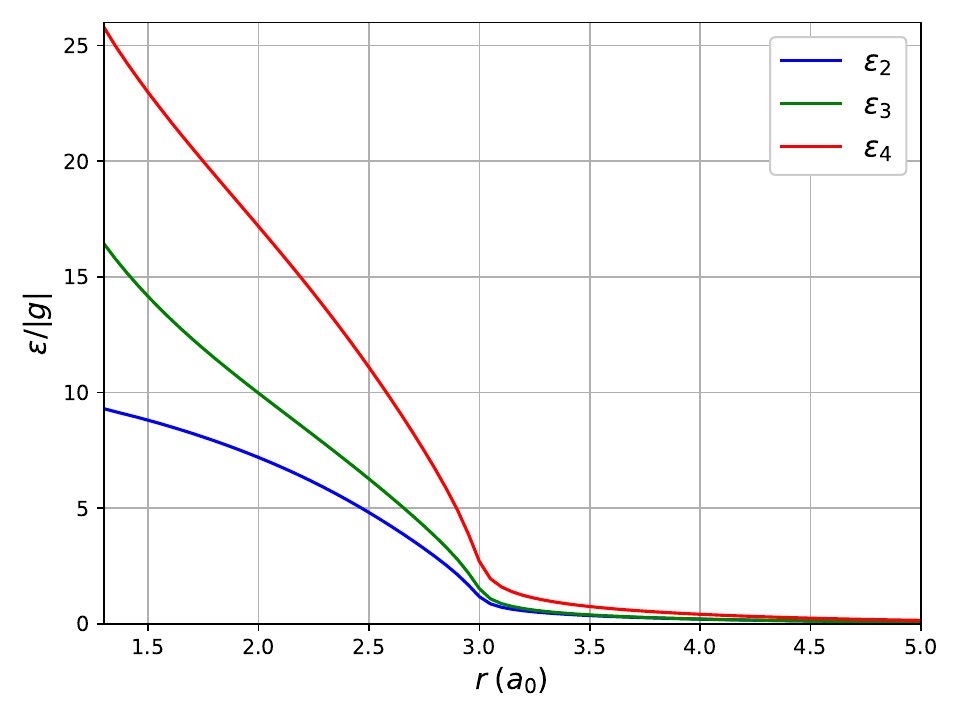} \hfill
		\includegraphics[width=0.49\textwidth]{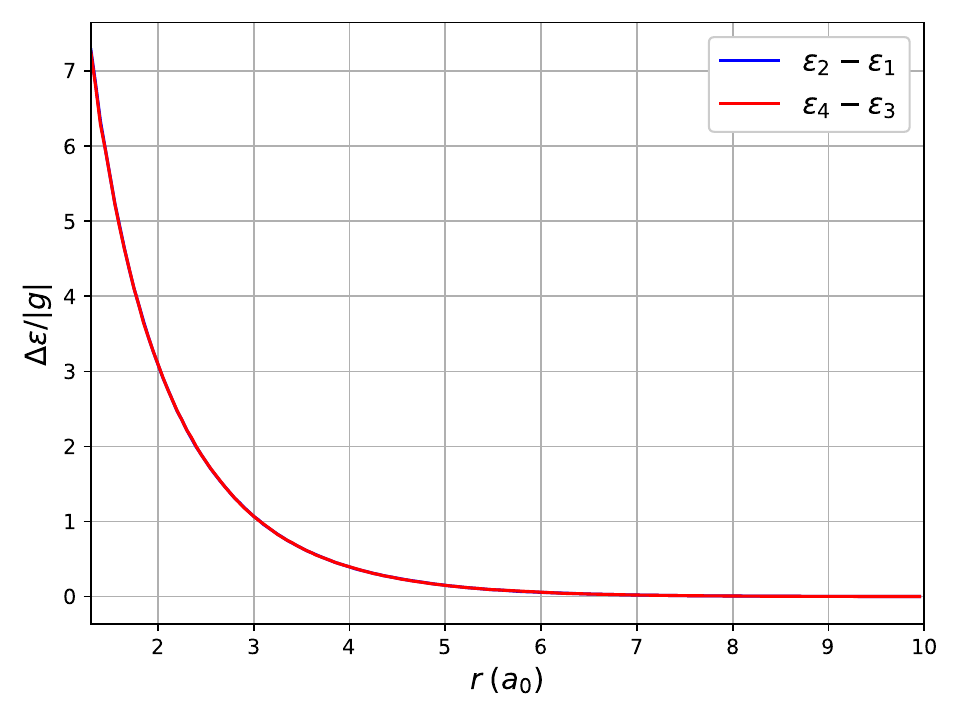}
	\end{subfigure}
	\caption{Left: Variational parameters optimized for 1100 RG state in units of $\vert g \vert$. $\varepsilon_1$ is set to zero. Right: Variational parameters optimized for 1010 RG state in units of $\vert g \vert$. The parameters separate into two sets that are well separated in energy, thus the difference between the parameters in each set is plotted, and the two curves are superimposed. In all cases $g<0$ (repulsive).}
	\label{fig:h4_rg_params}
\end{figure}
The optimal variational parameters found for the 1100 and 1010 RG states are shown in figure \ref{fig:h4_rg_params}. The parameters found for 1100 are quite different than those for 1010, so \emph{these two states are not eigenvectors of the same reduced BCS Hamiltonian}. The variational parameters for the 1100 RG state define a reduced BCS Hamiltonian. The 1010 RG state of this reduced BCS Hamiltonian \emph{is not} the 1010 RG state found variationally, and vice versa. The scale $g$ and reference $\varepsilon_1$ of the energy of the reduced BCS Hamiltonian can be changed with no effect on Richardson's equations and hence the RG states. The interaction $g$ is \emph{always} negative. The $\{\varepsilon\}$ found for the 1100 RG state are plotted relative to $\varepsilon_1$ in units of $\vert g \vert$.  They undergo a transition near $r=3.0\;a_0$ where there is an avoided crossing. They are close in energy relative to $g$. While the levels are not evenly spaced, this RG state is quite similar to the ground state of a PF model in the repulsive regime.

The variational parameters found for the 1010 RG state are quite different: they separate into two sets like in the VB type Hamiltonian \eqref{eq:H_VB}. The difference of the two $\{\varepsilon\}$ within each set are plotted in figure \ref{fig:h4_rg_params}. This RG state is well understood: one rapidity is trapped within each pair of near-degenerate $\{\varepsilon\}$. As the single-particle energies in a set get closer and closer, the RG pairs explicitly become bonding pairs \eqref{eq:bonding_pairs}, giving a GVB-PP state. For H$_6$, the optimal RG state found was 101010, while for H$_8$ it was 10101010. Again, we refer to these particular RG states as N\'{e}el RG states. The variational parameters $\{\varepsilon\}$ again arrange themselves into near-degenerate pairs well separated in energy relative to $g$. These systems are weakly interacting valence-bonds. The OO-DOCI orbitals shown in refs.\cite{limacher:2014a,ward_thesis} make this evident: they are bonding/anti-bonding pairs each localized on two neighboring hydrogens.

For the 1010 RG state, $\frac{\Delta \varepsilon}{\vert g \vert}$ seems to decay exponentially with $r$. Indeed, a plot of $\ln (\frac{\Delta \varepsilon}{\vert g \vert})$ is linear with $r$, and a least squares regression for each pair in H$_4$, H$_6$, H$_8$ is summarized in table \ref{table:regression}.
\begin{table}[ht!]
\centering
\begin{tabular}{l|c|c|c|c|c|c}
& $a$ & $b$ & $\sigma_a$ & $\sigma_b$ & $R^2$ & $r_0$   \\
\hline
H$_4$ (1) & -1.013 & 3.164 & 0.002 & 0.009 & 0.9996 & 3.122 \\
H$_4$ (2) & -1.013 & 3.165 & 0.002 & 0.009 & 0.9996 & 3.123 \\
H$_6$ (1) & -1.008 & 3.124 & 0.001 & 0.009 & 0.9997 & 3.101 \\
H$_6$ (2) & -1.008 & 3.125 & 0.001 & 0.009 & 0.9997 & 3.101 \\
H$_6$ (3) & -1.057 & 3.49  & 0.004 & 0.02  & 0.997  & 3.30 \\
H$_8$ (1) & -1.008 & 3.12  & 0.002 & 0.01  & 0.9996 & 3.10 \\
H$_8$ (2) & -1.009 & 3.13  & 0.002 & 0.01  & 0.9995 & 3.11 \\
H$_8$ (3) & -1.050 & 3.44  & 0.004 & 0.02  & 0.998  & 3.27 \\
H$_8$ (4) & -1.051 & 3.45  & 0.003 & 0.02  & 0.998  & 3.28 \\
\end{tabular} 
\caption{\label{table:regression} Linear regression of $\ln \Delta \varepsilon$ in units of $\vert g \vert$ as a function of $r$: $\ln \frac{ \Delta \varepsilon}{\vert g \vert} = a r + b$. Standard errors of the slope ($\sigma_a$), and y-intercept ($\sigma_b$) as well as the correlation coefficient ($R^2$) are reported. The x-intercept is computed as $r_0 = -\frac{b}{a}$.}
\end{table}
The x-intercept, $r_0$, represents where $\ln \frac{ \Delta \varepsilon}{\vert g \vert}=0$, or $\Delta \varepsilon = \vert g \vert$. At $r_0$, the gap is thus equal to the pairing strength. This point marks a transition \emph{for the particular pair} from weak to strong correlation. For the two pairs in H$_4$, this occurs around $r=3.1$, which is where the 1100 RG state performs the worst. The weakly correlated region is more or less RHF plus pairs, while the strongly correlated limit is GVB-PP. A 1010 RG state captures this transition smoothly. A 1100 RG state is a reasonable approximation in both limits but does not effectively describe the transition. From table \ref{table:regression}, one can see that this transition doesn't happen simultaneously for all pairs, but there is structure. Such patterns merit further study.

\begin{figure} [ht!]
	\begin{subfigure}{\textwidth}
		\centering
		\includegraphics[width=0.49\textwidth]{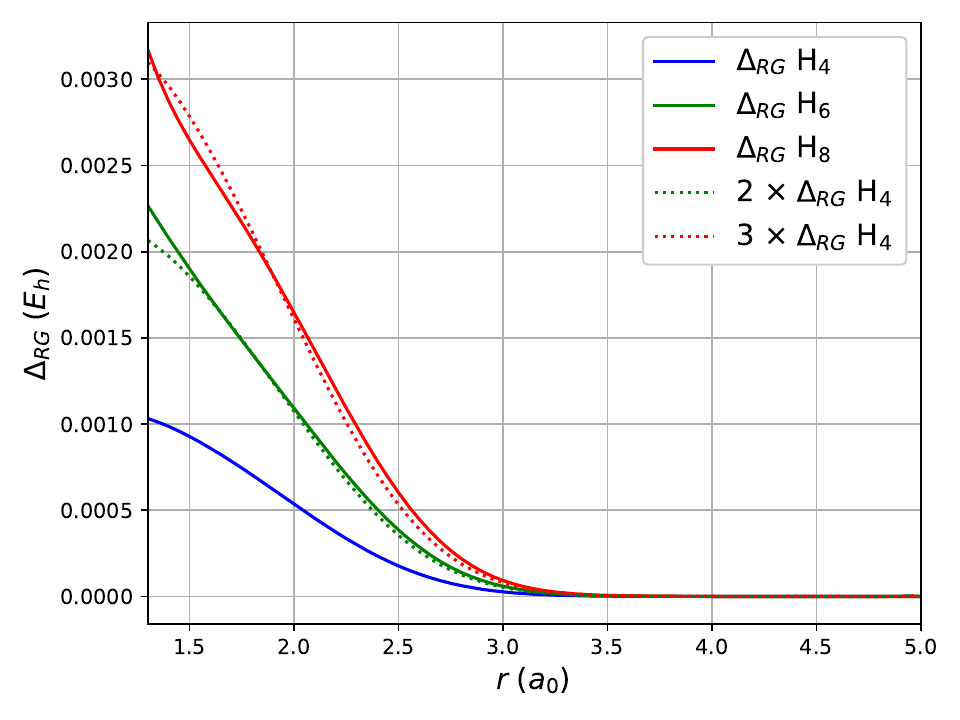} 
	\end{subfigure}
	\caption{Errors of variational RG treatment of linear hydrogen chains with respect to OO-DOCI. Results were computed with the OO-DOCI orbitals in the STO-6G basis. Dotted lines represent multiples of the error for H$_4$.}
	\label{fig:rg_h468}
\end{figure}

\begin{figure} [ht!]
	\begin{subfigure}{\textwidth}
		\centering
		\includegraphics[width=0.49\textwidth]{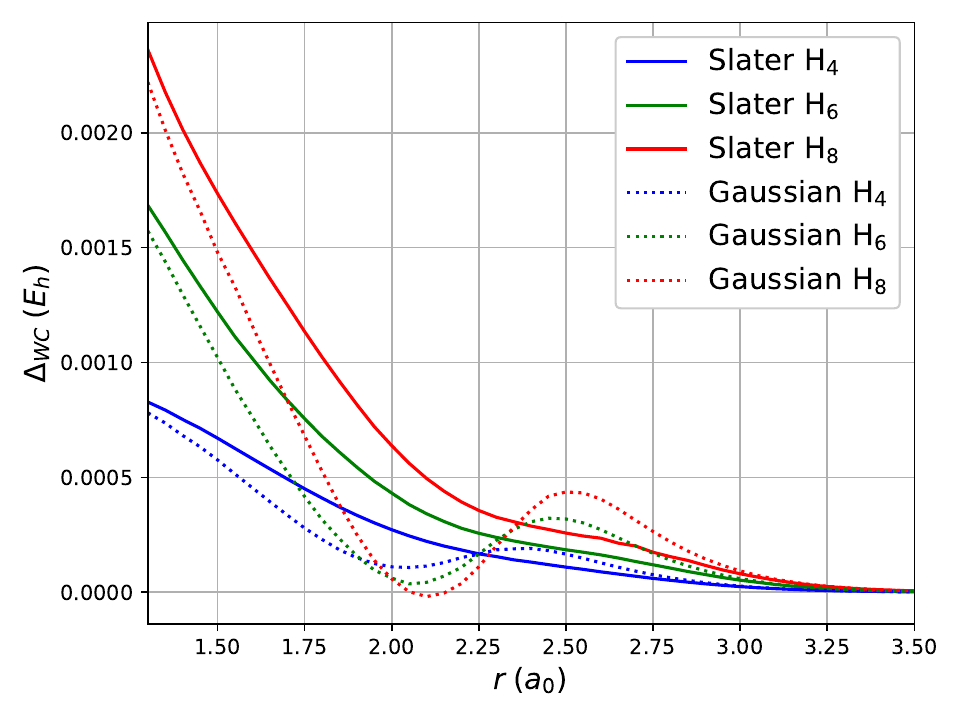} 
	\end{subfigure}
	\caption{Residual errors of natural orbital functional for seniority-zero weak correlation with Slater damping (solid lines) and Gaussian damping (dotted lines).}
	\label{fig:wc_h468}
\end{figure}

Now, given a variational RG approximation to OO-DOCI, can one improve systematically? First, the errors for N\'{e}el RG states
\begin{align}
\Delta_{RG} = E_{RG} - E_{OO-DOCI}
\end{align}
are shown for H$_4$, H$_6$, and H$_8$ in figure \ref{fig:rg_h468}.
As can be seen, $\Delta_{RG}$ seems to scale linearly with the number of pairs. This error is entirely in the seniority-zero sector and must represent weak correlation between the RG pairs. It is not unreasonable to approximate this error with natural orbital occupation number functionals. With the Generalized Natural Orbital Functional (GNOF),\cite{piris:2021,mitxelena:2022} Piris approximates this error as an update to the $P_{kl}$ contributions
\begin{align}
E_{WC} = - \sum^N_{k,l=1} {'} \sqrt{\gamma_k \gamma_l f(\gamma_k) f(\gamma_l)} V_{kkll},
\end{align}
in terms a damping function of the 1-RDM elements $f(\gamma_k)$. The primed summation leaves out diagonal terms, terms from the same VB pair, and terms with $\gamma_k$ and $\gamma_l$ near 1. While Piris employs a Gaussian, we have suggested\cite{johnson:2023} using a Slater exponential decay
\begin{align}
f^G(\gamma_k) &= \exp \left(- \left( \frac{\gamma_k}{0.02\sqrt{2}} \right)^2 \right) \\
f^S(\gamma_k) &= \exp \left(- \frac{\gamma_k}{0.02} \right)
\end{align}
as damping functions. From the residual errors, 
\begin{align}
\Delta_{WC} = E_{RG} + E_{WC} - E_{OO-DOCI} = \Delta_{RG} - \vert E_{WC} \vert
\end{align}
shown in figure \ref{fig:wc_h468},
we can see that neither functional is fantastic. We argued that the Slater-type is to be preferred as the error decays monotonically, whereas the Gaussian-type has oscillatory behavior and over-correlates for larger systems.

By variationally optimizing the parameters $\{\varepsilon\}$ for a particular RG state, we obtain a complete set of states. The direct way to account for the missing effects is to build the Coulomb Hamiltonian in the basis of RG states and diagonalize. From a given RG reference state, we define \emph{pair-singles} as RG bitstrings that differ from the reference by one 1 and one 0, \emph{pair-doubles} as RG bitstrings that differ from the reference by two 1s and two 0s etc. Looking at the expressions for the TDM elements, each state couples with each other RG state suggesting that a truncated CI approach is hopeless. However, computing the TDM elements numerically reveals that a given RG bitstring couples appreciably with its pair-singles and pair-doubles, but almost not at all with higher excitations.\cite{faribault:2010,johnson:2021} Thus it is reasonable to perform a CI with one RG state and its pair-singles, RGCIS, or one RG state and its pair-singles and pair-doubles, RGCISD.
\begin{figure} [ht!]
	\begin{subfigure}{\textwidth}
		\centering
		\includegraphics[width=0.49\textwidth]{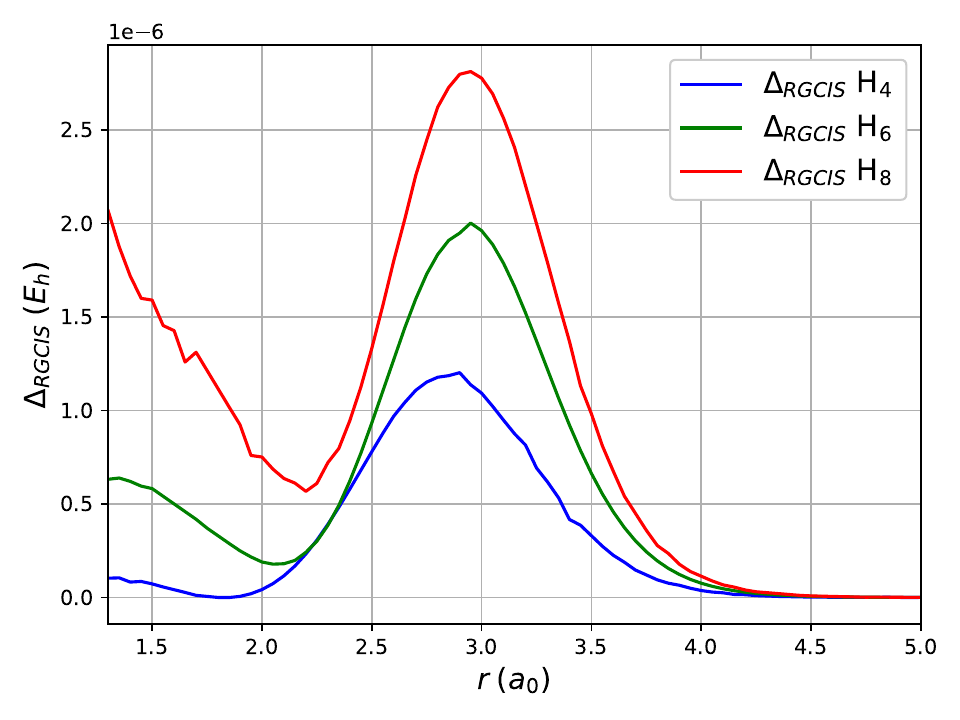} \hfill
		\includegraphics[width=0.49\textwidth]{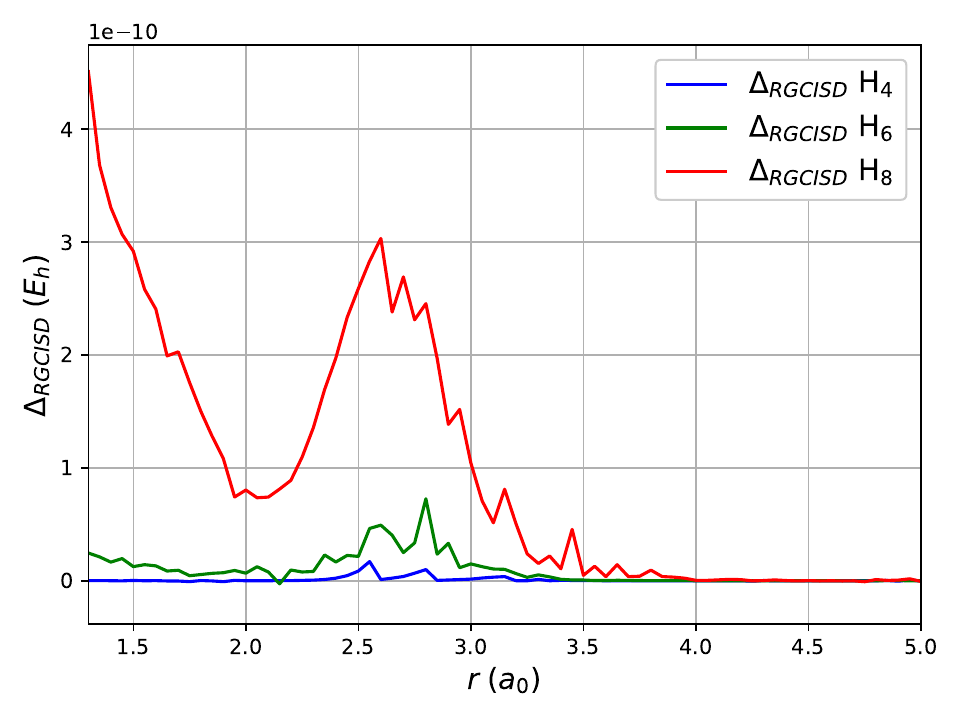}
	\end{subfigure}
	\caption{Errors of RGCI built from N\'{e}el RG state plus pair-singles (left) and N\'{e}el RG state plus pair-singles and pair-doubles (right) with respect to OO-DOCI. Results computed with the OO-DOCI orbitals in the STO-6G basis.}
	\label{fig:rgci_h468}
\end{figure}
The remaining errors
\begin{align}
\Delta_{RGCIS}  &= E_{RGCIS} - E_{OO-DOCI} \\
\Delta_{RGCISD} &= E_{RGCISD} - E_{OO-DOCI}
\end{align}
are shown in figure \ref{fig:rgci_h468}. One can see that RGCIS is a large improvement, while RGCISD is numerically OO-DOCI.

\section{Larger Context} \label{sec:context}
In this section we take a step back to show how RG states fit into the larger context. Generalizations are possible, and we describe only those restricted to the Lie algebra su(2). When required, RG states reduce to well known geminal wavefunction models of quantum chemistry. 
\subsection{Gaudin Algebra}
When presented with RG states, a natural question to ask is where the rational functions come from. Are there other types of solution? The situation is more or less completely classified when one introduces what is now called the Gaudin algebra\cite{dukelsky:2004,ortiz:2005} of objects $S^x(u)$, $S^y(u)$, and $S^z(u)$ defined by the structure
\begin{subequations}
\begin{align}
[S^x(u),S^y(v)] &= i (Y(u,v)S^z(u) - X(u,v) S^z(v)) \\
[S^y(u),S^z(v)] &= i (Z(u,v)S^x(u) - Y(u,v) S^x(v)) \\
[S^z(u),S^x(v)] &= i (X(u,v)S^y(u) - Z(u,v) S^y(v))
\end{align}
\end{subequations}
in terms of complex functions $X,Y,Z$. It is understood that 
\begin{align}
[S^x(u),S^x(v)] = [S^y(u),S^y(v)] = [S^z(u),S^z(v)] = 0.
\end{align}
For these objects to close a Lie algebra, the commutator must be associative, which forces 
\begin{align} \label{eq:YBE}
X(u,v) Y(v,w) + Y(w,u) Z(u,v) + Z(v,w) X(w,u) = 0.
\end{align}
Usually these are called Gaudin integrability conditions. They are a particular limit of quantum Yang-Baxter equations (YBE), so they are often referred to as YBE as well. Before looking at their solutions, we'll define the \emph{transfer matrix}
\begin{align}
S^2(u) = S^x(u) S^x(u) + S^y(u) S^y(u) + S^z(u) S^z(u).
\end{align}
With this structure, it is straightforward to verify that
\begin{align}
[S^2(u),S^2(v)] = 0
\end{align}
that is the transfer matrix commutes with itself regardless of the complex parameter. An integrable model may be built by choosing $N$ number of particular values for $u$.

There are a few different solutions of the YBE \eqref{eq:YBE}. First, Gaudin\cite{gaudin:1976} supposed that the functions were antisymmetric, only depending on the difference between two complex numbers, i.e. $X(u,v) = -X(v,u) = X(u-v)$. With this restriction, there are three types of solutions. First, if $X=Y=Z$, we get the \emph{rational} or isotropic solution
\begin{align}
X(u-v) = \frac{1}{u-v},
\end{align}
and the YBE become
\begin{align}
\frac{1}{u-v}\frac{1}{v-w} + \frac{1}{w-u}\frac{1}{u-v} + \frac{1}{v-w}\frac{1}{w-u} = 0
\end{align}
which is understood as partial fractions. These solutions of the YBE are typically called the \emph{rational} model and give the reduced BCS Hamiltonian in particular. 

For $X=Y \neq Z$, there are anisotropic solutions in terms of trigonometric or [hyperbolic] functions
\begin{align}
X(u-v) &= \frac{1}{\sin [h] (u-v)} \\
Z(u-v) &= \cot [h] (u-v).
\end{align}
The construction for the XXZ solutions are near-identical, but the physical models they describe are different from the rational models, and from each other. The hyperbolic version is the more studied of the two for the moment. It is usual to parametrize the solution as
\begin{align}
X(u-v) &= \frac{\sqrt{uv}}{u-v} \\
Z(u-v) &= \frac{u+v}{u-v}
\end{align}
which leads to Hamiltonians of the type
\begin{align} \label{H_XXZ}
\hat{H}_{XXZ} = \frac{1}{2} \sum^N_{i=1} \eta_i \bar{\eta}_i \hat{n}_i - G \sum^N_{i,j=1} \eta_i \bar{\eta}_j S^+_i S^-_j
\end{align}
which are similar to the reduced BCS Hamiltonian, but with a factorizable interaction in terms of complex quantities $\{\eta\}$. These solutions of the YBE are generally known as \emph{XXZ RG models}.\cite{ibanez:2009,dunning:2010,rombouts:2010,dukelsky:2011,marquette:2013,vanraemdonck:2014,claeys:2015,claeys:2016} 

If we insist $X\neq Y\neq Z$, the YBE has a solution in terms of elliptic functions 
\begin{align}
X(u-v) &= \frac{1 + \kappa \text{sn}^2(u-v)}{\text{sn}(u-v)} \\
Y(u-v) &= \frac{1 - \kappa \text{sn}^2(u-v)}{\text{sn}(u-v)} \\
Z(u-v) &= \frac{\text{cn}(u-v) \text{dn}(u-v)}{\text{sn}(u-v)}
\end{align}
where $\kappa$ is the elliptic modulus and $\text{sn}$, $\text{cn}$, and $\text{dn}$ are Jacobi elliptic functions. These models are less explored and are much more fragile. RG states exist, but they are much more complicated and are not necessarily the complete set of eigenvectors.\cite{takhtadzhan:1979,sklyanin:1996,sklyanin:1999,babujian:1999,gould:2002,esebbag:2015,relano:2016}

There are also solutions of the YBE that don't require the functions to be differences of two complex numbers
\begin{align}
X(u,v) &= \frac{\sqrt{u+j_x}\sqrt{v+j_y}\sqrt{v+j_z}}{v-u} \\
Y(u,v) &= \frac{\sqrt{v+j_x}\sqrt{u+j_y}\sqrt{v+j_z}}{v-u} \\
Z(u,v) &= \frac{\sqrt{v+j_x}\sqrt{v+j_y}\sqrt{u+j_z}}{v-u}.
\end{align}
Three arbitrary parameters $j_x,j_y,j_z$ can be chosen to give different physical models. The classification and solution of these \emph{non-skew-symmetric} models is an ongoing project of Skrypnyk.\cite{skrypnyk:2015,skrypnyk:2016a,skrypnyk:2016b,skrypnyk:2017a,skrypnyk:2017b,skrypnyk:2018,skrypnyk:2019,claeys:2019,shen:2020,skrypnyk:2021a,skrypnyk:2021b,skrypnyk:2022a,skrypnyk:2022b,skrypnyk:2022c,skrypnyk:2023a,skrypnyk:2023b}

\subsection{Conserved Charges} \label{sec:conserved_charges}
The ABA can be applied to the transfer matrix $S^2(u)$ in the exact same manner as for the reduced BCS Hamiltonian. In both the XXX and XXZ cases, particle number will be conserved, and we transform the Gaudin algebra objects
\begin{align}
S^{\pm}(u) = S^x(u) \pm i S^y(u).
\end{align}
The structure of the Gaudin algebra with these objects becomes
\begin{subequations}
\begin{align}
[S^+(u),S^-(v)] &= 2X(u,v) (S^z(u) -S^z(v)) \\
[S^z(u), S^{\pm}(v)] &= \pm (X(u,v)S^{\pm}(u) - Z(u,v) S^{\pm}(v)). 
\end{align}
\end{subequations}
The RG states are built with $S^+(u)$ acting on the vacuum $\ket{\theta}$, which is itself an eigenvector of $S^z(u)$
\begin{align}
S^z(u) \ket{\theta} = \alpha (u) \ket{\theta}.
\end{align}
The vacuum eigenvalue $\alpha(u)$ is a free parameter, and choosing it gives particular physical models. $\alpha(u)$ is the same parameter that appears in the matrix \eqref{eq:slv_mat} of Slavnov's theorem for the scalar products in terms of rapidities. By acting with $S^2(u)$ on  the RG states \eqref{eq:rg_state}, and using the Gaudin algebra structure to move $S^2(u)$ to the right, one finds they are eigenvectors of $S^2(u)$ provided that for each of the rapidities
\begin{align} \label{eq:abstract_rich}
\alpha(u_a) + \sum^M_{b (\neq a)=1} Z(u_b,u_a) = 0
\end{align}
there are non-linear equations to solve.

We will now obtain a complete set of conserved quantities and show that they produce the reduced BCS Hamiltonian. The approach we follow is that of Sklyanin.\cite{sklyanin:1989} For another approach, see ref.\cite{cambiaggio:1997} With a set of physical representations $S^+_i,S^-_i,S^z_i$ of su(2) and parameters $\varepsilon_i$, we can commit to a local representation of the Gaudin algebra
\begin{align}
S^+(u) &= \sum^N_{i=1} \frac{S^+_i}{u-\varepsilon_i} \\
S^-(u) &= \sum^N_{i=1} \frac{S^-_i}{u-\varepsilon_i} \\
S^z(u) &= \frac{1}{g} - \sum^N_{i=1} \frac{S^z_i}{u-\varepsilon_i}.
\end{align}
It is straightforward to show that this choice satisfies the same Gaudin algebra relations using partial fractions to separate the product into a sum. With this representation, the vacuum eigenvalue $\alpha(u)$ becomes
\begin{align}
\alpha(u) = \frac{1}{g} + \frac{1}{2} \sum^N_{i=1} \frac{1}{u - \varepsilon_i}
\end{align}
and hence the conditions \eqref{eq:abstract_rich} become Richardson's equations. The transfer matrix becomes
\begin{align} \label{eq:local_tmat}
S^2(u) = \frac{1}{g^2} - \frac{2}{g} \sum^N_{i=1} \frac{\hat{R}_i}{u-\varepsilon_i} + \sum^N_{i=1} \frac{S^2_i}{(u-\varepsilon_i)^2},
\end{align}
where $S^2_i$ \emph{is} the Casimir operator for the $i$th copy of su(2)
\begin{align}
S^2_i = S^z_i S^z_i + \frac{1}{2}(S^+_iS^-_i + S^-_i S^+_i).
\end{align}
Its action is a constant, as is the first term in \eqref{eq:local_tmat}. The remaining terms involve the \emph{conserved charges}
\begin{align}
\hat{R}_i = S^z_i - g \sum^N_{j (\neq i)=1} \frac{S^z_iS^z_j + \frac{1}{2}(S^+_iS^-_j + S^-_iS^+_j)}{\varepsilon_i - \varepsilon_j}.
\end{align}
It is tedious but straightforward to show that these objects commute
\begin{align}
[\hat{R}_i,\hat{R}_j] = 0, \quad \forall i,j
\end{align}
so that we can find their mutual eigenvectors. Further, taking the linear combination
\begin{align}
\sum^N_{i=1} \varepsilon_i \hat{R}_i = \hat{H}_{BCS} + \frac{g}{2} \left( \left(M - \frac{N}{2}\right) \left( M- \frac{N}{2} - 1 \right) - \frac{3}{4} N \right)
\end{align}
gives the reduced BCS Hamiltonian plus a constant to be ignored. If $\{v\}$ are a solution of Richardson's equations, then $\ket{\{v\}}$ is an eigenvector of the conserved charges
\begin{align}
\hat{R}_i \ket{\{v\}} = \frac{1}{2} \left( V_i - 1 - \frac{g}{2} \sum^N_{j (\neq i)=1} \frac{1}{\varepsilon_i - \varepsilon_j}  \right) \ket{\{v\}}
\end{align}
where $V_i$ are the EBV \eqref{eq:ebv}. The name \emph{eigenvalue-based variables} comes from their inclusion in the eigenvalues of the conserved charges.

\subsection{Geminal Wavefunctions and RG States}
In this section we'll classify the traditional geminal wavefunctions of quantum chemistry as particular RG states. This classification is more than mere stamp collecting: it implies that these wavefunctions are eigenvectors of model Hamiltonians. Their weak excitations can be constructed so that perturbation theories or CI treatments can be performed.

\subsubsection{Generalized Valence-Bond / Perfect Pairing}
We've used the term ``valence-bond'' quite liberally throughout the text so we should precise about the relationship between GVB-PP and RG states. The GVB-PP wavefunction\cite{goddard:1967,hay:1972,hunt:1972,goddard:1973,goddard:1978,beran:2005,small:2009,lawler:2010,small:2011,small:2012,small:2017,small:2018,wang:2019} factors into core, pairs, and unpaired contributions
\begin{align}
\ket{\text{GVB-PP}} = \Psi^{\dagger}_{core} \Psi^{\dagger}_{pairs} \Psi^{\dagger}_{unpaired} \ket{\theta}.
\end{align}
The core is often frozen, giving a Slater determinant of doubly occupied core orbitals. The unpaired part is usually a Slater determinant of unpaired electrons, but could be any configuration-state function. The pairs are localized entirely in two spatial orbitals each
\begin{align}
\Psi^{\dagger}_{pairs} = \prod^M_{a=1} \left( C_{a_1} S^+_{a_1} + C_{a_2} S^+_{a_2} \right).
\end{align}
The corresponding RG bitstring is constructed: each of the pairs contributes a substring of `10', and the remaining virtual orbitals each contribute a `0'. For $M_p$ pairs, and $M_v$ virtual orbitals that contribute to the core, pairs, or unpaired portions, GVB-PP corresponds to the RG state $(10)^{M_p} 0^{M_v}$ acting on the vacuum $\ket{\tilde{\theta}}= \Psi^{\dagger}_{core}\Psi^{\dagger}_{unpaired}\ket{\theta}$. The core can be unfrozen. For $M_c$ core pairs, we get the RG state $1^{M_c} (10)^{M_p} 0^{M_v}$ acting on the vacuum $\ket{\tilde{\theta}} = \Psi^{\dagger}_{unpaired}\ket{\theta}$. For these RG states to become GVB-PP, the single-particle energies $\{\varepsilon\}$ have to arrange themselves accordingly: for each VB pair, there are two $\{\varepsilon\}$ close in energy relative to $g$, while the gap between pairs is large compared with $g$. The $\{\varepsilon\}$ for the occupied and virtual orbitals are non-degenerate, and the spacing between them should be large compared with $g$. 
Once GVB-PP has been identified with an RG state, its weak excitations are clear: there are pair-singles obtained by exchanging a 1 with a 0, pair-doubles obtained by exchanging two 1s with two 0s, etc. From the RGCI treatment of the 1D hydrogen chains, and references\cite{faribault:2010,johnson:2021,johnson:2023}, it is clear that the dominant contributions are from pair-singles, while pair-doubles give a non-zero contribution. Pair-triples and above give no discernible benefit. Unlike Slater determinants, there are distinct types of excitation: \emph{swaps} corresponding to $(10)\rightarrow(01)$ and \emph{transfers} between pairs corresponding to $(10)(10)\rightarrow(11)(00)$. From the limited results we have, it appears that transfers are important while swaps are not. This will be studied in detail. 

The antisymmetrized product of strongly orthogonal geminals (APSG)\cite{hurley:1953,kutzelnigg:1964,nicely:1971} is a similar wavefunction except that each of the pairs in $\Psi^{\dagger}_{pairs}$ has more than two spatial orbitals. Like in GVB-PP, each spatial orbital appears in only one pair. For an APSG pair containing $k$ spatial orbitals, the corresponding RG state would have a string of one 1 and $(k-1)$ 0s. APSG has been used to describe bond-dissociation processes in chemistry for a long time.\cite{kutzelnigg:2010,kobayashi:2010,kutzelnigg:2012,surjan:2012,zoboki:2013,pernal:2014,jeszenszki:2014,pastorczak:2015,margocsy:2018,pernal:2018,pastorczak:2018,pastorczak:2019} The natural orbital functional PNOF5 is equivalent to APSG.\cite{piris:2011,pernal:2013} More recently, PNOF7\cite{piris:2017,piris:2019,mitxelena:2020a,mitxelena:2020b,rodriguez:2021,piris:2021} attempts to include correlation between APSG pairs. It is not N-representable, but can be qualitatively understood as APSG pairs interacting with one another like AGP pairs.\cite{moisset:2022a} The notion of strong-orthogonality, that each spatial orbital belongs in one specific pair, can be relaxed systematically, though the cost increases immediately.\cite{cassam:2006,cassam:2010,cassam:2012}

\subsubsection{Antisymmetrized Geminal Power}
AGP is a pair-wavefunction in which all of the pairs are the same. We have already seen the ground state of the reduced BCS Hamiltonian in the attractive limit is extreme AGP \eqref{eq:eAGP}. The first excited states are products of $M-1$ collective pairs \eqref{eq:collective_pair} with a single bonding pair \eqref{eq:bonding_pairs}. The second excited states are products of $M-2$ collective pairs with two bonding pairs etc. The only difficulty is that the bonding pairs are linearly dependent. This problem has a simple general solution that we will present as part of an upcoming contribution.

General AGP, whose coefficients $\{C\}$ are arbitrary
\begin{align}
\ket{\text{AGP}} = \left( \sum^N_{i=1} C_i S^+_i \right)^M \ket{\theta},
\end{align}
is not an eigenvector of the reduced BCS Hamiltonian. It is however the ground state of the XXZ Hamiltonian \eqref{H_XXZ} with
\begin{align}
C_i = -\frac{1}{\bar{\eta}_i}
\end{align} 
at the Moore-Read\cite{moore:1991,read:2000} point $G^{-1} = N-M+1$, where the rapidities of the RG ground state all collapse to zero. As this is the ground state, its bitstring is $1^M 0^{N-M}$ and its single and double excitations are obtained by exchanging 1s with 0s in the bitstring. While the ground state has all rapidities identically equal to zero, none of the excited states have \emph{any} rapidities equal to zero. Rather, they are obtained from the corresponding set of Richardson's equations for $\{u\}$
\begin{align}
N + \sum^N_{i=1} \frac{u_a + \eta_i \bar{\eta}_i}{u_a - \eta_i \bar{\eta}_i} + 2 \sum^M_{b (\neq a)=1} \frac{u_b + u_a}{u_b - u_a}.
\end{align}

AGP is a model wavefunction with a long history in quantum chemistry,\cite{coleman:1965,coleman:1989,coleman:1997} as well as nuclear structure\cite{rowe_book,rowe:1991,chen:1995,otsuka:2001,rowe:2001,lu:2022} and condensed matter theory.\cite{schrieffer_book,dukelsky:2016} AGP is not size-consistent\cite{linderberg:1980} which would severely limit its use in repulsive Coulomb systems. However, it is efficient to build upon with Jastrow factors, which can correct this deficiency.\cite{neuscamman:2012,neuscamman:2013,neuscamman:2016} It has long been known that AGP is a vacuum for the particle-hole random-phase approximation.\cite{linderberg:1977,ohrn:1979,ortiz:1981,sarma:1989,dukelsky:2019} Presently, the group of Scuseria is developing post-AGP methods and applying them to spin models in particular.\cite{henderson:2019,khamoshi:2019,henderson:2020,khamoshi:2020,dutta:2020,khamoshi:2021,dutta:2021,liu:2023}

\subsubsection{Antisymmetrized Product of 1-reference orbital Geminals / pair-Coupled Cluster Doubles}
The antisymmetrized product of 1-reference orbital geminals (AP1roG)\cite{limacher:2013} is a pair wavefunction incorporating an occupied/virtual separation of orbitals. It is thus tied to the basis of Slater determinants. Each pair has a contribution from one occupied orbital, and all of the virtual orbitals
\begin{align}
\ket{\text{AP1roG}} = \prod^M_{a=1} \left( S^+_a + \sum_{i \in \text{virt}} C_{ai} S^+_i \right) \ket{\theta}.
\end{align}
As each spatial orbital can contain at most one pair, AP1roG is equivalent to pair-coupled cluster doubles (pCCD)\cite{stein:2014}
\begin{align}
\ket{\text{pCCD}} = \exp \left( \sum_{\substack{i \in \text{occ} \\ v \in \text{virt} }} t^v_i S^+_v S^-_i \right) \ket{\text{HF}}.
\end{align}
This wavefunction is not feasible variationally, and as it is built in a basis of Slater determinants it does not have a compact expression in terms of RG states. AP1roG/pCCD is the method to beat at the moment. It is cheap to solve by projection and has been applied to many bond-breaking processes.\cite{limacher:2014a,limacher:2014b,henderson:2014a,henderson:2014b,boguslawski:2014a,boguslawski:2014b,boguslawski:2014c,tecmer:2014} While it is difficult to define systematic corrections, methods built on AP1roG/pCCD are promising.\cite{boguslawski:2015,degroote:2016,boguslawski:2016a,boguslawski:2016b,boguslawski:2017,boguslawski:2019,nowak:2019,nowak:2021,boguslawski:2021,marie:2021,kossoski:2021,baran:2021} 

\section{Conclusion}
This chapter is meant as an overview of RG states keeping the complications to an absolute minimum. In the end, they are simple objects. Limiting cases of RG states are presented first to give an intuition before defining them formally. The best method to solve Richardson's equations is presented along with small examples demonstrating the evolution of the EBV and the rapidities. This method is simple to implement. Density matrix elements of RG states are shown to be simple to compute from the EBV (or the rapidities). Prototypical strongly correlated systems are shown to have a simple description in terms of a single RG state. This treatment is systematically improvable by including excited RG states hierarchically. 

At the same time, this chapter cannot be a complete presentation. For a full appreciation of the details we refer the interested reader to refs.\cite{moisset:2022a,faribault:2022} Open problems we are currently addressing, in particular degenerate $\{\varepsilon\}$, RG states with non-zero seniority, and variational optimization are not discussed. These problems have answers, but they are not yet complete. 

Ultimately, RG states are analogous to HF Slater determinants for weakly correlated electrons. They represent mean-field behavior and are systematically improvable. Anything that can be done with an HF reference can be adapted to an RG reference. 

\section{Acknowledgements}
The author thanks Paul W. Ayers, Peter A. Limacher, Stijn De Baerdemacker, Patrick Bultinck, Dimitri Van Neck, Alexandre Faribault, A. Eugene DePrince III, Thomas Henderson and Gustavo E. Scuseria for many discussions over the course of this project. This work was supported by the Natural Sciences and Engineering Research Council of Canada and the Digital Research Alliance of Canada.

\bibliography{chapter}
\bibliographystyle{unsrt}

\end{document}